\documentclass[a4paper,UKenglish,cleveref, autoref, thm-restate]{lipics-v2021}

\title{Proof-Relevant Interpolation for
Circular Proofs in Linear Logic}

\author{Daniel Osorio-Valencia}{Dipartimento di Informatica, Università di Torino, Italy}{}{}{}
\author{Alexis Saurin}{IRIF, Université Paris Cité \& CNRS and INRIA Picube, France}{}{}{}

\authorrunning{D. Osorio-Valencia and A. Saurin}

\Copyright{Daniel Osorio-Valencia and Alexis Saurin}

\ccsdesc[500]{Theory of computation~Proof theory}

\keywords{Circular Proofs, Cut-elimination, Interpolation, Linear Logic, Mu-calculus} 

\hideLIPIcs
\nolinenumbers

\usepackage{packages}
\usepackage{macros}

\begin{document}

\maketitle

\begin{abstract}
First introduced for proving Craig's interpolation theorem, Maehara's method provides a constructive, proof-theoretic procedure for extracting interpolants in sequent calculus proofs. Building on this method, Saurin developed a ``proof-relevant'' interpolation theorem for first-order linear logic, in the sense that the interpolated proof is obtained by cut-introduction steps that can be reversed via cut-elimination. In this work, we extend Saurin's proof-relevant interpolation to linear logic proofs with cuts as well as to circular proofs in linear logic with fixed-points. We specifically show how Maeheras's method can be extended in a proof-relevant way to work on non-cut-free and non-well-founded (but regular) proof objects, and guarantee that the interpolation procedure preserves strong validity.
We also study conditions under which one can interpolate system L terms that are not necessarily in normal form.
\end{abstract}

% !TEX root = main.tex
\section{Introduction}

Logical results can often be considered at various levels, depending on whether one has a model-theoretic or proof-theoretic 
perspective. The basic focus of logic, the study of logical consequence,  can indeed be considered from the point of view
%  This expresses the fact that the basic focus or logical studies, logical consequence, can be considered at the level
 (i) of models -- considering validity statement -- or (ii) of provability -- existence of a proof in a given deductive system -- possibly (iii) of more specific provability statements -- existence of cut-free or focused proofs typically -- but also at the level (iv) of proofs themselves or even (v) of the proofs' dynamics. Such perspectives are typical in type theory but also prove to be common in proof theory. 
In particular, when focusing on the deductive side, those distinctions partly amount to whether one consider a proof-relevant or proof-irrelevant versions of the results under consideration. 

\noindent{\bf Interpolation.}
Craig's Interpolation theorem~\cite{DBLP:journals/jsyml/Craig57,DBLP:journals/jsyml/Craig57a} -- a central result of first-order logic which found many applications both in mathematical logic~\cite{DBLP:journals/synthese/Feferman08} and in computer science~\cite{mcmillan-abstractions-from-proofs04,McMillan18,DBLP:journals/synthese/Lavalette08} -- is typically such a result that can be considered at various levels, with semantical or deductive versions of the statement of interpolation (often connected when one has a completeness theorem) and even a cut-free provability version that is the heart of Maehara's celebrated method. Indeed, Maehara's method~\cite{Maehara} happens to turn Craig's interpolation as a direct consequence of cut-admissibility, and is thus applicable to a wide range of contexts~\cite{Mints01,schutte,KOWALSKI_ONO_2017,Fitting84,Kuznets18,DBLP:conf/tbillc/AfshariL19,shamkanov2014,DBLP:journals/synthese/Feferman08,Vaananen08} which may explain why this method is popular among proof-theorists.

\noindent{\bf From Maehara to Proof-relevant interpolation.}
The idea behind Maehara's method is to consider a cut-free proof of the entailment to interpolate and to show that each sequent of the derivation can be interpolated, generating an interpolant for the conclusion sequent. 

Maehara's method was recently revisited by Saurin~\cite{Saurin} who gave a series of proof-relevant interpolation theorems based on cut-introduction, by identifying the structure of proof-relevant interpolation situation, \PRIS. (See \Cref{sec:proofRelIntLL} in the background.)
Those results prove, for several logics, that if $\pi$ is a cut-free sequent proof of  $A\vdash B$, one can find an interpolant formula $I$ in the common vocabulary of $A$ and $B$ and interpolating proofs $\pi_1,\pi_2$ of $A\vdash I$ and $I\vdash B$ respectively such that $\pi_1$ composed with $\pi_2$ cut-reduces to $\pi$. 
Saurin's result can precisely be viewed as the proof-relevant version of Maehara's method and it generalizes previous results by \v Cubri\'c~\cite{Cubricphd,DBLP:journals/aml/Cubric94}. In the following we shall indistinctly speak of proof-relevant or \emph{Maehara-\v Cubri\' c interpolation} for this style of interpolation results.
Similarly to Saurin's approach, \v Cubri\'c' result is a proof-relevant version of Prawitz' proof of the interpolation theorem that adapted Maehara's method for natural deduction~\cite{Prawitz1965-PRANDA}. In this way,  \v Cubri\'c' result provides an interpolation result for simply typed $\lambda$-terms in normal form\footnote{Namely: if $t$ is \emph{normal} and such that $x:A\vdash t: B$, there exists a type $I$ of which the atomic types occur both in $A$ and $B$ as well as two normal terms $u,v$ such that: (i) $x: A \vdash u: I$, (ii) $\vdash v: I\rightarrow B$ and (iii) $(v)u \longrightarrow_\beta^\star t$.}.

We  stress this fundamental limitation of  \v Cubri\'c' approach to $\lambda$-terms in normal forms. 

\noindent{\bf The assumptions of Maehara.}
Indeed, despite being general and flexible, Maehara's original method strongly relies on several structural assumptions: 
\begin{enumerate}
\item %(iii) 
the inductive character of
sequent proofs, to reason by induction on the subderivations.
\item %(i) 
cut-admissibility, to reason on cut-free proofs and benefit from the subformula property,  
\end{enumerate}

The restriction to cut-free proofs is unsatisfactory from both a logical and a computational point of view, indeed 
(i) the logic under study may not satisfy a full cut-elimination theorem ({\it e.g.} cut-admissibility up to analytic cuts), cut-elimination induces a size blow-up of proofs, resulting in much larger interpolants. 
(ii) from the computational perspective of \v Cubri\' c and Saurin's works, this restriction is a severe limitation: for instance, \v Cubri\' c result applies to normal forms only and therefore requires to first normalise $\lambda$-terms before interpolating.

\noindent{\bf Why interpolating circular proofs with cuts.}
In the present paper we are interested in strengthening interpolation results for fixed-point logics by proving their proof-relevant counterpart. In order to do that, we will show how one can weaken the two assumptions mentioned in the previous paragraph in order to do proof-relevant interpolation in presence of cuts as well as for circular proofs. Additionally, proof-relevant interpolation over circular proofs must also preserve the validity condition specified by the proof system, which ensures the soundness of circular proofs. This is not automatic, and addressing it is one of the main technical points of this paper. 
The focus on circular proofs adds a further motivation for interpolating proofs with cuts. Indeed, while \muLLi has a cut-elimination theorem, regularity of a proof is not preserved by cut-elimination in general\footnote{To understand why this is so, simply think that there is a circular proof in \muLLi that takes a natural number $n$ and returns a stream enumerating all natural number from $n$: when cuting this proof with the proof encoding $0$, we have a circular proof with cuts that has, as a normal form, the cut-free proof encoding the stream of natural numbers, which is of course not regular.}.

\noindent{\bf Related works.}
To the best of our knowledge, we present the first results on (i) proof-relevant interpolation  (ii) for circular proofs (iii) with cuts altogether. On the other hand, each of these three topics has been investigated independently in various prior works:
\begin{description}
\item[(i) Proof-relevant interpolation:] the most closely related work in this direction is of course Saurin's proof-relevant interpolation theorems for a variety of sequent calculi (and term-calculi for sequent calculus)~\cite{Saurin} that we extend here.
In the early 90s, \v Cubri\'c exploited Prawitz' interpolation theorem in natural deduction~\cite{Prawitz1965-PRANDA} to prove an interpolation result for the simply typed $\lambda$-calculus as well as a corresponding factorisation result for bicartesian closed categories~\cite{Cubricphd,DBLP:journals/aml/Cubric94} later extended by
Matthes~\cite{matthes-interpolation-01} to a natural deduction with general elimination rules. 
Recently, Veltri and Wan proved a proof-relevant interpolation theorem for the non-associative Lambek-calculus, formalized in Agda~\cite{VeltriW25}.
\item[(ii) Interpolation for fixed-point logics and circular proofs:]
After the seminar results by Agostino {\it et al}~\cite{DBLP:journals/jsyml/DAgostinoH00,DBLP:journals/japll/DAgostinoL06} on interpolation for the modal $\mu$-calculus, several works contributed to interpolation for the modal $\mu$-calculus using circular, or cyclic, proof systems both for Craig and Lyndon's interpolation~\cite{DBLP:conf/tbillc/AfshariL19,DBLP:conf/lics/KloibhoferV25} and Uniform interpolation~\cite{DBLP:conf/tableaux/AfshariLT21}. The techniques extending Maehara's method to circular proofs are due to Shamkanov in the setting of {G}ödel-{L}öb provability logic~\cite{shamkanov2014}.
Compared to our work, the absence of proof-relevance allow quite drastic modifications on proofs, for instance regularisation. 
\item[(iii) Interpolation for proofs with cuts:]
While several works aimed at extending interpolation theorems to incorporate some form of cuts~\cite{Carbone97,HetzlJalali24,KOWALSKI_ONO_2017,DBLP:conf/aiml/OnoS22,iris24lecturenotes,lolic24cuts,DBLP:conf/ictac/CorbardL25}, typically to prove interpolation in presence of analytic cuts and semi-analytic rules~\cite{JalaliTabatabaiJ25}, most did not directly investigate  how the cut \emph{inference} itself  impacts directly the interpolation process itself (that is the synthesis of the interpolant formula). Among the works on interpolation theorems in the presence of cuts, we should refer to Hetzl and Jalali's work on completeness of interpolation algorithms~\cite{HetzlJalali24}, van der Giessen's lecture notes on interpolation which prove interpolation in the presence of analytic cuts~\cite{iris24lecturenotes}, a short abstract by Lolic that reports some results interpolation with cuts~\cite{lolic24cuts} and of course Fitting's book that investigate interpolation for modal and intuitionistic logic and cuts~\cite{Fitting83}. 
\end{description}

\noindent{\bf Contributions and organisation of the paper.}
 \Cref{sec:preliminaries} provides some background on \muLL circular proofs and on proof-relevant interpolation. Then, we extend \LL{} proof-relevant interpolation to proofs with cuts in \Cref{sec:interpolate-cuts} thus providing the first proof-relevant treatment of proofs with cuts and generalising several Craig's interpolation methods with cuts; we also transfer this result to a computational system L generalising Saurin's result on normal system L terms. Last, we consider in \Cref{sec:proofRelIntCircProof} the problem of proof-relevant interpolation for circular proofs with cuts in propositional \muLLi. We restrict to a better behaved (but less general) class of circular proofs called \emph{strongly valid}, which we show to be closed under our interpolation procedure. This result provides a first step towards an interpolation theorem for all valid proofs in first order \muLLi. We conclude in \Cref{sec:conclusion}.

% !TEX root = main.tex
\section{Background}
\label{sec:preliminaries}

\subsection{Circular proofs for linear logic with least and greatest fixed-points}

For simplicifying the background presentation, we first introduce a common framework, \fomuLLi, in which our bacground notions are set. The following sections will focus on two fragments of this logic: the fixed-point-free fragment, \foLL, and the FO quantifier free, \muLLip respectively.

We assume a first-order language $\calL$ without equality nor function symbols for simplicity. The language of first-order linear logic formulas with least and greatest fixed-points (or \defname{\muLL formulas}) and its deduction system, are defined in the following way:

\begin{definition}[\muLL formulas]
\label{def:LLformulas}
\label{def:muLLformulas}
We assume to be given countable sets of (i) atomic formulas equipped with an involution inducing a partitioning between \defname{positive atoms}, written $a (\vec{x})$ (or just $a$), and \defname{negative atoms}, written $a^\perp (\vec{x})$ (or just $a^\perp$) and of (ii) fixed-point variables, written $X, Y, Z, \dots$ such that $X^\perp = X$. 
\muLL formulas are defined inductively as:
\\

$\begin{array}{lrll}
    F &::=\  & a^\perp(\vec{x}) ~\mid~\perp~~~\mid~\top ~\mid F\parr F ~\mid F\with F ~
    \mid \wn F
    \mid \forall x F
    &\textbf{neg. form.}\\
    & \mid& a(\vec{x}) ~\mid ~1~~\mid ~0~\mid F\tensor F ~\mid F\oplus F ~
    \mid \oc F \mid \exists x F
    &\textbf{pos. form.}\ \\
&\mid &X  \mid \mu X.  F \mid \nu X. F & \textbf{fixed-point formulas}\ \\
%&\mid &\wn F &\mid \oc F &&&& \textbf{exponential formulas}\ \\
\end{array}$

  A \defname{\LL formula} is a formula build using only the first two lines of the above grammar. \defname{Linear negation} is defined as extending the \LL involution on \LL formulas (changing $a^\perp$, $\bot$, $\top$, $\parr$, $\with$,
  $\forall$,
  $\wn$ into $a$,  $\llone$, $\llzero$, $\tensor$, $\oplus$,
  $\exists$,
  $\oc$ respectively) to fixed-points constructs: $(\mu X.F)^{\bot} = \nu X.F^{\bot}$ (note that $\mu, \nu$ bind fixed-point variables). A \defname{one-sided \muLL sequent} is an ordered list of \muLL formulas usually written $\vdash \Gamma$.
\end{definition}

 % !TEX root = main.tex
\def\tinfex{\begin{prooftree}
\Hypo{\vdash \Gamma, G\tikzmark{tex2}, F, \Delta}
\Infer{1}[\rex]{\vdash \Gamma, F\tikzmark{tex1}, G, \Delta}
\end{prooftree}}
  
\def\tinfcontr{\begin{prooftree}
\Hypo{\vdash ? F\tikzmark{tqc2}, ? F, \Gamma}
\Infer{1}[\rcontr]{\vdash ? F\tikzmark{tqc1}, \Gamma}
\end{prooftree}}

\def\tinfweak{\begin{prooftree}
\Hypo{\vdash\tikzmark{tqw2} \Gamma}
\Infer{1}[\rweak]{\vdash ? F,\tikzmark{tqw1} \Gamma}
\end{prooftree}}

\def\tinfder{\begin{prooftree}
\Hypo{\vdash F\tikzmark{tqd2}, \Gamma}
\Infer{1}[\rder]{\vdash ? F\tikzmark{tqd1}, \Gamma}
\end{prooftree}}

\def\tinfprom{\begin{prooftree}
\Hypo{\vdash F\tikzmark{tbp2}, ?\Gamma}
\Infer{1}[\rprom]{\vdash ! F\tikzmark{tbp1}, ?\Gamma}
\end{prooftree}}

\def\tinfparr{\begin{prooftree}
\Hypo{\vdash F,\tikzmark{tpa2} G, \Gamma}
\Infer{1}[\rparr]{\vdash F\parr\tikzmark{tpa1} G, \Gamma}
\end{prooftree}}

\def\tinftensor{\begin{prooftree}
\Hypo{\vdash F\tikzmark{tta2},\Gamma}
\Hypo{\vdash G\tikzmark{tta3},\Delta}
\Infer{2}[\rotimes]{\vdash F\tensor G, \tikzmark{tta1}\Gamma,\Delta}
\end{prooftree}}

\def\tinfplus{\begin{prooftree}
\Hypo{\vdash F_i\tikzmark{tpla2}, \Gamma}
\Infer{1}[\roplusi]{\vdash F_1\oplus\tikzmark{tpla1} F_2, \Gamma}
\end{prooftree}}

\def\tinfwith{\begin{prooftree}
\Hypo{\vdash F\tikzmark{twa2},\Gamma}
\Hypo{\vdash G\tikzmark{twa3},\Gamma}
\Infer{2}[\rwith]{\vdash F\with G, \tikzmark{twa1}\Gamma}
\end{prooftree}}

\def\tinfbot{\begin{prooftree}
\Hypo{\vdash \Gamma\tikzmark{tbotb2}}
\Infer{1}[\rbot]{\vdash \bot, \Gamma\tikzmark{tbotb1}}
\end{prooftree}}

\def\tinftop{\begin{prooftree}
\Hypo{}
\Infer{1}[\rtop]{\vdash \top, \Gamma}
\end{prooftree}}

\def\tinfone{\begin{prooftree}
\Hypo{}
\Infer{1}[\rone]{\vdash \llone}
\end{prooftree}}

\def\tinfmu{\begin{prooftree}
\Hypo{ \vdash F[\mu X.F/X]\tikzmark{tma2}, \Gamma}
\Infer{1}[\rmu]{ \vdash \mu X.\tikzmark{tma1}F, \Gamma}
\end{prooftree}}

\def\tinfnu{\begin{prooftree}
\Hypo{\vdash G[\nu X.G/X]\tikzmark{tna2}, \Gamma}
\Infer{1}[\rnu]{\vdash \nu X.\tikzmark{tna1}G, \Gamma}
\end{prooftree}}

\def\tinfforall{\begin{prooftree}
\Hypo{ \vdash F\tikzmark{tfa2}, \Gamma}
\Infer{1}[\rforall]{ \vdash \forall x\tikzmark{tfa1}F, \Gamma}
\end{prooftree}}

\def\tinfexists{\begin{prooftree}
\Hypo{\vdash G\{t/x\}\tikzmark{tea2}, \Gamma}
\Infer{1}[\rexists]{\vdash \exists x\tikzmark{tea1}G, \Gamma}
\end{prooftree}}

\def\tinfax{
\begin{prooftree}
%\Hypo{F \equiv G}
\Infer{0}[\rax]{\vdash F, F^\perp}
\end{prooftree}}

\def\tinfcut{
  \begin{prooftree}
  \Hypo{\vdash \Gamma,\tikzmark{tca2} F}
  \Hypo{\vdash F^\perp,\tikzmark{tca3} \Delta}
\Infer{2}[{\rcut}]{\vdash \Gamma,\tikzmark{tca1} \Delta}
\end{prooftree}}

\begin{figure}[t]
\begin{center}
\scalebox{.85}{
$
\arraycolsep=1.2pt\def\arraystretch{2}
%\begin{array}{cc}
%\tinfax & \tinfcut \\  
%  \tinfmu & \tinfnu  \\
%\tinfbot  & \tinfone  \\
% \tinfparr  & \tinftensor \\
% \tinftop  & \text{(no rule for 0)}  \\
% \tinfwith  & \tinfplus \\
%\end{array}
%\begin{array}{cccc}
%\tinfax & \tinfcut&  \tinfmu & \tinfnu  \\
% \tinfparr  & \tinftensor &\tinfbot  & \tinfone  \\
% \tinfwith  & \tinfplus &  \tinftop  & \text{(no rule for 0)}  \\
%\tinfder & \tinfcontr & \tinfweak & \tinfprom\\
%\end{array}
\begin{array}{ccccc}%|rc}
% \tinfder& \tinfprom&\tinfweak &\tinfcontr
% (a)& \tinfax & \tinfcut && \tinfex &(b)\qquad& \tinfder\\[5pt]
% &\tinfparr  & \tinftensor & \tinfbot & \tinfone&& \tinfprom \\[5pt]
% &\tinfwith  & \tinfplus & \tinftop & \text{(no rule for 0)}  &&\tinfweak\\[5pt]
%  &\tinfnu & \tinfmu
%&\tinfexists & \tinfforall &\text{(in \rforall, $x\not\in \mathsf{FV}(\Gamma)$)} &&&\tinfcontr
(a) & \tinfax & \tinfone & \tinftop & \tinfbot \\[5pt]
    & \tinfex & \tinfcut & \tinfparr & \tinftensor \\[5pt]
    & & \tinfwith  &  \tinfplus & \\[5pt]
\hline 
(b) &\tinfder& \tinfweak &\tinfcontr &\tinfprom \\[5pt]
\hline 
(c) & & \tinfforall &\tinfexists& \\[5pt]
\hline
(d) & &\tinfnu & \tinfmu & \\
\end{array}
\begin{tikzpicture}[overlay,remember picture,-,line cap=round,line width=0.1cm]
  
%%contraction
%   \draw[rounded corners, smooth=2,green, opacity=.4] ([xshift=-2mm] pic cs:tqc1) to ([xshift=-3mm,yshift=2mm] pic cs:tqc2);
%   \draw[rounded corners, smooth=2,green, opacity=.4] ([xshift=-2mm] pic cs:tqc1) to ([xshift=3mm,yshift=2mm] pic cs:tqc2);
%   \draw[rounded corners, smooth=2,cyan, opacity=.4] ( [xshift=2mm]pic cs:tqc1) to ([xshift=9mm,yshift=2mm] pic cs:tqc2);

%%weakening
%   \draw[rounded corners, smooth=2,cyan, opacity=.4] ([xshift=1mm] pic cs:tqw1) to ([xshift=1mm, yshift=2mm] pic cs:tqw2);  
%

% %dereliction
%   \draw[rounded corners, smooth=2,green, opacity=.4] ([xshift=-2mm] pic cs:tqd1) to ([xshift=-2mm, yshift=2mm] pic cs:tqd2);
%   \draw[rounded corners, smooth=2,cyan, opacity=.4] ( [xshift=2.5mm]pic cs:tqd1) to ([xshift=2.5mm, yshift=2mm] pic cs:tqd2);

%      %promotion
%   \draw[rounded corners, smooth=2,green, opacity=.4] ([xshift=-2mm] pic cs:tbp1) to ([xshift=-2mm, yshift=2mm] pic cs:tbp2);
%   \draw[rounded corners, smooth=2,cyan, opacity=.4] ( [xshift=3mm]pic cs:tbp1) to ([xshift=3mm, yshift=2mm] pic cs:tbp2);

      %echange
   \draw[rounded corners, smooth=2,green, opacity=.4] ([xshift=-2mm] pic cs:tex1) to ([xshift=3mm, yshift=2mm] pic cs:tex2);
   \draw[rounded corners, smooth=2,green, opacity=.4] ([xshift=3mm] pic cs:tex1) to ([xshift=-2mm, yshift=2mm] pic cs:tex2);
   \draw[rounded corners, smooth=2,cyan, opacity=.4] ( [xshift=7mm]pic cs:tex1) to ([xshift=7mm, yshift=2mm] pic cs:tex2);
   \draw[rounded corners, smooth=2,cyan, opacity=.4] ( [xshift=-5mm]pic cs:tex1) to ([xshift=-5mm, yshift=2mm] pic cs:tex2);

   %bot
   \draw[rounded corners, smooth=2,cyan, opacity=.4] ([xshift=-1mm] pic cs:tbotb1) to ([xshift=-1mm, yshift=2mm] pic cs:tbotb2);

%parr
   \draw[rounded corners, smooth=2,green, opacity=.4] ([xshift=-1mm] pic cs:tpa1) to ([xshift=-3mm,yshift=2mm] pic cs:tpa2);
   \draw[rounded corners, smooth=2,green, opacity=.4] ([xshift=-1mm] pic cs:tpa1) to ([xshift=1mm,yshift=2mm] pic cs:tpa2);
   \draw[rounded corners, smooth=2,cyan, opacity=.4] ( [xshift=5mm]pic cs:tpa1) to ([xshift=5mm,yshift=2mm] pic cs:tpa2);
   
%tensor
   \draw[rounded corners, smooth=2,green, opacity=.4] ([xshift=-6mm] pic cs:tta1) to ([xshift=-1mm,yshift=2mm] pic cs:tta2);
   \draw[rounded corners, smooth=2,green, opacity=.4] ([xshift=-6mm] pic cs:tta1) to ([xshift=-1mm,yshift=2mm] pic cs:tta3);
   \draw[rounded corners, smooth=2,cyan, opacity=.4] ([xshift=1mm] pic cs:tta1) to ([xshift=2mm,yshift=2mm] pic cs:tta2);
   \draw[rounded corners, smooth=2,cyan, opacity=.4] ([xshift=5mm] pic cs:tta1) to ([xshift=3mm,yshift=2mm] pic cs:tta3);

   %plus
   \draw[rounded corners, smooth=2,green, opacity=.4] ([xshift=-2mm] pic cs:tpla1) to ([xshift=-3mm,yshift=2mm] pic cs:tpla2);
%   \draw[rounded corners, smooth=2,green, opacity=.4] ([xshift=-1mm] pic cs:tpla1) to ([xshift=1mm] pic cs:tpla2);
   \draw[rounded corners, smooth=2,cyan, opacity=.4] ( [xshift=6.5mm]pic cs:tpla1) to ([xshift=2.5mm,yshift=2mm] pic cs:tpla2);
   
%with
   \draw[rounded corners, smooth=2,green, opacity=.4] ([xshift=-6mm] pic cs:twa1) to ([xshift=-1mm,yshift=2mm] pic cs:twa2);
   \draw[rounded corners, smooth=2,green, opacity=.4] ([xshift=-6mm] pic cs:twa1) to ([xshift=-1mm,yshift=2mm] pic cs:twa3);
   \draw[rounded corners, smooth=2,cyan, opacity=.4] ([xshift=1mm] pic cs:twa1) to ([xshift=2mm,yshift=2mm] pic cs:twa2);
   \draw[rounded corners, smooth=2,cyan, opacity=.4] ([xshift=1mm] pic cs:twa1) to ([xshift=3mm,yshift=2mm] pic cs:twa3);

%forall   
   \draw[rounded corners, smooth=2,green, opacity=.4] (pic cs:tfa1) to ([xshift=-1mm,yshift=2mm] pic cs:tfa2);
   \draw[rounded corners, smooth=2,cyan, opacity=.4] ([xshift=5mm] pic cs:tfa1) to ([xshift=2mm,yshift=2mm] pic cs:tfa2);
   
%exists   
   \draw[rounded corners, smooth=2,green, opacity=.4] (pic cs:tea1) to ([xshift=-4mm,yshift=2mm] pic cs:tea2);
   \draw[rounded corners, smooth=2,cyan, opacity=.4] ([xshift=5mm] pic cs:tea1) to ([xshift=2mm,yshift=2mm] pic cs:tea2);

%mu   
   \draw[rounded corners, smooth=2,green, opacity=.4] (pic cs:tma1) to ([xshift=-18mm,yshift=2mm] pic cs:tma2);
   \draw[rounded corners, smooth=2,cyan, opacity=.4] ([xshift=5mm] pic cs:tma1) to ([xshift=2mm,yshift=2mm] pic cs:tma2);
   
%nu   
   \draw[rounded corners, smooth=2,green, opacity=.4] (pic cs:tna1) to ([xshift=-18mm,yshift=2mm] pic cs:tna2);
   \draw[rounded corners, smooth=2,cyan, opacity=.4] ([xshift=5mm] pic cs:tna1) to ([xshift=2mm,yshift=2mm] pic cs:tna2);

%cut
   \draw[rounded corners, smooth=2,cyan, opacity=.4] ([xshift=-2mm] pic cs:tca1) to ([xshift=-2mm] pic cs:tca2);
   \draw[rounded corners, smooth=2,cyan, opacity=.4] ([xshift=1mm] pic cs:tca1) to ([xshift=1mm] pic cs:tca3);

%contraction
   \draw[rounded corners, smooth=2,green, opacity=.4] ([xshift=-2mm] pic cs:tqc1) to ([xshift=-3mm,yshift=2mm] pic cs:tqc2);
   \draw[rounded corners, smooth=2,green, opacity=.4] ([xshift=-2mm] pic cs:tqc1) to ([xshift=3mm,yshift=2mm] pic cs:tqc2);
   \draw[rounded corners, smooth=2,cyan, opacity=.4] ( [xshift=2mm]pic cs:tqc1) to ([xshift=9mm,yshift=2mm] pic cs:tqc2);

%weakening
   \draw[rounded corners, smooth=2,cyan, opacity=.4] ([xshift=1mm] pic cs:tqw1) to ([xshift=1mm, yshift=2mm] pic cs:tqw2);  

   %dereliction
   \draw[rounded corners, smooth=2,green, opacity=.4] ([xshift=-2mm] pic cs:tqd1) to ([xshift=-2mm, yshift=2mm] pic cs:tqd2);
   \draw[rounded corners, smooth=2,cyan, opacity=.4] ( [xshift=2.5mm]pic cs:tqd1) to ([xshift=2.5mm, yshift=2mm] pic cs:tqd2);

      %promotion
   \draw[rounded corners, smooth=2,green, opacity=.4] ([xshift=-2mm] pic cs:tbp1) to ([xshift=-2mm, yshift=2mm] pic cs:tbp2);
   \draw[rounded corners, smooth=2,cyan, opacity=.4] ( [xshift=3mm]pic cs:tbp1) to ([xshift=3mm, yshift=2mm] pic cs:tbp2);
\end{tikzpicture}
$
}
\end{center}

  \caption{\muLLi inferences: (a) \label{fig:mall-seq}Propositional \MALL inferences (no rule for 0); 
(b) \label{fig:ll-seq}\LL exponential inferences; 
(c) First-order inferences (in \rforall, $x\not\in \mathsf{FV}(\Gamma)$);
(d) Fixed-point inferences.
}
\end{figure}

\begin{definition}[\muLLip and \LL sequent calculus]
	The inference rules of \muLLip (resp. \LL) sequent calculus are given by the rows $a,b$ and $d$ in Figure \ref{fig:ll-seq} (resp. rows $a,b$ and $c$ in Figure \ref{fig:ll-seq}). The colored formulas in each inference rule define the \defname{ancestor (or subocurrence) relation} between the conclusion and the premises of the rules.
\end{definition}

\begin{remark}
    For $\muLLi$ proofs, we restrict ourselves to the propositional fragment $\muLLip$, since the cut-elimination procedure for circular proofs in \cite{cutElimInf}, that is used in the Theorem \ref{the:main}, considers only propositional formulas. For a complete treatment of quantifiers, one should use parametrized fixed-point variables and respective unfolding rules, as in \cite{FOCyclic}. We leave this for future work.
\end{remark}

We represent the tree structure of \muLLip derivations using \defname{derivation graphs}, after %Brotherston 
\cite{brotherston}.

\begin{definition}[\muLLip derivation graph]
A \muLLip derivation graph is a tuple $\mathcal{D} = (V, s, r, p)$ where $V$ is a set of vertices, $s: V \rightarrow Seqs$ is a total function labelling vertices with sequents in \muLL, $r: V \rightarrow Rules$ is a partial function labelling vertices with inference rules in \muLLip, $p: \mathbb{N} \times V \rightarrow V$ is a partial function that maps a number $n$ and a vertex $v$, with a vertex $v'$, such that $v'$ is labeled by the $n^{\text{th}}$ premise of the rule $r(v)$ over the sequent $s(v)$, and for every $v \in V$, 
            \adjustbox{valign=c, scale=0.75}{\begin{prooftree}
                            \hypo{\vdash s(p_1(v)) \quad \dots \quad \vdash s(p_m(v))}
                            \infer1[$(r(v))$]{\vdash s(v)}
                        \end{prooftree}} 
is an instance of the rule $r(v)$ in \muLLip, with $m$ premises. A vertex $B \in V$ such that $r(B)$ is undefined is called a \defname{bud} node. The set of bud nodes of 
%a derivation graph
 $\mathcal{D}$, is written as $Bud(\mathcal{D})$. If 
 %the set
  $V$ is finite, we say that $\mathcal{D}$ is a \defname{finite} derivation graph.
\end{definition}

A \muLLip derivation graph $\mathcal{D}$ can be seen as a directed graph $(V, E)$ where the set of edges is defined as $E = \{(v, p_j(v)) \ | \ v \in V \ \text{and $p_j(v)$ is defined} \}$. Moreover, $\mathcal{D}$ is said to be a \defname{derivation tree}, if its directed graph is a tree. We name the root of the tree as $root(\mathcal{D}) \in V$ and $s(root(\mathcal{D}))$ the conclusion of the tree.
Note that the ancestor relation can be extended by transitivity to formulas of non-consecutive sequents of a derivation graph.

\begin{definition}[\LL proof]
An \defname{\LL open proof} $\pi$ of a sequent $\vdash \Gamma$ (written as $\pi \vdash \Gamma$) is a finite derivation tree with conclusion $\vdash \Gamma$ such that all its formulas (resp. inferences)  are \LL formulas (resp. inferences).
An \defname{\LL proof} $\pi$ is an open proof such that 
$Bud(\mathcal{\pi}) = \emptyset$.
\end{definition}

%==================================================================================
%==================================================================================
%==================================================================================
%==================================================================================

Due to the recursive nature of the fixed-point rules, it is possible to construct infinite derivation trees even in the absence of cuts and contractions. Such infinite derivation trees are referred to as \defname{\muLLip pre-proofs}. In some cases, these infinite trees contain only finitely many distinct subtrees. When this occurs, we call the derivation trees \defname{regular}. Such regular infinite trees can be finitely represented as trees with \defname{back-edges} that form cycles, yielding what are known as \defname{$\muLLop$ pre-proofs}.

\begin{definition}[\muLLo pre-proof graph] 
A \defname{\muLLo pre-proof graph} (or simply pre-proof) of the sequent $\vdash \Gamma$, is a tuple $\mathcal{P} = (\mathcal{D}, \mathcal{R})$, where $\mathcal{D} = (V, s, r, p)$ is a $\muLLi$ finite derivation tree and $\mathcal{R}: Bud(\mathcal{D}) \rightarrow V$ is a total function, that maps each one of the bud nodes $B \in Bud(\mathcal{D})$ to an internal node on the path from $root(\mathcal{D})$ to $B$, called a \defname{companion}, satisfying that if $\mathcal{R}(B) = C$, then $r(C)$ is defined and $s(B) = s(C)$. We say that $\mathcal{P}$ is an \defname{open pre-proof} if $\mathcal{R}$ is not defined for all $B \in Bud(\mathcal{D})$.
\end{definition}

The \defname{graph of a pre-proof} $\mathcal{P} = (\mathcal{D}, \mathcal{R})$, written as $\mathcal{G}_{\mathcal{P}}$, is the graph obtained after adding the edges induced by $\mathcal{R}$ and deleting the bud vertices. Specifically, $\mathcal{G}_{\mathcal{P}} = (V', s, r, p')$ where, $V' = V \setminus Bud(\mathcal{D})$ and $p'$ is defined for any $j$ as: $p'_j(v) = \mathcal{R}(p_j(v))$ if $p_j(v) \in Bud(\mathcal{D})$, or $p'_j(v) = p_j(v)$ otherwise.
Soundness of  \muLLop pre-proofs is ensured by imposing a correctness criterion known as the  \defname{progress} or \defname{validity} or \defname{global trace condition}:

\begin{definition}[Traces and validity]
\label{def:traceValid}
Let $\mathcal{P}$ be a \muLLo pre-proof. A \defname{trace} over a (possibly infinite) path $(v_i)_{i\in \omega}$ in the graph $\mathcal{G}_{\mathcal{P}}$, is defined as a sequence of formulas $(F_i)_{i\in \omega}$, such that for each index $i$, we have $F_i \in s(v_i)$ 
and either $F_i \rightarrow F_{i+1}$ or $F_i = F_{i+1}$, where the relation $\rightarrow$ denotes the \defname{Fischer-Ladner subformula relation}, defined as follows:
\begin{align*}
(F \star G) \rightarrow F \qquad
(F \star G) \rightarrow G \qquad
(\Delta F) \rightarrow F \qquad
(F^\perp) \rightarrow F \qquad
(\sigma X. F) \rightarrow F[\sigma X. F / X]
\end{align*}
With $\star \in \{\parr, \with, \otimes, \oplus \}$, $\Delta \in \{ \forall x, \exists x, ?, ! \}$, and $\sigma \in \{ \mu, \nu \}$. Let $F = \nu X. F'$. A \defname{$F$-progress point} of a trace, corresponds to the vertex $v_j$ in $(v_i)_{i\in \omega}$ such that $F_j = F$, $F_{j+1} = F'[\nu X. F'/X]$ and $r(v_j) = \nu$. We say that an infinite path in $\mathcal{G}_{\mathcal{P}}$ is \defname{valid} if it has a trace $\uptau = (F_i)_{i\in \omega}$ such that the minimal formula in the set of infinitely recurring formulas\footnote{Due to first-order, we use in fact the propositional skeleton of formulas here, defined as usual: $sk(F \star G) = sk(F) \star sk(G)$ with $\star \in \{\parr, \with, \otimes, \oplus \}$, $sk(\Delta F) = \Delta sk(F)$ with $\Delta \in \{  ?, ! \}$, $sk(u)=u$ with $u\in\{ 1,\top,0,\bot\}$, $sk(\sigma X. F) = \sigma X. sk(F)$ with $\sigma \in \{ \mu, \nu \}$, $sk(\Delta F) = sk(F)$ with $\Delta \in \{ \forall x, \exists x \}$ and $sk(a(\vec x))= a$.}, denoted $\textsf{min}(\textsf{Inf}(\uptau))$, is a $\nu$-formula, i.e. of the form $F = \nu X. F'$, where $F$ is minimal with respect to the usual subformula ordering, and $\uptau$ infinitely often visits a $F$-progress point. $\mathcal{P}$ is said to be a \defname{valid proof} or just a \defname{proof} if and only if every infinite path in $\mathcal{G}_{\mathcal{P}}$ is valid.
\end{definition}
In addition, we consider a stronger notion of validity, known as \defname{strong validity}. 
\begin{definition}[Strong validity]
    Let $\mathcal{P}$ be a \muLLo pre-proof. An infinite path $(v_i)$ in $\mathcal{G}_{\mathcal{P}}$ is said to be strongly valid if there exists a valid trace $\uptau = (F_i)$ over $(v_i)$, and there exists an index $k$ such that for all $h, i \geq k$, whenever $v_h = v_i$, it follows that $F_h = F_i$ (as formula occurrences of the considered sequent). A pre-proof $\mathcal{P}$ is said to be a \defname{strongly valid proof} if and only if every infinite path in $\mathcal{G}_{\mathcal{P}}$ is strongly valid.
\end{definition}

This ensures that $\uptau$ exhibits some stability or convergence along the path, and can be represented as a function $\uptau(v)$ mapping vertices $v$ of the pre-proof to occurrences in $s(v)$.

%==================================================================================
%==================================================================================
%==================================================================================
%==================================================================================
%==================================================================================
%==================================================================================
%==================================================================================
%==================================================================================

% !TEX root = main.tex
\subsection{Proof-relevant interpolation for cut-free proofs in linear logic}
\label{sec:proofRelIntLL}

We now recall Saurin's proof-relevant interpolation result for cut-free linear logic~\cite{Saurin}. To state the result, we first need the definition of \defname{vocabulary} of a formula:

\begin{definition}[Vocabulary of a formula]
Let $F$ be a $\LL$ formula. $\voc (F)$ is defined as the set of all atoms occurring in $F$, together with their linear negations, while $\voc^+ (F)$ (resp. $\voc^- (F)$) is the set of positive (resp. negative) atoms occurring in $F$. This notion is naturally lifted to list of formulas $\Gamma$, written as $\voc (\Gamma)$.
\end{definition}

The interpolation theorem for \LL is stated as:

\begin{theorem}[Saurin \cite{Saurin}, Theorem 5]
    \label{the:interpol1}
    Let $\Gamma, \Delta$ be lists of \LL formulas and $\pi \vdash \Gamma, \Delta$ a cut-free proof. There exists a $\LL$ formula $I$ such that $\voc^+ (I) \subseteq \voc^- (\Gamma) \cap \voc^+ (\Delta)$, $\voc^- (I) \subseteq \voc^+ (\Gamma) \cap \voc^- (\Delta)$, and two cut-free proofs $\pi^L, \pi^R$ of $\vdash \Gamma, I$ and $\vdash I^{\bot}, \Delta$ respectively such that \adjustbox{valign=c, scale=0.75}{
    \begin{prooftree}
            \hypo{\pi^L}
            \infer1[]{\vdash \Gamma, I}
            \hypo{\pi^R}
            \infer1[]{\vdash I^{\bot}, \Delta}
            \infer2[\textup{\textsf{(Cut)}}]{\vdash \Gamma, \Delta}
        \end{prooftree}
    } $\cutred^* \pi$.
\end{theorem}

The proof of this theorem amounts to a cut-introduction process synthesizing the \defname{interpolant} 
%(which is the formula $I$ in Theorem \ref{the:interpol1}) 
in two phases, called the $\defname{ascending}$ and $\defname{descending}$ phase.

\smallskip
\noindent{\bf Ascending phase.}
Interpolation works with decorated proofs where the formulas of each sequents are colored or split. 
Given a  \defname{split} of the conclusion $\vdash \Gamma$ of $\pi$ into a pair of sublists $(\Gamma', \Gamma'')$ (often represented by coloring the formulas of $\Gamma$ red or blue), the 
ascending phase consists in traversing $\pi$ upward from conclusion to axioms, while splitting each sequent $\vdash \Delta$ into a split $(\Delta', \Delta'')$ using the ancestor relation, propagating the split of the conclusion. 
A proof  decorated in coherence with the ancestor relation is called \defname{coherently decorated}.

% !TEX root = main.tex

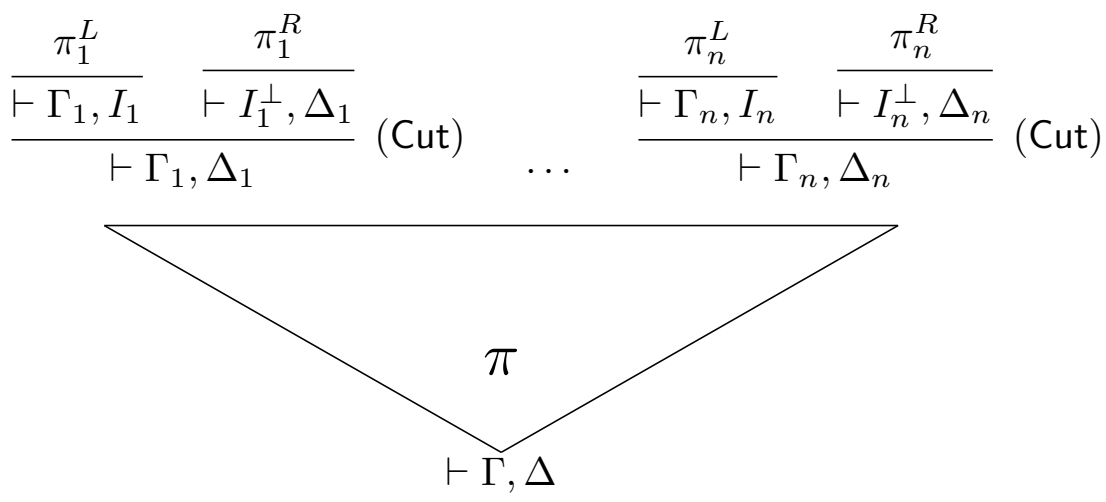
\begin{figure}
  \centering
\scalebox{.75}{
\begin{prooftree}
\Hypo{\scalebox{1}{$\pi^L_1$}}
\Infer{1}{\vdash \Gamma_1, I_1}
\Hypo{\scalebox{1}{$\pi^R_1$}}
\Infer{1}{\vdash I_1^\perp, \Delta_1}
\Infer 2[\rcut]{\vdash \Gamma_1,\Delta_1}
\Hypo{\cdots}
\Hypo{\scalebox{1}{$\pi^L_n$}}
\Infer{1}{\vdash \Gamma_n, I_n}
\Hypo{\scalebox{1}{$\pi^R_n$}}
\Infer{1}{\vdash I_n^\perp, \Delta_n}
\Infer{2}[\rcut]{\vdash \Gamma_n,\Delta_n}

\Infer[no rule]{3}{
\begin{tikzpicture}
\node at (0,2.6){};
\draw (0,0.5) -- (3.5,2.5);
\draw (0,0.5) -- (-3.5,2.5);
\draw (3.5,2.5) -- (-3.5,2.5);
%\node at (.9,1.2){$\vdash \Gamma,\Delta$};
%\node at (-.9,1.2){$\vdash \Gamma,\Delta$};
\node at (0,1.3){\scalebox{1.5}{$\pi$}};
\node at (0,.3) {$\vdash \Gamma,\Delta$};
\end{tikzpicture}
}
%\pi}
%\Infer 1{\vdash \Gamma, \Delta}
\end{prooftree}
}
\caption{A \PRIS -- Proof-relevant Interpolation Situation.}
\label{fig:pris1}
\end{figure}
\noindent{\bf Descending phase.}
Once every sequent in the proof has been decorated during the ascending phase, the descending phase consists in applying \defname{cut-introductions rules}, starting from the leaves of the derivation tree, until one reaches the conclusion. 
This cut-introduction process is described by Saurin~\cite{Saurin} by using \PRIS, or \defname{proof-relevant interpolation situation} that are represented in Figure \ref{fig:pris1}.
    A $\PRIS$ for $(\Gamma, \Delta)$ is the data of 
    (i) \textbf{The goal},  a cut-free $\LL$ decorated proof $\pi$ of split conclusion $\vdash \Gamma, \Delta$, with open (split) premises $(\vdash \Gamma_i, \Delta_i)_{1 \leq i \leq n}$, 
    (ii) \textbf{The partial interpolants}, that is for each open premise $\vdash \Gamma_i, \Delta_i$, a formula $I_i$ st. $\voc^+({I_i}) \subseteq \voc^-({\Gamma_i}) \cap \voc^+({\Delta_i})$ and $\voc^-({I_i}) \subseteq \voc^+({\Gamma_i}) \cap \voc^-({\Delta_i})$ and (iii) \textbf{The partial solutions}, that is for each open premise $\vdash \Gamma_i, \Delta_i$, derivations $\pi^L_i$ (resp. $\pi^R_i$) of conclusions $\vdash \Gamma_i, I_i$ (resp. $\vdash  I^{\bot}_i, \Delta_i$).
Saurin singles out three types of \PRIS: initial, solved and elementary \PRIS, which are represented in \Cref{fig:pris2}.

The interpolation problem is then rephrased as follows: Given an initial $\PRIS$, is it possible to produce a solved $\PRIS$ using only cut-introduction rules? 
% !TEX root = main.tex

\begin{figure}[t]
\centering
\begin{subfigure}{0.15\textwidth}
\centering
\textbf{(a)}
\begin{prooftree}
    \hypo{\pi}
    \infer1{\vdash \Gamma, \Delta}
\end{prooftree}

%\caption{}
\label{fig:a}
\end{subfigure}
\hfill
\begin{subfigure}{0.25\textwidth}
\centering
\textbf{(b)}
\scalebox{.75}{
\begin{prooftree}
    \hypo{\pi^L}
    \infer1{\vdash \Gamma, I}

    \hypo{\pi^R}
    \infer1{\vdash I^{\bot}, \Delta}

    \infer2[(\textsf{Cut})]{\vdash \Gamma, \Delta}
\end{prooftree}
}
%\caption{}
\label{fig:b}
\end{subfigure}
\hfill
\begin{subfigure}{0.57\textwidth}
\centering
\textbf{(c)}
\scalebox{.75}{
\begin{prooftree}
    \hypo{\pi_1^L}
    \infer1{\vdash \Gamma_1, I_1}
    \hypo{\pi_1^R}
    \infer1{\vdash I^{\bot}_1, \Delta_1}
    \infer2[(\textsf{Cut})]{\vdash \Gamma_1, \Delta_1}

    \hypo{\cdots}

    \hypo{\pi_n^L}
    \infer1{\vdash \Gamma_n, I_n}
    \hypo{\pi_n^R}
    \infer1{\vdash I^{\bot}_n, \Delta_n}
    \infer2[(\textsf{Cut})]{\vdash \Gamma_n, \Delta_n}

    \infer3[(r)]{\vdash \Gamma, \Delta}
\end{prooftree}
}
%\caption{}
\label{fig:c}
\end{subfigure}
\caption{(a) Initial \PRIS, (b) Solved \PRIS, (c) Elementary \PRIS.}
\label{fig:pris2}
\end{figure}
The key ingredient is elementary \PRIS where there remains a unique inference rule to solve the interpolation problem: Theorem \ref{the:interpol1} follows directly from solving that case.

\begin{remark}
In the following we will focus on Craig interpolation rather than Lyndon as it is more natural to consider with cuts; from now on we neglect the polarity of the vocabulary.
\end{remark}
% !TEX root = main.tex
\section{Interpolation with Cuts}
\label{sec:interpolate-cuts}

In the present section, we extend Saurin's result on proof-relevant interpolation for \LL sequent proofs~\cite{Saurin} to non-cut-free proofs, meaning that we aim at performing interpolation directly on proofs containing cuts. 
Note that the results of this section readily apply to intuitionistic and classical sequent calculi as well. 
Before proceeding, a number of observations is in order: 
\begin{enumerate}
\item 
Since Saurin's {\it proof-relevant interpolation}  procedure only relies on cut-introduction rules, the presence of cuts should not prevent the dynamics of the cut-elimination process to work: indeed, only the vocabulary of the synthesized candidate-interpolant formula depends on the cut-freeness assumption, a key condition for interpolation to hold though. 
\item %
Instead of designing an ad hoc constraint on the cut-formulas, one can consider a proof-relevant {\it weak interpolation process} that handles every proof with cuts, at the cost of relaxing the control on the interpolant vocabulary (since the cut-formulas can contribute to the interpolant vocabulary): 
the cut-introduction performed in solving \PRIS can indeed be extended to handle the cut-case, as long as the cut formulas are split themselves and  
the interpolant is allowed to use a vocabulary exploiting the language of the cut-formulas.
\item Last, our strategy will be to analyze  conditions over the cut-formulas, that ensure an actual interpolation results. We will therefore obtain sufficient conditions 
so that the synthesized formula is an interpolant of the conclusion sequent. 
\end{enumerate}

\begin{remark}
There is a structuring principle, often exploited in the theory of linear logic proof nets correctness criteria, that states that a cut inference on  $C, C^\perp$ has the same geometry as a $\tensor$ rule introducing $C\tensor C^\perp$ (or the corresponding conjunction when working in classical or intuitionistic logic), followed by some form of hiding operation. This works fine when considering \MLL (where proof nets best behave) but it gets more complex in other fragments of \LL (with additives, exponentials, etc.) 
While one could carry formally the above simulation of the cut inference  on  $C, C^\perp$ for the various fragments of \LL\footnote{For instance for full \LL, one would consider  formula $\quest \exists \vec{x}. (C\tensor C^\perp)$.}, we will instead introduce a special treatment of the cuts in \PRIS, for two reasons. First, our treatment will be more uniform than considering specific encoding for each specific logical fragment/system that one considers.
The second and deeper reason amounts to the fact that the crucial part of our analysis (detailed in the following section) relies in identifying the cut-vocabulary and allowing to express adequate constraints on the cut vocabulary.
\end{remark}

\subsection{CPRIS: Proof-relevant Interpolation Situations with Cuts}

First, one needs the following lemma that is straightforwardly proved by induction
that all \LL proofs (with cuts) can be coherently decorated, given an initial splitting of their conclusion, as well as a \defname{splitting of every cut-inference ocurrence}, i.e. for each cut inference $i$, we assign an arbitrary color for the cut formulas $C_i$ and $C_i^\perp$, that we use in the ascending phase.

\begin{lemma}
    \label{lem:coherentProof}
    For any $\LL$ proof $\pi$  
    and any splittings
     of its conclusion sequent and 
      of its cut inference  occurrences, the ascending phase terminates with a coherently decorated proof.
\end{lemma}

One can now propose the adequate generalisation of the structures of~\cite{Saurin} allowing to solve \PRIS in presence of cuts and allowing to take into account how cut formulas can contribute to the vocabulary of the interpolant: 
we introduce \CPRIS for that purpose that generalise \PRIS  with the notion of cut vocabularies.\footnote{
In what follows, we adapt the method in the previous paragraphs to Craig interpolation, as the polarity constraints induced by Lyndon interpolation cannot be satisfied in the presence of cuts.} 

\begin{definition}[\PRIS with Cuts]
\label{def:cutpris}
A \CPRIS for $(\Gamma, \Delta)$ is the data of: 
\begin{itemize}
\item \defname{the goal}, that is, an \LL coherently decorated proof $\pi$ of $\vdash \spl{\Gamma}, \spr{\Delta}$ with $n\geq 0$ open premises $(\vdash \spl{\Gamma_i},\spr{\Delta_i})_{1\leq i \leq n}$  such that for each $1\leq i \leq n$
the formulas in $\Gamma_i$ (resp. $\Delta_i$) are ancestors of formulas in $\Gamma$ (resp. $\Delta$) or ancestors of a cut-formula; 
\item \defname{the cut vocabularies}: for $1\leq i \leq n$, a pair $(\mathbb{C}_i^L, \mathbb{C}_i^R)$ of sets of literals;
\item \defname{the partial interpolants,} that is for each $1\leq i \leq n$, a formula $I_i$ st. 
$\voc{(I_i)}\subseteq (\voc {(\Gamma_i)}\cup \mathbb{C}_i^L)\cap (\voc {(\Delta_i)}\cup  \mathbb{C}_i^R)$
and;
\item \defname{the partial solutions}, that is, for each $1\leq i \leq n$, derivations $\pi_i^L$ (resp. $\pi_i^R$) of conclusion $\vdash \Gamma_i, I_i$ (resp. $\vdash I_i^\perp, \Delta_i$) such that if ${\cal C}^L$ (resp. ${\cal C}^R$) is the set of cut-formulas 
in $\pi_i^L$ (resp. in  $\pi_i^R$), then $\bigcup_{C\in{\cal C}^\epsilon}\voc{(C)} = \mathbb{C}_i^\epsilon, \epsilon\in\{L,R\}$.
\end{itemize}
\end{definition}

With the above definition, the notions of initial, solved, and elementary \CPRIS are trivially  adapted from the previous section.

\subsection{Weak proof-relevant interpolation theorem with cuts}

\begin{lemma}
\label{main-lemma}
For any elementary \CPRIS $\pi$
there exist $I, \pi^L,\pi^R$ such that 
%\scalebox{.9}{
$\pi'=\begin{prooftree}
\Hypo{\pi^L}
\Infer[no rule]1{\vdash \spl\Gamma, I}
\Hypo{\pi^R}
\Infer[no rule]1{\vdash  I^\perp, \spr\Delta}
\Infer2[\rcut]{\vdash \spl\Gamma,\spr\Delta}
\end{prooftree}
$
%}
is a solved \CPRIS and 
$\pi \longleftarrow^\star_{\mathsf{cut}}\pi'$.
\end{lemma}

\begin{proof}[Proof sketch]
By case analysis on the last rule of the elementary \CPRIS. The extension is trivial for all inferences but the cut rule.
In the case of an elementary \CPRIS concluded with a cut-inference, the extension directly follows for the case of the $\tensor$ rule, the only difference is that instead of being part of the vocabulary of the context formulas, the cut-formulas contribute to the cut vocabularies. (Details are provided in \Cref{app-main-lemma}.)
\end{proof}

\Cref{main-lemma} implies by an easy induction the following theorem, using a \CPRIS obtained by splittings consistent with $\spl{\Gamma},\spr{\Delta}$ and a partitioning of cut formulas as $\spl{\{(C_i,C_i^\perp), i\in L\}}$, $\spr{\{(C_i, C_i^\perp), i\in R\}}$.
Indeed, each application of \Cref{main-lemma} decreases the size of the goal of the \CPRIS by one which straightforwardly ensures the result.

\begin{theorem}[Weak interpolation theorem with cuts]
\label{thm:proof-rel-int-cut-split}
Let $\pi$ be an \LL proof of $\vdash \Gamma, \Delta$.
Let $\calC= \{(C_1, C_1^\perp), \dots, (C_n, C_n^\perp)\}$
be the cut formulas of $\pi$, let $L,R$ be any
partitioning of $\{1,\dots,n\}$ and $\mathbb C^L = \VOC{C_i, i\in L}$ and  $\mathbb C^R = \VOC{C_i, i\in R}$.
Then there exists a \LL formula $I$ such that 
$\VOC{I}\subseteq (\VOC {\Gamma} \cup \mathbb C^L) \cap (\VOC {\Delta}\cup \mathbb C^R)$
and two \LL proofs
$\pi_1,\pi_2$ of $\vdash \Gamma, I$ and $\vdash I^\perp, \Delta$ respectively such that 
\quad %
{
$\begin{prooftree}
\Hypo{\pi_1}
\Infer[no rule] 1{\vdash \Gamma, I}
\Hypo{\pi_2}
\Infer[no rule] 1{\vdash I^\perp, \Delta}
\Infer 2[\rcut]{\vdash \Gamma,\Delta}
\end{prooftree}
~\longrightarrow^\star_{\mathsf{cut}}~
\pi.$
}
\end{theorem}

As such, the above result does not satisfy the vocabulary constraints required by Craig's interpolation theorem as the interpolant formula $I$ may resort to the cut-vocabulary which is why we speak of a weak interpolation result. On the other hand, the above results provide us a toolkit to analyse situations where interpolation may be achieved in the presence of cuts: indeed, in~\Cref{thm:proof-rel-int-cut-split}, 
we have the choice of the partitioning of the set of cut-formulas $\calC$.

\subsection{Proof-relevant interpolation with cuts}

In the present section, we want to express conditions on the cut inference occurring in a proof that allow to get a full proof-relevant interpolation result from the above weak interpolation theorem.
To motivate this, let us consider the simpler case where the proof $\pi$ contains a single occurrence of the cut inference on formula $C$ that is split to the left. 
\begin{description}
\item[If $\VOC C \subseteq \VOC {\Gamma}$]
, then 
$\VOC{I}\subseteq (\VOC {\Gamma} \cup \VOC C) \cap \VOC {\Delta} = \VOC {\Gamma}  \cap \VOC {\Delta}$: $I$ is indeed a Craig interpolant  (same if $C$ is on the right and $\VOC C \subseteq \VOC{\Delta}$).
%(resp. $\VOC{I}\subseteq \VOC {\Gamma}  \cap (\VOC {\Delta}\cup \VOC C)$)

\item[If $\VOC C \cap \VOC {\Delta} = \emptyset$]
, then 
$\VOC{I}\subseteq (\VOC {\Gamma} \cup \VOC C) \cap \VOC {\Delta} =  \VOC {\Gamma}  \cap \VOC {\Delta}$:
 $I$ is also a Craig interpolant (same if $C$ is on the right and $\VOC C \cap \VOC {\Gamma} = \emptyset$).
\end{description}

Exploiting the second remark of the above analysis, one easily sees that 
if the set of cut formulas $\calC$ can be split in $\calC^L$ and $\calC^R$ (of respective vocabulary $\mathbb C^L$ and  $\mathbb C^R$) such that $(\mathbb C^L\setminus \VOC \Gamma) \cap (\VOC \Delta \cup \mathbb C^R)=\emptyset$ and 
$(\mathbb C^R\setminus \VOC \Delta) \cap (\VOC \Gamma \cup \mathbb C^L)=\emptyset$,
then the interpolation vocabulary constraint is satisfied in \Cref{thm:proof-rel-int-cut-split}. 

This allows us to state the following proof-relevant interpolation theorem with cuts:  

\begin{theorem}[Interpolation theorem with cuts]
\label{thm:proof-rel-int-cut}
Let $\pi$ be an \LL proof of $\vdash \Gamma, \Delta$.
Let $\calC= \{(C_1, C_1^\perp), \dots, (C_n, C_n^\perp)\}$
be the cut formulas of $\pi$, let $L,R$ be any partitioning of $\{1,\dots,n\}$ 
and $\mathbb C^L = \VOC{C_i, i\in L}$ and  $\mathbb C^R = \VOC{C_i, i\in R}$ 
such that 
$(\mathbb C^L\setminus \VOC \Gamma) \cap (\VOC \Delta \cup \mathbb C^R)=\emptyset$ and 
 $(\mathbb C^R\setminus \VOC \Delta) \cap (\VOC \Gamma \cup \mathbb C^L)=\emptyset$.

Then there exists an \LL formula $I$ such that 
$\VOC{I}\subseteq \VOC {\Gamma} \cap \VOC {\Delta}$
and two \LL proofs
$\pi_1,\pi_2$ of $\vdash \Gamma, I$ and $\vdash I^\perp, \Delta$ respectively such that 
\quad 
{
$\begin{prooftree}
\Hypo{\pi_1}
\Infer[no rule] 1{\vdash \Gamma, I}
\Hypo{\pi_2}
\Infer[no rule] 1{\vdash I^\perp, \Delta}
\Infer 2[\rcut]{\vdash \Gamma,\Delta}
\end{prooftree}
~\longrightarrow^\star_{\mathsf{cut}}~
\pi.$
}
\end{theorem}

\begin{remark}  
Note first that the above result also holds for \LJ and \LK, either by mimicking the construction here which goes through in these logics, or by exploiting the above theorem together with standard embeddings of those logics into linear logic, as in~\cite{Saurin}. 
\end{remark}

Of course, when each cut inference with cut formula $C$ is (semi-)analytic, then 
either $\VOC{C}\subseteq\VOC{\Gamma}$ or  $\VOC{C}\subseteq\VOC{\Delta}$, the condition \Cref{thm:proof-rel-int-cut} is met, resulting in the following corollary:

\begin{corollary}[Proof-relevant interpolation theorem with (semi-)analytic cuts]
\label{cor:proof-rel-int-analityc-cut}
Let $A,B$ be \LL formulas and $\pi$ be an \LL proof of $\vdash A, B$ such that 
every cut-inference is (semi-)analytic.
Then there exists a \LL formula $I$ such that 
$\voc{(I)}\subseteq \voc {(A)}\cap \voc {(B)}$
and two \LL proofs
$\pi_1,\pi_2$ of $\vdash A, I$ and $\vdash I^\perp, B$ respectively such that 
\quad 
%\myscale
{
$\begin{prooftree}
\Hypo{\pi_1}
\Infer[no rule] 1{\vdash A, I}
\Hypo{\pi_2}
\Infer[no rule] 1{\vdash I^\perp, B}
\Infer 2[\rcut]{\vdash A, B}
\end{prooftree}
~\longrightarrow^\star_{\mathsf{cut}}~
\pi.$
}
\end{corollary}

\begin{remark} 
The above conditions are, to our knowledge, more general than alternative treatments of (non proof-relevant) interpolation theorems in the presence of cuts that can be found in the literature~\cite{iris24lecturenotes,HetzlJalali24}. 
Since we take a proof-relevant perspective, it makes sense to state conditions on the cut inferences based not only on the cut formulas but also on how those cut-formulas are used in the proof itself. The result can then be refined and strengthened by noting that the interpolant only depends on atoms used in formulas involved in axiom rules: only the subformulas of cut formulas involved in axioms need to be considered to check whether the generated interpolant falls in the adequate language.  
\end{remark}

\medskip

A first benefit of the above results is that performing cut elimination can induce a blow up in the size of the proof and therefore of the interpolant and it directly affects the efficiency of the procedure for computing an interpolant.
Another one is that one can hope to apply this proof-relevant interpolation technique to proof systems having only a partial cut-elimination result (for instance when cuts may be stuck on non logical axioms for instance).

Last, the extension of the results to non normal proof is important for the computational interpretation as already stressed in the introduction: we shall discuss this in what follows. 

\subsection{Computational interpretations}
\label{sec:computational}

While we leave the task of extending \v Cubri\'c' result to non-evaluated $\lambda$-terms to future work, we explain below how our results already provide computational results in System  $\mathsf{L}$,
which is a term calculus due to Curien and Herbelin~\cite{CurienH00,CM10,Munch13PhD} with a typing system in correspondence with the sequent calculus system, instead of the natural deduction system.
Our results are indeed readily applicable to Saurin's computational analysis of interpolation in System $\mathsf{L}$ \cite[Section 5]{Saurin} where
he shows an analogous to \v Cubri\'c' result for the multiplicative and additive typing of System L but terms must be reduced to normal forms for the procedure to be applicable\footnote{Interestingly even in that case, Saurin has to take care of trivial cuts as in this calculus, the construction associated with the cut rule corresponds to filling an evaluation context with a term and no all such constructions are reducible.}.    
\Cref{thm:proof-rel-int-cut} readily allows to strengthen the above result (due to space constraints, we refer the reading to \cite[Section 5]{Saurin} for the notations and definitions that we adopt below). The following theorem is proved in \Cref{app:systemL}:

\def\VocGammaDeltaAL{\VOC{\Gamma_1,\Delta_1,A}\cap \VOC{\Gamma_2,\Delta_2}}
\def\VocGammaDeltaAR{\VOC{\Gamma_1,\Delta_1}\cap \VOC{\Gamma_2,\Delta_2,A}}
\def\VocGammaDelta{\VOC{\Gamma_1,\Delta_1}\cap \VOC{\Gamma_2,\Delta_2}}

\begin{theorem}
\label{thm:print-cut-systemL}
Let $c: (\spl {\Gamma_1},\spr{\Gamma_2} \vdash  \spl {\Delta_1},\spr{\Delta_2})$ be a decorated L-command and let $\calC$ be the set of types of all the $t',e'$ such that $\langle t' \mid e' \rangle$ is a  subcommand of $c$. 
Let us assume that $\calC$ is split into $\calC^L,\calC^R$ and $\mathbb C^L = \VOC{\calC^L}$ and  $\mathbb C^R = \VOC{\calC^R}$ such that 
$(\mathbb C^L\setminus \VOC {\Gamma_1,\Delta_1}) \cap (\VOC  {\Gamma_2,\Delta_2} \cup \mathbb C^R)=\emptyset$ and 
 $(\mathbb C^R\setminus \VOC  {\Gamma_2,\Delta_2}) \cap (\VOC  {\Gamma_1,\Delta_1} \cup \mathbb C^L)=\emptyset$.
Then, there exist a type $I\in \VocGammaDelta$ and $t,e$ such that
        $\spl {\Gamma_1} \vdash t: \spl I \mid \spl {\Delta_1}$ and 
        $\spr{\Gamma_2} \mid e: \spr I \vdash  \spr{\Delta_2}$, and $\langle t \mid e \rangle \longrightarrow^\star c$.     
\end{theorem}

\begin{proof}[Proof Sketch]
While the proof by Saurin \cite{Saurin} relies on a long case analysis on both terms, evaluation contexts and commands, it is sufficient to adapt the case of commands as the other cases are straightforward by a simple check that the propagation of the cut-vocabulary constraint works. 
Regarding the case of commands, while Saurin considers only cases  $\langle x \mid e \rangle $  and $\langle t \mid \alpha \rangle $ (due to the assumption on normal terms), we now have full fledged commands which are interpolated in a way close to the tensor case in \cite[Thm 20]{Saurin}. 

\end{proof}

% !TEX root = main.tex
\section{Proof-relevant interpolation for circular proofs}
\label{sec:proofRelIntCircProof}

In this section, we extend the proof-relevant interpolation procedure for \LL proofs presented in the Section \ref{sec:proofRelIntLL}, for \muLLo proofs (restricted to the propositional setting for simplicity). We will allow derivations with back-edges and cuts. While the cuts will be subject to conditions inherited from the previous section, the structure of the back-edges may or may not have an impact on the interpolation result. Indeed, while any pre-proof can be interpolated in a proof-relevant way (this is an easy corollary of the construction of the following section), it is meaningful to wonder whether a derivation $\pi$ that satisfies the progress condition can be interpolated with two proofs $\pi_1,\pi_2$ which also satisfy the progress condition. 
Intuitively, given the structure of proof-relevant interpolation, the reader should have in mind that each cycle of $\pi$ will give rise to both a cycle in $\pi_1$ and in $\pi_2$, while the inferences of $\pi$ will get dispatched between $\pi_1$ and $\pi_2$ depending on whether they correspond to a left-split or right-split formula. A consequence is that if a given cycle is validated by a trace on a formula $F$ that is left-split, $\pi_1$ will inherited the progress of $\pi$ but not $\pi_2$: the interpolant must therefore be built in such a way as to restore progress to the corresponding cycle in $\pi_2$.
The following paragraphs explain why and how this is possible.  

\medskip 
To accomplish this goal, we first need to introduce some additional notions on circular proofs, adapting to \muLLo proofs various notions due to Brotherston \cite{brotherston} and then we will respectively extend the ascending and descending phase for circular pre-proofs. Without loss of generality, we will always assume that the bud-companion relation $\mathcal{R}$ of a pre-proof $\mathcal{P}$ is injective, as explained in the Appendix \ref{appendix:injectiveRel}.

% !TEX root = main.tex
\subsection{(Ordered) trace manifolds for circular pre-proofs}

Intuitively, \defname{(ordered) trace manifolds} are finite structures containing finite segments of valid infinite traces that can be stitched together to generate every progressing trace for each infinite path in the pre-proof.  
 We first introduce some preliminary definitions and lemmas.

\begin{definition}[Induced subgraph]
    Let $G = (V, s, r, p)$ be a derivation graph and let $V' \subseteq V$. Then the \defname{subgraph} of $G$ induced by $V'$ is defined as $S = (V', s, r, p')$ where for some $j \in \mathbb{N}$, $p'_j(v) = p_j(v)$ if $v \in V'$ and $p_j(v) \in V'$, and undefined otherwise. We write $S \subseteq G$ to say that $S$ is a subgraph of $G$. If $V' = \emptyset$ we say that the subgraph is \defname{trivial}. 
\end{definition}

%\begin{definition}[Strong/weak graph connectivity]
%    A directed graph $G = (V, E)$ is said to be \defname{strongly connected} if for any $v, v' \in V$, there is a path from $v$ to $v'$. We say that $G$ is \defname{weakly connected} if it is strongly connected after adding the edges in the converse of the relation $E$.
    %, that is, for any two vertices there is a path between them in the set of edges $E \cup \{(v',v) \ | \ (v,v') \in E \}$. $G$ is said to be \textit{non-trivial} if $|V| > 1$ or if it contains only one vertex $v$ and there is an edge $(v,v) \in E$.
%\end{definition}

Every strongly connected subgraph of a circular pre-proof can be characterized by the cycles obtained from the bud-companion relationships. These are \defname{basic cycles}.

\begin{definition}[Basic cycle]
    \label{def:BaiscCycle}
    Let $\mathcal{P} = (\mathcal{D}, \mathcal{R})$ be a \muLLo pre-proof and let $B \in Bud(\mathcal{D)}$. Then the \defname{basic cycle} $\mathcal{C}_{B}$ in $\mathcal{G}_{\mathcal{P}}$ is obtained from the unique path from $\mathcal{R}(B)$ to $B$ by replacing the unique edge $(v, B)$ in $\mathcal{D}$ for the edge $(v, \mathcal{R}(B))$ in $\mathcal{G}_{\mathcal{P}}$. Since $\mathcal{C}_{B}$ is a cycle in $\mathcal{G}_{\mathcal{P}}$, it is also a non-trivial strongly connected subgraph in $\mathcal{G}_{\mathcal{P}}$.
\end{definition}

We can characterize how basic cycles intersect with each other by extracting the structural connectivity of the bud nodes.

\begin{definition}[Structural connectivity]
    \label{def:structConnect}
    Let $\mathcal{P} = (\mathcal{D}, \mathcal{R})$ be a \muLLo pre-proof. The relation $\leq_{\mathcal{P}}$ on $Bud(\mathcal{D})$ is defined as: $B_2 \leq_{\mathcal{P}} B_1$ iff $\mathcal{R}(B_2)$ appears on the basic cycle $\mathcal{C}_{B_1} \in \mathcal{G}_{\mathcal{P}}$.
\end{definition}

The union of basic cycles, whose bud nodes are weakly $\leq_{\mathcal{P}}$-connected, give rise to strongly connected subgraphs in a pre-proof. 

\begin{lemma}[Brotherston \cite{brotherston}, Lemma 7.1.7]
    \label{lem:SCCAsUnion} 
    Let $\mathcal{P} = (\mathcal{D}, \mathcal{R})$ be a \muLLo pre-proof and let $S$ be a non-trivial subgraph of $\mathcal{G}_{\mathcal{P}}$. Then $S$ is strongly connected iff there exists a non-empty set of buds $\mathcal{B}_{S} \subseteq Bud(\mathcal{D})$ such that $\mathcal{B}_{S}$ is weakly $\leq_{\mathcal{P}}$-connected and $S = \bigcup_{B \in \mathcal{B}_{S}} \mathcal{C}_{B}$.
\end{lemma}

Using the previous constructions, we define \defname{trace manifolds} in the setting of valid \muLLo pre-proofs. We therefore adapt Brotherston’s original definition of trace manifolds (\cite{brotherston}, Definition 7.2.1) to accommodate to the requirements of valid proofs in our system.

\begin{definition}[Trace manifold]
%    Let $\mathcal{P} =(\mathcal{D}, \mathcal{R})$ be a \muLLo pre-proof. 
A \defname{trace manifold} for  a \muLLo pre-proof $\mathcal{P} =(\mathcal{D}, \mathcal{R})$ is a set of traces
%    \begin{align*}
        $\{\uptau_{S,B_i} \ | \ S = \bigcup_{B \in \mathcal{B}_S} \mathcal{C}_{B} \text{ where } \mathcal{B}_S \subseteq Bud(\mathcal{D}) \text{ weakly $\leq_{\mathcal{P}}$-connected, $B_i \in \mathcal{B}_S$} \}$
%    \end{align*}
s.t.:
%    satisfying:
    \begin{enumerate}
        \item For all $S$ and $B_i \in \mathcal{B}_S$, $\uptau_{S,B_i}$ is a finite trace following the basic cycle $\mathcal{C}_{B_i}$ in $\mathcal{G}_{\mathcal{P}}$ and furthermore, it uses the same formula at both instances of $\mathcal{R}(B_i)$ on $C_{B_i}$. (Equivalently, $\uptau_{S,B_i}$ is a trace following the unique path in $\mathcal{D}$ from $\mathcal{R}(B_i)$ to $B_i$ and uses the same formula at $\mathcal{R}(B_i)$ and $B_i$).
        
        \item For all $S$, if $B_i,B_j \in \mathcal{B}_{S}$ and $B_j \leq_{\mathcal{P}} B_i$ then $\uptau_{S,B_j}(\mathcal{R}(B_j)) = \uptau_{S,B_i}(\mathcal{R}(B_j))$.

        \item For every $S$, let $\uptau_{S,B_{i_1}}, \uptau_{S,B_{i_2}}, ... \uptau_{S,B_{i_n}}$ be the traces  associated to $S$ in the manifold. Then there is a $k$ such that $\uptau_{S,B_{i_k}}$ has at leat one $F$-progress point and $F$ is minimal in $\bigcup^{n}_{j=1} \textsf{Form}(\uptau_{S,B_{i_j}})$ with respect to the subformula relation, where $\textsf{Form}(\uptau)$ is the set of formulas that appear in $\uptau$.
    \end{enumerate}
\end{definition}

The notion of \defname{ordered trace manifolds} defined below allows us to group further the valid traces of a circular pre-proof, according to an \defname{induction order}.

\begin{definition}[Induction order]
    \label{def:indOrder}
    Let $\mathcal{P} =(\mathcal{D}, \mathcal{R})$ be a \muLLo pre-proof. A (non-strict) partial order $(Bud(\mathcal{D}), \vartriangleleft)$ is said to be an \defname{induction order} for $\mathcal{P}$ if: 
    \begin{itemize}
        \item it is \defname{forest-like}: $B \vartriangleleft B_1$ and $B \vartriangleleft B_2$ implies $B_1 = B_2$ or $B_1 \vartriangleleft B_2$ or $B_2 \vartriangleleft B_1$ 
        %(i.e. $\vartriangleleft$ is \defname{forest-like})
        \item Every weakly $\leq_{\mathcal{P}}$-connected set $\mathcal{B} \subseteq Bud(\mathcal{D})$ has a $\vartriangleleft$-greatest element, i.e. an element $B_{max} \in \mathcal{B}$ such that $B \vartriangleleft B_{\max}$ for all $B \in \mathcal{B}$. (Note that in particular we then have $B_1 \leq_{\mathcal{P}} B_2$ implies $B_1 \vartriangleleft B_2$ or $B_2 \vartriangleleft B_1$)
    \end{itemize}
\end{definition}

Again, we modify Brotherston's original definition of ordered trace manifold (\cite{brotherston}, Definition 7.2.6) to accommodate to our notion of validity.

\begin{definition}[Ordered trace manifold]
    \label{def:ordTraceMan}
    Let $\mathcal{P} =(\mathcal{D}, \mathcal{R})$ be a \muLLo pre-proof and let $\vartriangleleft$ be an induction order for $\mathcal{P}$. An \defname{ordered trace manifold} with respect to $\vartriangleleft$ is a set of traces
%    \begin{align*}
$        \{\uptau_{B_j,B_i} \ | \ B_j, B_i \in Bud(\mathcal{D}), B_i \vartriangleleft B_j \}
$
%    \end{align*}
    satisfying the following:
    \begin{enumerate}
        \item For all $B_i, B_j \in Bud(\mathcal{D})$ with $B_i \vartriangleleft B_j$, $\uptau_{B_j,B_i}$ is a trace following the basic cycle $\mathcal{C}_{B_i}$ in $\mathcal{G}_{\mathcal{P}}$ and furthermore, it uses the same formula at both instances of $\mathcal{R}(B_i)$ on $\mathcal{C}_{B_i}$. (Equivalently, $\uptau_{B_j,B_i}$ is a trace following the unique path in $\mathcal{D}$ from $\mathcal{R}(B_i)$ to $B_i$ and uses the same formula at $\mathcal{R}(B_i)$ and $B_i$).

        \item For all $B_i,B_j,B_k \in Bud(\mathcal{D})$, if $B_j,B_i \vartriangleleft B_k$ and $B_j \leq_{\mathcal{P}} B_i$, then $\uptau_{B_k,B_j}(\mathcal{R}(B_j)) = \uptau_{B_k,B_i}(\mathcal{R}(B_j))$.

        \item For every $B_k \in Bud(\mathcal{D}$), let $\uptau_{k,B_{i_1}}, \uptau_{k,B_{i_2}}, ... \uptau_{k,B_{i_n}}$ be the traces  associated to $B_k$ in the manifold. Then $\uptau_{B_k,B_k}$ has at least one $F$-progress point and $F$ is minimal in $\bigcup^{n}_{j=1} \textsf{Form}(\uptau_{B_k,B_{i_j}})$ with respect to the subformula relation.
    \end{enumerate}
\end{definition}

Following Brotherston's proof ideas (\cite{brotherston}, Proposition 7.2.3 and Lemma 7.2.8), we can prove that, for a given pre-proof, having a trace manifold implies validity, and that having a trace manifold is equivalent to having an ordered trace manifold. The proofs of Lemmas \ref{lemm:traceManifold} and \ref{lemm:ordTraceManifold} are included in the Appendix \ref{appendix:traceManifolds}.

\begin{lemma}
    \label{lemm:traceManifold}
    A \muLLo pre-proof is valid if it has a trace manifold.
\end{lemma}

\begin{lemma}
    \label{lemm:ordTraceManifold}
    A \muLLo pre-proof $\mathcal{P}$ has a trace manifold if and only if there exists an induction order $\vartriangleleft$ for $\mathcal{P}$, and $\mathcal{P}$ has an ordered trace manifold with respect to $\vartriangleleft$.
\end{lemma}

\subsection{Ascending phase for circular proofs}

First, we adapt the definition of decorated and coherently decorated proofs to circular pre-proofs:

\begin{definition}[Decorated and coherent $\mu\textsf{LL}^{\omega}$ pre-proofs]
    Let $\mathcal{P} = (\mathcal{D}, \mathcal{R})$ be a \muLLo pre-proof. $\mathcal{P}$ is a \defname{decorated pre-proof} if its derivation tree $\mathcal{D}$ is decorated. $\mathcal{P}$ is a \defname{coherently decorated pre-proof} if (i) its derivation tree $\mathcal{D}$ is coherent and (ii) for all bud node $B \in Bud(\mathcal{D})$, then the split of $s(B)$ is the same split of its companion node $s(\mathcal{R}(B))$.
\end{definition}

It is not immediate to generalise Lemma \ref{lem:coherentProof} to circular proofs since the coherence property is not necessarily maintained by the relation $\mathcal{R}$ (see example in Appendix \ref{appendix:exampleNonCoh}). Up to some unfoldings of back-edges, this can however be restored by preserving validity:

\begin{lemma}
\label{lem:splitInvariant}
Let $\mathcal{P}_0 = (\mathcal{D}_0, \mathcal{R}_0)$ be a valid $\muLLo$ proof with a trace manifold, with a split conclusion $\vdash \Gamma_0, \Delta_0$, and injective $\mathcal{R}$. There exists a $\muLLo$ proof $\mathcal{P} = (\mathcal{D}, \mathcal{R})$ with a trace manifold, having the same infinite unfolding as $\mathcal{P}_0$, such that the split conclusion $\vdash \Gamma_0, \Delta_0$ produces a coherently decorated proof after the ascending phase.
\end{lemma}

\begin{proof}[Proof sketch]
The proof of the lemma amounts to the finiteness of the number of splittings of a given sequent: by unfolding one can ensure that every bud can be associated with a companion decorated similarly. 
In Appendix \ref{appendix:correctAlgo1} the proof goes by introducing an algorithm (Algorithm \ref{algo:splitInv}) that actually 
builds for any \muLLo pre-proof $\mathcal{P} = (\mathcal{D}, \mathcal{R})$ with injective $\mathcal{R}$, an equivalent pre-proof via unfolding, that maintains coherence after the ascending phase.
\end{proof}

\begin{remark}
The previous lemma can be generalised to ensure the existence of a valid $\muLLo$ proof $\mathcal{P}$ that can be coherently decorated w.r.t. to any split. 
\end{remark}
 
%=======================================================================
%=======================================================================
%=======================================================================
%=======================================================================
%=======================================================================

\subsection{Tree-compatibility}

Before the definition of the descending phase for circular proofs, we need to make sure that the circular proof that we want to interpolate satisfies the \defname{tree-compatibility} property.

\begin{definition}[Tree-Compatibility]
    An induction order $(Bud(\mathcal{D}), \vartriangleleft)$ is said to be \defname{tree-compatible} if $B_1 \leq_{\mathcal{P}} B_2$ and $B_2 \nleq_{\mathcal{P}} B_1$ implies $B_1 \vartriangleleft B_2$, for any $B_1, B_2 \in Bud(\mathcal{D})$.
\end{definition}

Fortunately, we can transform any trace manifold of a valid circular proof that can be coherently decorated, into an ordered trace manifold using Lemma \ref{lemm:ordTraceManifold}. Additionally, this proof can be furthered transformed into an equivalent proof that can be also coherently decorated, with a new ordered trace manifold that has induction order that is tree-compatible.

\begin{lemma}
    \label{lemm:treeComp}
Let $\mathcal{P}_0 = (\mathcal{D}_0, \mathcal{R}_0)$ be a valid $\mu\textsf{LL}^{\omega}$ proof with an ordered trace manifold with induction order $(Bud(\mathcal{D}_0),\vartriangleleft_0)$, that can be coherently decorated. 
There is a valid $\mu\textsf{LL}^{\omega}$ proof $\mathcal{P}$ with an ordered trace manifold with a tree-compatible induction order $(Bud(\mathcal{D}),\vartriangleleft)$, that can be coherently decorated.
\end{lemma}

\begin{proof}[Proof sketch]
The proof (see Appendix \ref{appendix:correctAlgo2}) generalises to our setting a similar property by Sprenger and Dam:  we introduce Algorithm \ref{algo:treecomp} that performs the required unfoldings of back-edges to the original pre-proof in order to make it tree-compatible. 
After executing Algorithm \ref{algo:treecomp}, relation $\mathcal{R}$ may not be injective anymore but  $\mathcal{P}$ can be transformed to maintain tree-compatibility while restoring injectivity of $\mathcal{R}$, as explained in Appendix \ref{appendix:injectiveRel}.
\end{proof}

%=======================================================================
%=======================================================================
%=======================================================================
%=======================================================================
%=======================================================================

\subsection{Descending phase for circular proofs}
\label{sub:desPhaseCircProofs}
We extend the descending phase to work on \muLLo pre-proofs $\mathcal{P} = (\mathcal{D}, \mathcal{R})$ with injective $\mathcal{R}$. 
Our general approach follows two steps: the first step is to complete the treatment of the missing derivations cases for the 
%to add to the case analysis done in the Appendix \ref{appendix:proofElemPris}, the unconsidered
 elementary \CPRIS 
 %obtained
  by including the bud nodes and the fixed-point rules, so we can use the descending phase over the derivation tree $\mathcal{D}$:
% !TEX root = main.tex
\begin{figure}[t!]

\begin{tabular}{cc}
\centering
\begin{tabular}{c@{\hspace{1mm}}c@{\hspace{1mm}}c}
\scalebox{0.75}{
\begin{prooftree}
\Hypo{\spl{\Gamma},\spr{\Delta}}
\end{prooftree}
}
&
\scalebox{0.75}{$\rightarrow$}
&
\scalebox{0.75}{
\begin{prooftree}
\Hypo{\vdash \spl{\Gamma}, X_{B_i}}
\Hypo{\vdash \spr{\Delta}, X_{B_i}}
\Infer2[\textsf{Cut}]{\vdash \spl{\Gamma},\spr{\Delta}}
\end{prooftree}
}
\end{tabular} &

%===================================

\begin{tabular}{c@{\hspace{1mm}}c@{\hspace{1mm}}c}
\scalebox{0.75}{
\begin{prooftree}
    \hypo{\pi^L_1}
    \infer1[]{\vdash \spl{F[\sigma X.F/X],\Gamma'},I_1}
    \hypo{\pi^R_1}
    \infer1[]{\vdash I_1^{\bot}, \spr{\Delta}}
    \infer2[(\textsf{Cut})]{\vdash \spl{F[\sigma X.F/X],\Gamma'}, \spr{\Delta}}
    \infer1[($\sigma$)]{\vdash \spl{\sigma X.F,\Gamma'}, \spr{\Delta}}
\end{prooftree}
}
&
\scalebox{0.75}{$\rightarrow$}
&
\scalebox{0.75}{
\begin{prooftree}
    \hypo{\pi^L_1}
    \infer1[]{\vdash \spl{F[\sigma X.F/X],\Gamma'}, I_1}
    \infer1[($\sigma$)]{\vdash \spl{\sigma X.F,\Gamma'}, I_1}
    \hypo{\pi^R_1}
    \infer1[]{\vdash I_1^{\bot}, \spr{\Delta}}
    \infer2[(\textsf{Cut})]{\vdash \spl{\sigma X.F,\Gamma'}, \spr{\Delta}}
\end{prooftree}
}
\end{tabular} \\
%===================================
 \textbf{(a)} & \textbf{(b)}
\end{tabular}
\caption{Elementary and solved \CPRIS (a)  for bud nodes, (b)  for the $\mu, \nu$ rules.}
\label{fig:newPRIS}
\end{figure}
\begin{itemize}
    \item \textbf{Bud nodes:} Let $Bud(\mathcal{D}) = \{ B_1, \cdots, B_n \}$. To each bud node $B_i$ with split sequent $\vdash \Gamma, \Delta$, we associate the fresh fixed-point variable $X_{B_i}$, and transform the leaf in $\mathcal{D}$, as done in the subfigure (a) of Figure \ref{fig:newPRIS}. 
    \item \textbf{Fixed-point rules:} Without loss of generality, consider a split $(\Gamma, \Delta)$ with $\Gamma =  \Gamma' \cup \{ \sigma X. F \}$. The elementary \CPRIS for the $\sigma$-rules, where $\sigma \in \{ \mu, \nu \}$, can be solved as shown in the subfigure (b) of Figure \ref{fig:newPRIS}.
\end{itemize}

The second step is to maintain the bud-companion relations of the original proof by transforming the companion nodes during the descending phase, while appropriately pairing each new bud with its corresponding companion:
\begin{itemize}
    \item \textbf{Companion nodes:} Suppose we reach a companion node $C$ with a split sequent $\vdash \Gamma, \Delta$, during the descending phase. Since $\mathcal{R}$ is injective, there is a unique bud node $B \in Bud(\mathcal{D})$ such that $\mathcal{R}(B) = C$, with an associated fixed-point variable $X_{B}$, as explained before. After solving the elementary \CPRIS for the node $C$ with rule $r(C)$, we are left with the solved \CPRIS \adjustbox{valign=c, scale=0.75}{
        \begin{prooftree}
                \hypo{\mathcal{P}^l}
                \infer1[]{\vdash \Gamma, I}
                \hypo{\mathcal{P}^r}
                \infer1[]{\vdash \Delta, I^{\bot}}
                \infer2[\textsf{(Cut)}]{\vdash \Gamma, \Delta}
            \end{prooftree} 
        } where $\mathcal{P}^l = (\mathcal{D}^l, \mathcal{R}^l)$ and $\mathcal{P}^r = (\mathcal{D}^r, \mathcal{R}^r)$ are partial pre-proofs, and $X_B$ is a free variable in $I$. We then transform the mentioned solved \CPRIS into:
%        \begin{center}
\qquad             \scalebox{.8}{
            \begin{prooftree}
                \hypo{\mathcal{P}^l[\sigma X_B. I/X_B]}
                \infer1[]{\vdash \Gamma, I[\sigma X_B. I/X_B]}
                \infer1[$(\sigma)$]{\vdash \Gamma, \sigma X_B. I}

                \hypo{\mathcal{P}^r[\sigma^{\bot} X_B.I^{\bot}/X_B]}
                \infer1[]{\vdash \Delta, I^{\bot}[\sigma^{\bot} X_B.I^{\bot}/X_B]}
                \infer1[$(\sigma^{\bot})$]{\vdash \Delta, \sigma^{\bot} X_B.I^{\bot}}
                
                \infer2[\textsf{(Cut)}]{\vdash \Gamma, \Delta}
            \end{prooftree}
            }
%        \end{center}
        where $\mathcal{P}^l[\sigma X_B. I/X_B]$ (resp. $\mathcal{P}^r[\sigma^{\bot} X_B.I^{\bot}/X_B]$) updates all occurrences of $X_B$ in $\mathcal{P}^l$ (resp. $\mathcal{P}^r$) with $\sigma X_B. I$ (resp. $\sigma^{\bot} X_B.I^{\bot}$). Hence, this substitution also updates the bud node $B^l$ labelled with $\vdash \Gamma, X_B$ (resp. $B^r$ labelled with $\vdash \Delta, X_B$) in $\mathcal{D}^l$ (resp. $\mathcal{D}^r$) into $\vdash \Gamma, \sigma X_B. I$ (resp. $\vdash \Delta, \sigma^{\bot} X_B.I^{\bot}$). Consequently, in the function $\mathcal{R}^l$ (resp. $\mathcal{R}^r$), we pair the bud node $B^l$ (resp. $B^r$) with the left (resp. right) premise of the cut. Additionally, $\sigma \in \{\mu, \nu \}$ with $\mu^{\bot} = \nu$ and $\nu^{\bot} = \mu$. 
        While the choice of the value to assign for $\sigma$ is irrelevant if we were simply interpolating pre-proof (disregarding validity), it will on the contrary happen to be crucial to interpolate while preserving progress as mentioned at the start of the section. This condition will be studied in the proof of the Theorem \ref{the:main}. 
\end{itemize}

\begin{remark}
\label{remark:cut-intro-circ}
Two  crucial remarks are in order at this point: 
(i) while the process described above (that does not cope yet with the question of progress) does a global update of the proof when  companions are encountered during the descending phase -- via the substitution of the interpolating formula -- the sequence of proofs corresponds locally to a cut-introduction: as a consequence, once all the substitution are introduced when the descending phase terminates, reversing the proof sequence provides a \defname{cut-elimination sequence}. Moreover,
(ii) the above-mentioned cut-elimination sequence happens to synchronise on the buds: this is the reason why circularity is preserved by cut-elimination in the next theorem.
 \end{remark}
 
The following theorem provides a detailed description of the properties of the resulting pre-proof after the interpolation process.

\begin{theorem}[Proof-relevant interpolation with (semi-)analytic cuts for \muLLo]
\label{the:main}
Let $\mathcal{P}_0 = (\mathcal{D}_0, \mathcal{R}_0)$ be a valid $\mu\textsf{LL}^{\omega}$ proof with a trace manifold and a split conclusion $\vdash \Gamma, \Delta$, such that every cut-inference is (semi-)analytic. After the ascending and descending phase on $\mathcal{P}_0$, we obtain an interpolated pre-proof $\mathcal{P} = \adjustbox{valign=c, scale=0.75}{
        \begin{prooftree}
                \hypo{\mathcal{P}^l}
                \infer1[]{\vdash \Gamma, I}
                \hypo{\mathcal{P}^r}
                \infer1[]{\vdash \Delta, I^{\bot}}
                \infer2[\textsf{(Cut)}]{\vdash \Gamma, \Delta}
            \end{prooftree} 
        }$ satisfying that (i) $\mathcal{P}^l = (\mathcal{D}^l, \mathcal{R}^l)$ and $\mathcal{P}^r = (\mathcal{D}^r, \mathcal{R}^r)$ are two pre-proofs with conclusions $\vdash \Gamma, I$ and $\vdash \Delta, I^{\bot}$ respectively, where $\mathcal{R}^l$ (resp. $\mathcal{R}^r$) is injective and $\voc{(I)}\subseteq \voc {(\Gamma)}\cap \voc {(\Delta)}$, (ii) $\mathcal{P}$ is a valid proof, (iii) $\mathcal{P} \rightarrow^{\omega}_{\rcut} \mathcal{P}_0$.
\end{theorem}

\begin{proof}
    The full proof is given in Appendix \ref{appendix:proofMainTheorem}. We transform the trace manifold for $\mathcal{P}_0$ into an ordered trace manifold using the Lemma \ref{lemm:ordTraceManifold}. Then using Algorithm \ref{algo:splitInv} and \ref{algo:treecomp}, we obtain a new proof that can be coherently decorated, and has an ordered trace manifold with a tree-compatible induction order. After the descending phase, the structural connectivity and induction order of $\mathcal{P}_0$ are duplicated on both sides $\mathcal{P}^l$ and $\mathcal{P}^r$, while each trace of its ordered trace manifold is transferred to exactly one side (Lemma \ref{lemm:finInterpol}). The remaining copies of the cycles are validated by traces following the interpolant. It suffices to choose each $\sigma_i$ (resp. $\sigma^{\perp}_i$) as $\nu$ (resp. $\mu$) opposite to the transfer side, since tree-compatibility guarantees the minimality condition (Theorem \ref{the:preFinal0}).

For the last part of the theorem one can perform an (infinite) cut-elimination reduction. \Cref{remark:cut-intro-circ}.(i) ensures that it consists in a cut-elimination reduction sequence while \Cref{remark:cut-intro-circ}.(ii) ensures that the cut-reduction can be performed by unfolding the circular proofs and preserving the synchronisation of the two circular proofs so that at the limit, the proof object is regular and is the infinite unfolding of the original circular proof $\mathcal{P}_0$.
\end{proof}
 
\begin{remark}
Note that while usually \muLLo circular proofs are not closed by cut-elimination~\cite{baeldeDoumaneSaurin} (the limit of the cut-reduction sequence may be non-regular), the structure of interpolants and interpolating proofs is such that this property holds. Intuitively, this comes from the fact that the constructions we perform essentially works by "infinitary" cut-introduction. 
\end{remark}

% !TEX root = main.tex
\section{Conclusion}
\label{sec:conclusion}

\noindent{\bf Summary of the paper.}
In this paper we established  Maehara-\v Cubri\' c style interpolation lemmas and the respective interpolation theorems first for \LL proofs with cuts (under adequate assumptions on the vocabularies used by the cut formulas) and to \muLLi circular proofs (under the assumption that the circular proof has a trace manifold).
The proof-relevance requirement forced us to be extra careful in the management of the proofs during interpolation. For instance, one cannot exploit a general cut-free completeness theorem coupled with a non-wellfounded proof regularisation result to consider only cut-free circular proofs. This is the reason why our result is subject to some assumptions on the class of proofs that we can interpolate, providing a further motivation for considering cuts in proofs during interpolation.

To our knowledge, our management of cuts extends the various existing (non proof-relevant) treatments of interpolation with cuts. Moreover, we discussed that proof-relevance allows to foresee generalisations of the results that exploit more precise information on how the cut-formulas are actually used in the proof. 

\medskip
\noindent{\bf Future works.}
First, we plan to  analyse under which condition on the use of cut-formulas in axioms we can preserve the interpolation theorems of this paper.

Second, a crucial future work is about achieving this refinement of \v Cubri\'c' results.
In \Cref{sec:computational}, we discussed the computational meaning of our results and how our treatment of cuts could allow extending \v Cubri\'c' results in the simply-typed $\lambda$-calculus to non normal terms.
We conjecture that our results 
%on proof-relevant interpolation with cuts are transferable 
 can be used to strengthen \v Cubri\'c results for the simply typed $\lambda$-calculus. 
 Still, extending our analysis of proof-relevant interpolation for non cut-free proofs to the simply-typed $\lambda$-calculus and natural deduction is not direct due to the fact that 
 %. Indeed, while in the sequent calculus the existence of the cut-inference at the object level makes it easy to track, the fact that
  the notion of redexes lives at the meta-level
  %, in lambda-calculus and natural deduction 
  which makes it more complex to handle\footnote{{\it E.g.} redexes may hide other redexes, like in $(\lambda x. \lambda y. t) u v$ where arguments $u$ is passed through $x$ and $v$ through $y$. Therefore, when the subformula property does not hold, we would need to express our constraints on \emph{potential-cuts} (consider for instance $((\lambda x. x) \lambda y.u) v$).}.
 
 Still, to support and illustrate our conjecture we show how it works on a simple example. Assume that we are given types $A,B,C$ such that the atomic types of $C$ all occurs in $B$, and assume that we have $x:A \vdash \lambda y. u: C \rightarrow B$ and $x:A \vdash v: C$ both normal. We easily see that one can interpolate $(\lambda y. u)v$ without needing to normalize first: by \v Cubri\'c result there are (i) types $I$ with atomic types occurring both in $A$ and $B$ and  $J$ with atomic types occurring both in $A$ and $C$, (ii) normal terms $x: A\vdash u_1: I$ and $z': I \vdash u_2: C \rightarrow B$ such that $u_2\{u_1/z'\} \longrightarrow_\beta^\star u$ and 
(iii) normal terms $x: A\vdash v_1: J$ and $z'': J \vdash v_2: C$ such that $v_2\{v_1/z''\} \longrightarrow_\beta^\star v$.
By setting $K= I\times J$,  $t_1 = (u_1,v_1)$ and $t_2 = \lambda z^K. (u_2\{\pi_1(z)/z'\}) v_2\{\pi_2(z)/z''\}$, one easily sees that 
(i) the atomic types of $K$ occur both in $A$ and $B$, (ii)  $x:A \vdash t_1: K$, (iii) $\vdash t_2: K \rightarrow B$ and (iv) 
$(t_2)t_1 \longrightarrow_\beta^\star (\lambda y. u)v$.
The recent work of Lennon-Bertrand and Saurin who revisited \v Cubri\'c' results using bidirectional typing systems to control the subformula property and better state the inductive assumption at the core of \v Cubri\'c' proof may be a useful tool in this direction in order to control the language of the types involved in redexes.

\bibliography{references}

\newpage
\appendix

% !TEX root = main.tex
\section{Appendix on Section \ref{sec:preliminaries}}

\subsection{\LL cut-elimination reduction}
\label{app:ll-cut-red}

\newcommand{\extllder}{
\begin{prooftree}
\Hypo{\vdash \Sigma, C}
\Hypo{\vdash C^\perp, \Delta, F}
\Infer{1}[\rder]{\vdash C^\perp, \Delta, ?F} 
\Infer{2}[\rcut]{\vdash \Sigma, \Delta, ?F}
\end{prooftree}
&\underset{r}{\longrightarrow}& 
\begin{prooftree}
\Hypo{\vdash \Sigma, C}
\Hypo{\vdash C^\perp, \Delta,F}
\Infer{2}[\rcut]{\vdash \Sigma, \Delta, F}
\Infer{1}[\rder]{\vdash \Sigma, \Delta, ?F}
\end{prooftree}
\\[25pt]
}

\newcommand{\extllcont}{
\begin{prooftree}
\Hypo{\vdash \Sigma, C}
\Hypo{\vdash C^\perp, \Delta,?F,?F}
\Infer{1}[\rcontr]{\vdash C^\perp, \Delta, ?F} 
\Infer{2}[\rmcutpar]{\vdash \Sigma, \Delta,?F}
\end{prooftree}
&\underset{r}{\longrightarrow}&
\begin{prooftree}
\Hypo{\vdash \Sigma, C}
\Hypo{\vdash C^\perp, \Delta,?F, ?F}
\Infer{2}[\rcut]{\vdash \Sigma, \Delta,?F,?F}
\Infer{1}[\rcontr]{\vdash \Sigma, \Delta,?F}
\end{prooftree}
\\[25pt]
}

\newcommand{\extllweak}{
\begin{prooftree}
\Hypo{\vdash \Sigma, C}
\Hypo{\vdash C^\perp,\Delta}
\Infer{1}[\rweak]{\vdash C^\perp, \Delta, ?F} 
\Infer{2}[\rmcutpar]{\vdash \Sigma, \Delta, ?F}
\end{prooftree}
&\underset{r}{\longrightarrow}&
\begin{prooftree}
\Hypo{\vdash \Sigma, C}
\Hypo{\vdash C^\perp, \Delta}
\Infer{2}[\rcut]{\vdash \Sigma, \Delta}
\Infer{1}[\rweak]{\vdash \Sigma, ?F}
\end{prooftree}
\\[25pt]
}

\newcommand{\extllprom}{
  \begin{prooftree}
     \Hypo{\vdash F, ? \Gamma, \wn C}
      \Infer{1}[\rprom]{\vdash ! F, ?\Gamma, \wn C}
% \Hypo{\C C}
      \Hypo{\vdash \oc C^\perp, \wn \Sigma}      
%      \Infer{1}[\rprom~$1\leq i \leq n$]{\vdash  ! G_i^\bot,  ? \Delta_i}
%     \Hypo{\scriptsize 1\leq i \leq n}
      %\Hypo{\vdash  G_1^\bot,  ? \Delta_1}      \Infer{1}[\rprom]{\vdash  ! G_1^\bot,  ? \Delta_1}
      %\Hypo{\dots}
      %\Hypo{\vdash  G_n^\bot,  ? \Delta_n}      \Infer{1}[\rprom]{\vdash  ! G_n^\bot,  ? \Delta_n}
      \Infer{2}[\rmcutpar]{\vdash ! F, \wn \Gamma, ? \Sigma}
  \end{prooftree}
  %&&
%\\[20pt] \hspace{4cm}
&
\underset{r}{\longrightarrow}
&
%\\
  \begin{prooftree}
    \Hypo{\vdash F, ? \Gamma}
%\Hypo{\vdash G, ?\Delta}
      \Hypo{\vdash \oc C^\perp, \wn \Sigma}
 %    \Hypo{1\leq i \leq n}
%      \Hypo{\vdash  G_1^\bot,  ? \Delta_1}      \Infer{1}[\rprom]{\vdash  ! G_1^\bot,  ? \Delta_1}
%      \Hypo{\dots}
%      \Hypo{\vdash  G_n^\bot,  ? \Delta_n}      \Infer{1}[\rprom]{\vdash  ! G_n^\bot,  ? \Delta_n}
      \Infer{2}[\rcut]{\vdash F, \wn \Gamma, ?\Sigma}
      \Infer{1}[\rprom]{\vdash !F, \wn \Gamma, ? \Sigma}
    \end{prooftree}
    %&&
\\[25pt]
}

\newcommand{\extllpromold}{
  \begin{prooftree}
     \Hypo{\vdash F, ? \Gamma,  ?G_1, \dots, ?G_n}
      \Infer{1}[\rprom]{\vdash ! F, ?\Gamma,  ?G_1, \dots, ?G_n}
% \Hypo{\C C}
      \Hypo{\vdash  G_i^\bot,  ? \Delta_i}      
      \Infer{1}[\rprom~$1\leq i \leq n$]{\vdash  ! G_i^\bot,  ? \Delta_i}
%     \Hypo{\scriptsize 1\leq i \leq n}
      %\Hypo{\vdash  G_1^\bot,  ? \Delta_1}      \Infer{1}[\rprom]{\vdash  ! G_1^\bot,  ? \Delta_1}
      %\Hypo{\dots}
      %\Hypo{\vdash  G_n^\bot,  ? \Delta_n}      \Infer{1}[\rprom]{\vdash  ! G_n^\bot,  ? \Delta_n}
      \Infer{2}[\rmcutpar]{\vdash ! F, ? \Gamma, ? \Delta_1,\dots ? \Delta_n}
  \end{prooftree}
  %&&
%\\[20pt] \hspace{4cm}
&
\underset{r}{\longrightarrow}
&
%\\
  \begin{prooftree}
    \Hypo{\vdash F, ? \Gamma,  ?G_1, \dots, ?G_n}
%\Hypo{\vdash G, ?\Delta}
      \Hypo{\vdash  G_i^\bot,  ? \Delta_i}      
      \Infer{1}[\rprom~$1\leq i \leq n$]{\vdash  ! G_i^\bot,  ? \Delta_i}
 %    \Hypo{1\leq i \leq n}
%      \Hypo{\vdash  G_1^\bot,  ? \Delta_1}      \Infer{1}[\rprom]{\vdash  ! G_1^\bot,  ? \Delta_1}
%      \Hypo{\dots}
%      \Hypo{\vdash  G_n^\bot,  ? \Delta_n}      \Infer{1}[\rprom]{\vdash  ! G_n^\bot,  ? \Delta_n}
      \Infer{2}[\rmcutpar]{\vdash F, ? \Gamma, ? \Delta_1,\dots ? \Delta_n}
      \Infer{1}[\rprom]{\vdash !F, ? \Gamma, ? \Delta_1,\dots ? \Delta_n}
    \end{prooftree}
    %&&
\\[25pt]
}

%{\scriptsize\ensuremath{\mathsf{mcut(\iota, \perp\!\!\!\perp')}

\newcommand{\intllder}{
  \begin{prooftree}
    \Hypo{
    %\pi}
    %\Infer{1}{
    \vdash F, \Gamma}
      \Infer{1}[\rder]{\vdash ? F, \Gamma}
 \Hypo{%\pi'}
    %\Infer{1}{
    \vdash F^\bot, ? \Delta}
      \Infer{1}[\rprom]{\vdash ! F^\bot, ? \Delta}
      \Infer{2}[\rmcutpar]{\vdash \Gamma, \wn \Delta}
  \end{prooftree}
%\\[20pt] \hspace{4cm}

 & \underset{r}{\longrightarrow} &
  \begin{prooftree}
    \Hypo{%\pi}
    %\Infer{1}{
    \vdash F, \Gamma}
 \Hypo{%\pi'}
    %\Infer{1}{
    \vdash F^\bot, ? \Delta}
      \Infer{2}[\rcut]{\vdash \Gamma, \wn \Delta}
  \end{prooftree}
\\[25pt]
}
\newcommand{\intllcont}{
  \begin{prooftree}
% \Hypo{\pi'}
%    \Infer{1}{\vdash F^\bot, ? \Delta}
%      \Infer{1}[\rprom]{\vdash ! F^\bot, ? \Delta}
            \Hypo{\pi}
            \Infer 1[\rprom] {\vdash {\oc F^\perp}, \wn \Delta}
    \Hypo{%\pi}
    %\Infer{1}{
    \vdash  ? F, ? F, \Gamma}
      \Infer{1}[\rcontr]{\vdash ? F, \Gamma}
      \Infer{2}[\rmcutpar]{\vdash\wn \Delta, \Gamma}
  \end{prooftree}
%\\[20pt] \hspace{4cm}
& \underset{r}{\longrightarrow} &
  \begin{prooftree}
%\Hypo{\pi'}
%    \Infer{1}{\vdash F^\bot, ? \Delta}
%    \Infer{1}[\rprom]{\vdash ! F^\bot,  ? \Delta}
%      \Infer{2}[\rcut]{\vdash ?^\bullet F, \Gamma,  ?^\bullet \Delta}
  \Hypo{\pi}
            \Infer 1[\rprom] {\vdash {\oc F^\perp}, \wn \Delta}%\Hypo{\pi'}
%    \Infer{1}{\vdash F^\bot, ? \Delta}
%    \Infer{1}[\rprom]{\vdash ! F^\bot,  ? \Delta}
  \Hypo{\pi}
            \Infer 1[\rprom] {\vdash {\oc F^\perp}, \wn \Delta}
            \Hypo{%\pi}
    %\Infer{1}{
    \vdash  ? F, ? F, \Gamma}
      \Infer{2}[\rcut]{\vdash\wn \Delta, ? F, \Gamma}
      \Infer{2}[\rcut]{\vdash\wn \Delta, \wn \Delta,\Gamma}
      \Infer{1}[\rcontr$^\star$]{\vdash \wn \Delta, \Gamma}
    \end{prooftree}
\\[25pt]
}

\newcommand{\intllweak}{
  \begin{prooftree}
  \Hypo{\pi}
            \Infer 1[\rprom] {\vdash {\oc F^\perp}, \wn \Delta}     \Hypo{
     %\pi}
    %\Infer{1}{
    \vdash \Gamma}
      \Infer{1}[\rweak]{\vdash ? F, \Gamma}
% \Hypo{\pi'}
%    \Infer{1}{\vdash F^\bot, ? \Delta}
%      \Infer{1}[\rprom]{\vdash ! F^\bot, ? \Delta}
      \Infer{2}[\rmcutpar]{\vdash  \wn \Delta, \Gamma}
  \end{prooftree}
%\\[20pt] 
  & \underset{r}{\longrightarrow} &
   \begin{prooftree}
     \Hypo{%\pi}
    %\Infer{1}{
    \vdash \Gamma}
      \Infer{1}[\rweak$^\star$]{\vdash \wn \Delta, \Gamma}
    \end{prooftree}
\\[25pt]
}

\def\rmcutpar{\rcut}
 \begin{definition}%[External reduction rules]
  \label{def:ExtReductions}
\defname{Cut commutations} or \defname{External reductions} are defined in \Cref{fig:external-muLLi}. 
\end{definition}

\begin{definition} \label{def:IntReductions}
\defname{\LL cut-reduction} are the reduction rules 
  given in  \Cref{fig:external-muLLi}, \Cref{fig:principal-muLLi} together with the usual cut-elimination rule for first-order quantifiers and the following two reductions (together with their symmetrical cases):
  \begin{itemize}
    \item
      the $\rcut/\rcut$ commutation
      $$\begin{array}{l}
        \scalebox{0.8}{\begin{prooftree}
          \Hypo{\vdash \Sigma, G}
          \Hypo{\vdash \Delta,F, G^\bot}
          \Hypo{\vdash \Gamma,F^\bot}
          \Infer{2}[\rcut]{\vdash G^\perp, \Delta,\Gamma}     
          \Infer{2}[\rcut]{\vdash \Sigma, \Delta,\Gamma}
        \end{prooftree}}
        \\[10pt]
        \qquad\qquad\qquad
        \ \underset{r}{\longrightarrow}\qquad
        \scalebox{0.8}{\begin{prooftree}
          \Hypo{\vdash \Sigma, G}
          \Hypo{\vdash \Delta,F, G^\bot}
          \Infer{2}[\rcut]{\vdash  \Sigma, \Delta,F}     
          \Hypo{\vdash \Gamma,F^\bot}
          \Infer{2}[\rcut]{\vdash \Sigma, \Delta,\Gamma}
        \end{prooftree}}
      \end{array}$$
    \item
      the $\rcut/\rax$ reduction 
      $$\begin{array}{l}
        \scalebox{0.8}{\begin{prooftree}
          \Infer{0}[\rax]{\vdash F, F^\bot}
          \Hypo{\pi}
          \Infer 1{\vdash F, \Gamma}    
          \Infer{2}[\rcut]{\vdash F, \Gamma}
        \end{prooftree}}
        \\[10pt]
        \qquad\quad\qquad\qquad\qquad
        \ \underset{r}{\longrightarrow}\ \qquad
     \scalebox{0.8}{\begin{prooftree}
             \Hypo{\pi}
          \Infer 1{\vdash F, \Gamma}    
         \end{prooftree}}
      \end{array}$$

\end{itemize}
\end{definition}

\begin{figure*}[htpb!]
  %\begin{framed}
  \begin{center}
%\scalebox{.95}{
$\begin{array}{clc}
\extllder
\extllcont
\extllweak
\extllprom
\begin{prooftree}
  \Hypo{\vdash \Sigma, C^\perp}
\Hypo{\vdash C, \Delta,F}
\Hypo{\vdash \Gamma,G}
\Infer{2}[\rtensor]{\vdash C, \Delta,\Gamma, F\tensor G} 
\Infer{2}[\rmcutpar]{\vdash \Sigma, \Delta, \Gamma, F\tensor G}
\end{prooftree}
&\underset{r}{\longrightarrow}& 
\begin{prooftree}
\Hypo{\vdash \Sigma, C^\perp}
\Hypo{\vdash C, \Delta,F}
\Infer{2}[\rcut]{\vdash \Sigma, \Delta, F}
\Hypo{\vdash \Gamma,G}
\Infer{2}[\rtensor]{\vdash \Sigma, \Delta, \Gamma, F\tensor G}
\end{prooftree}
\\[25pt]
\begin{prooftree}
\Hypo{\vdash \Sigma, C^\perp}
\Hypo{\vdash C, \Delta,F,G}
\Infer{1}[\rparr]{\vdash C, \Delta, F\parr G} 
\Infer{2}[\rmcutpar]{\vdash \Sigma, \Delta, F\parr G}
\end{prooftree}
&\underset{r}{\longrightarrow}&
\begin{prooftree}
\Hypo{\vdash \Sigma, C^\perp}
\Hypo{\vdash C, \Delta,F,G}
\Infer{2}[\rcut]{\vdash \Sigma, \Delta, F,G}
\Infer{1}[\rparr]{\vdash \Sigma, \Delta, F\parr G}
\end{prooftree}
\\[25pt]
 \begin{prooftree}
 \Hypo{\vdash \Sigma, C^\perp}
 \Hypo{\vdash C, \Delta,F}
 \Hypo{\vdash C, \Delta,G}
 \Infer{2}[\rwith]{\vdash C, \Delta, F\with G} 
 \Infer{2}[\rmcutpar]{\vdash \Sigma, \Delta, F\with G}
 \end{prooftree}
 &\underset{r}{\longrightarrow}& 
 \begin{prooftree}
 \Hypo{\vdash \Sigma, C^\perp}
 \Hypo{\vdash C, \Delta,F}
 \Infer{2}[\rcut]{\vdash \Sigma, \Delta, F}
 \Hypo{\vdash \Sigma, C^\perp}
 \Hypo{\vdash C, \Delta,G}
 \Infer{2}[\rcut]{\vdash \Sigma, \Delta, G}
 \Infer{2}[\rwith]{\vdash \Sigma, \Delta, F\with G}
 \end{prooftree}
 \\[25pt]
 \begin{prooftree}
 \Hypo{\vdash \Sigma, C^\perp}
 \Hypo{\vdash C, \Delta,F_i}
 \Infer{1}[\roplusi]{\vdash C, \Delta, F_1\oplus F_2} 
 \Infer{2}[\rmcutpar]{\vdash \Sigma, \Delta, F_1\oplus F_2}
 \end{prooftree}
 &\underset{r}{\longrightarrow}& 
 \begin{prooftree}
 \Hypo{\vdash \Sigma, C^\perp}
 \Hypo{\vdash C, \Delta,F_i}
 \Infer{2}[\rcut]{\vdash \Sigma,\Delta, F_i}
 \Infer{1}[\roplusi]{\vdash \Sigma, \Delta, F_1\oplus F_2}
 \end{prooftree}
 \\[25pt]
 \begin{prooftree}
 \Hypo{\vdash \Sigma, C^\perp}
 \Infer{0}[\rtop]{\vdash C, \Delta, \top} 
 \Infer{2}[\rmcutpar]{\vdash \Sigma, \Delta, \top}
 \end{prooftree}
 &\underset{r}{\longrightarrow}& 
 \begin{prooftree}
 \Infer{0}[\rtop]{\vdash \Sigma, \Delta, \top}
 \end{prooftree}
%\end{array}$
%}
\\[25pt]
\begin{prooftree}
\Hypo{\vdash \Sigma, C^\perp}
\Hypo{\vdash C, \Delta}
\Infer{1}[\rbot]{\vdash C, \Delta, \bot} 
\Infer{2}[\rmcutpar]{\vdash \Sigma, \Delta, \bot}
\end{prooftree}
& \underset{r}{\longrightarrow} &
\begin{prooftree}
\Hypo{\vdash \Sigma, C^\perp}
\Hypo{\vdash C, \Delta}
\Infer{2}[\rcut]{\vdash \Sigma,\Delta}
\Infer{1}[\rbot]{\vdash \Sigma, \Delta, \bot}
\end{prooftree}
\\[15pt]
\end{array}$
\end{center}
In the promotion commutation case, the premisse in $\vdash \oc C^\perp, \wn \Sigma $ is assumed to end with a promotion rule.\\ 
\caption{Cut-commutations rules, or External reduction rules,
  where $r=(\mathsf{ext},F)$ and $F$ is the principal occurrence.}
\label{fig:external-muLLi}
\end{figure*}

\begin{figure*}[htpb!] %% principal rules (fig:principal)
 % \begin{framed}
  \begin{center}
\scalebox{.92}{
$\begin{array}{clc}
\intllder
\intllcont
\intllweak
\begin{prooftree}
\Hypo{\vdash \Delta,F}
\Hypo{\vdash \Gamma,G}
\Infer{2}[\rtensor]{\vdash \Delta,\Gamma,F\tensor G}
\Hypo{\vdash \Theta,F^\bot,G^\bot}
\Infer{1}[\rparr]{\vdash \Theta,F^\bot\parr G^\bot} 
\Infer{2}[\rcut]{\vdash \Delta, \Gamma, \Theta}
\end{prooftree}
& \underset{r}{\longrightarrow} &
\begin{prooftree}
\Hypo{\vdash \Delta,F}
\Hypo{\vdash \Gamma,G}
\Hypo{\vdash \Theta,F^\bot,G^\bot}
\Infer{2}[\rcut]{\vdash \Gamma, \Theta, F^\bot}
\Infer{2}[\rcut]{\vdash \Delta, \Gamma, \Theta}
\end{prooftree}
\\[25pt]
 \begin{prooftree}
 \Hypo{\vdash \Delta, F_1}
 \Hypo{\vdash \Delta,F_2}
 \Infer{2}[\rwith]{\vdash \Delta,F_1\with F_2}
 \Hypo{\vdash \Gamma,F_i^\bot}
 \Infer{1}[\roplusi]{\vdash \Gamma,F_1^\bot\oplus F_2^\bot} 
 \Infer{2}[\rcut]{\vdash \Delta, \Gamma}
 \end{prooftree}
 & \underset{r}{\longrightarrow} &
 \begin{prooftree}
 \Hypo{\vdash \Delta,F_i}
 \Hypo{\vdash \Gamma,F_i^\bot}
 \Infer{2}[\rcut]{\vdash \Delta, \Gamma}
 \end{prooftree}
 \\[25pt]
\begin{prooftree}
  \Hypo{\vdash \Gamma}
  \Infer{1}[\rbot]{\vdash \Gamma,\bot}
  \Infer{0}[\rone]{\vdash \llone}
  \Infer{2}[\rmcutpar]{\vdash \Gamma}
\end{prooftree}
  & \underset{r}{\longrightarrow} &
\begin{prooftree}
  \Hypo{\vdash \Gamma}
\end{prooftree}
  \end{array}
  $}
  \end{center}
 % \end{framed}
\caption{Key cut-reduction rules, or Principal reductions,
where $r=(\mathsf{princ},\{F,F'^\bot\})$ with $\{F, F'^\bot\}$ the principal occurrences that have been reduced.}
\label{fig:principal-muLLi}
\end{figure*}

% !TEX root = main.tex
\section{Appendix on \Cref{sec:interpolate-cuts}}

\subsection{
Details on the Proof of the Main Lemma}
\label{app-main-lemma}

We now provide the full proof of \Cref{main-lemma}.

\subsubsection{Axiom case}

\begin{itemize}
\item If $\pi = \scalebox{0.8}{\begin{prooftree}
\Infer{0}[\rax]{\vdash \spl{F}, \spr{F^\perp}}\end{prooftree}}$, 
one simply takes $I = F^\perp$, $\pi^l_1 = \scalebox{0.8}{\begin{prooftree}
\Infer{0}[\rax]{\vdash \spl{F}, {F^\perp}}\end{prooftree}}$ and $\pi^r_1 = \scalebox{0.8}{\begin{prooftree}
\Infer{0}[\rax]{\vdash {F}, \spr{F^\perp}}\end{prooftree}}$. 
 
 The cut between $\pi^l_1$ and $\pi^r_1$ reduces to $\pi$ by one cut-axiom reduction step.\\ ~
    \item If $\pi = \scalebox{0.8}{\begin{prooftree}
\Infer{0}[\rax]{\vdash \spr{F}, \spl{F^\perp}}\end{prooftree}}$, 
the case is symmetrical taking $I = F$.\\ ~
%, $\pi^l_1 = \begin{prooftree}
%\Infer{0}[\rax]{\vdash \spr{F}, {F^\perp}}\end{prooftree}$ and $\pi^r_1 = \begin{prooftree}
%\Infer{0}[\rax]{\vdash {F}, \spl{F^\perp}}\end{prooftree}$. 
%   The cut between $\pi^l_1$ and $\pi^r_1$ simply reduces to $\pi$ by a cut-axiom reduction case.    
   \item If $\pi = \scalebox{0.8}{\begin{prooftree}
\Infer{0}[\rax]{\vdash \spl{F}, \spl{F^\perp}}\end{prooftree}}$,  
one takes  $I = \bot$, 
    $\pi^l_1 = \scalebox{0.8}{\begin{prooftree}\Hypo{\pi}\Infer 1[\rbot] {\vdash \spl{F}, \spl{F^\perp}, \bot}\end{prooftree}}$ and 
    $\pi^r_1 = \scalebox{0.8}{\begin{prooftree}\Hypo{}\Infer{1}[\rone]{\vdash \llone
}
\end{prooftree}}$. 

The cut of $\pi^l_1$ and $\pi^r_1$ reduces to $\pi$ by a key $\llone / \bot$ case.\\ ~
   \item If $\pi = \scalebox{0.8}{\begin{prooftree}
\Infer{0}[\rax]{\vdash \spr{F}, \spr{F^\perp}}\end{prooftree}}$,  
the case is symmetrical to the previous one, taking  $I = \llone$:
    $\pi^l_1 = \scalebox{0.8}{\begin{prooftree}\Hypo{}\Infer{1}[\rone]{\vdash \llone}
\end{prooftree}}$ and    $\pi^r_1 = \scalebox{0.8}{\begin{prooftree}\Hypo{\pi}\Infer 1[\rbot] {\vdash \spr{F}, \spr{F^\perp}, \bot}\end{prooftree}}$. 
%The situation is symmetric to the above case.
\end{itemize}

\subsubsection{Cut rules}

\begin{description}

\item[If the last rule is $\rcut$], with cut formulas \spr{in the right splitting} ie.
if 

%Keep this space otherwise the proof below moves out of margin

$\pi= \scalebox{0.8}{\begin{prooftree}
\Hypo{\pi^L_1}
\Infer[no rule] 1{\vdash \spl{\Gamma_1}, I_1}
\Hypo{\pi^R_1}
\Infer[no rule] 1{\vdash I_1^\perp, \spr{\Delta_1, C}}
\Infer 2[\rcut]{\vdash \spl{\Gamma_1},\spr{\Delta_1, C}}
\Hypo{\pi^L_2}
\Infer[no rule] 1{\vdash \spl{\Gamma_2}, I_2}
\Hypo{\pi^R_2}
\Infer[no rule] 1{\vdash I_2^\perp, \spr{\Delta_2, C^\perp}}
\Infer 2[\rcut]{\vdash \spl{\Gamma_2},\spr{\Delta_2, C^\perp}}
\Infer 2[\rcut]{\vdash \spl{\Gamma_1,\Gamma_2}, \spr{\Delta_1,\Delta_2}}
%, \spr{A\tensor B}}
\end{prooftree}}
$, then setting $I=I_1\tensor I_2$, 
$\pi^L=\scalebox{0.8}{\begin{prooftree}
\Hypo{\pi^L_1}
\Infer [no rule]1{\vdash \spl{\Gamma_1}, I_1}
\Hypo{\pi^L_2}
\Infer[no rule] 1{\vdash \spl{\Gamma_2}, I_2}
\Infer 2[\rtensor]{\vdash \spl{\Gamma_1,\Gamma_2}, I_1\tensor I_2}
\end{prooftree}}
$
and $\pi^R=$
$
\scalebox{0.8}{
\begin{prooftree}
\Hypo{\pi^R_1}
\Infer[no rule] 1{\vdash I_1^\perp, \spr{\Delta_1, C}}
\Hypo{\pi^R_2}
\Infer [no rule]1{\vdash I_2^\perp, \spr{\Delta_2, C^\perp}}
\Infer 2[\rcut]{\vdash  I_1^\perp,  I_2^\perp,  \spr{\Delta_1,\Delta_2}}
\Infer 1[\rparr]{\vdash  (I_1 \tensor I_2)^\perp,  \spr{\Delta_1,\Delta_2}}
\end{prooftree}
}
$
one gets a solved \CPRIS $\pi'$ (indeed, the constraint on the interpolant vocabulary is preserved as formulas $C,C^\perp$ move from the \spr{right} context to the right cut vocabulary $\mathbb{C}^R$) such that $\pi \longleftarrow^\star_{\mathsf{cut}} \pi'$ by a commutative reduction of $\rcut$ and a key $\rtensor/\rparr$ case. The case when the cut formulas are colored \spl{left} is symmetrical:

if $\pi= \scalebox{0.8}{\begin{prooftree}
\Hypo{\pi^L_1}
\Infer[no rule] 1{\vdash \spl{\Gamma_1, C}, I_1}
\Hypo{\pi^R_1}
\Infer[no rule] 1{\vdash I_1^\perp, \spr{\Delta_1}}
\Infer 2[\rcut]{\vdash \spl{\Gamma_1, C},\spr{\Delta_1}}
\Hypo{\pi^L_2}
\Infer[no rule] 1{\vdash \spl{\Gamma_2, C^\perp}, I_2}
\Hypo{\pi^R_2}
\Infer[no rule] 1{\vdash I_2^\perp, \spr{\Delta_2}}
\Infer 2[\rcut]{\vdash \spl{\Gamma_2, C^\perp},\spr{\Delta_2}}
\Infer 2[\rcut]{\vdash \spl{\Gamma_1,\Gamma_2}, \spr{\Delta_1,\Delta_2}}
%, \spr{A\tensor B}}
\end{prooftree}}
$, then setting $I=I_1\parr I_2$, 
$\pi^L=
\scalebox{0.8}{
\begin{prooftree}
\Hypo{\pi^L_1}
\Infer[no rule] 1{\vdash  \spl{\Gamma_1, C}, I_1}
\Hypo{\pi^L_2}
\Infer [no rule]1{\vdash  \spl{\Gamma, C^\perp}, I_2}
\Infer 2[\rcut]{\vdash    \spl{\Gamma_1,\Gamma_2}, I_1,  I_2}
\Infer 1[\rparr]{\vdash    \spl{\Delta_1,\Delta_2}, I_1 \parr I_2}
\end{prooftree}
}
$
and $\pi^R=$

$
\scalebox{0.8}{\begin{prooftree}
\Hypo{\pi^R_1}
\Infer [no rule]1{\vdash  I^\perp_1,\spr{\Delta_1}}
\Hypo{\pi^R_2}
\Infer[no rule] 1{\vdash  I^\perp_2, \spr{\Delta_2}}
\Infer 2[\rtensor]{\vdash (I_1\parr I_2)^\perp, \spr{\Delta_1,\Delta}}
\end{prooftree}}
$ one gets a solved \CPRIS $\pi'$ (indeed, the constraint on the interpolant vocabulary is preserved as formulas $C,C^\perp$ move from the \spl{left} context to the left cut vocabulary $\mathbb{C}^L$) such that $\pi \longleftarrow^\star_{\mathsf{cut}} \pi'$ by a commutative reduction of $\rcut$ and a key $\rtensor/\rparr$ case.
\end{description}

\subsubsection{Logical rules}

We analyze the possible cases for a logical rule involved in an elementary \CPRIS. Note that in each case, the principal formula may be part of the left or right part of the splitting; we treat only one case each time since the other is symmetrical be taking the dual interpolant and exchanging $\pi^L$ and $\pi^R$.

\begin{description}

\item[If the last rule is $\rparr$], that is 
if $\pi = \scalebox{0.8}{\begin{prooftree}
\Hypo{\pi_1^L}
\Infer[no rule] 1{\vdash \spl{\Gamma}, I}
\Hypo{\pi_1^R}
\Infer[no rule] 1{\vdash I^\perp, \spr{\Delta,A,B}}
\Infer 2[\rcut]{\vdash \spl{\Gamma},\spr{\Delta, A, B}}
\Infer 1[\rparr]{\vdash \spl\Gamma, \spr{\Delta, A\parr B}}
\end{prooftree}}
$ then taking $I'=I$, $\pi^L = \pi_1^L$ and $\pi^R = 
 \scalebox{0.8}{\begin{prooftree}
\Hypo{\pi_1^R}
\Infer[no rule] 1{\vdash I^\perp, \spr{\Delta,A,B}}
\Infer 1[\rparr]{\vdash I^\perp, \spr{\Delta, A\parr B}}
\end{prooftree}}$
we obtain a solved \CPRIS $\pi'$ such that $\pi \longleftarrow_{\mathsf{cut}} \pi'$ by a commutative reduction of $\rcut$ (Indeed, $\mathbb{C}^L,\mathbb{C}^R$ are unchanged).
\item[If the last rule is $\rotimes$], that is if \\
$\pi= \scalebox{0.8}{\begin{prooftree}
\Hypo{\pi^L_1}
\Infer[no rule] 1{\vdash \spl{\Gamma_1}, I_1}
\Hypo{\pi^R_1}
\Infer[no rule] 1{\vdash I_1^\perp, \spr{\Delta_1, A}}
\Infer 2[\rcut]{\vdash \spl{\Gamma_1},\spr{\Delta_1, A}}
\Hypo{\pi^L_2}
\Infer[no rule] 1{\vdash \spl{\Gamma_2}, I_2}
\Hypo{\pi^R_2}
\Infer[no rule] 1{\vdash I_2^\perp, \spr{\Delta_2, B}}
\Infer 2[\rcut]{\vdash \spl{\Gamma_2},\spr{\Delta_2, B}}
\Infer 2[\rtensor]{\vdash \spl{\Gamma_1,\Gamma_2}, \spr{\Delta_1,\Delta_2}, \spr{A\tensor B}}
\end{prooftree}}
$, then setting $I=I_1\tensor I_2$, 
$\pi^L=\scalebox{0.8}{\begin{prooftree}
\Hypo{\pi^L_1}
\Infer [no rule]1{\vdash \spl{\Gamma_1}, I_1}
\Hypo{\pi^L_2}
\Infer[no rule] 1{\vdash \spl{\Gamma_2}, I_2}
\Infer 2[\rtensor]{\vdash \spl{\Gamma_1,\Gamma_2}, I_1\tensor I_2}
\end{prooftree}}
$
and $\pi^R=$

\scalebox{0.8}{
\begin{prooftree}
\Hypo{\pi^R_1}
\Infer[no rule] 1{\vdash I_1^\perp, \spr{\Delta_1, A}}
\Hypo{\pi^R_2}
\Infer [no rule]1{\vdash I_2^\perp, \spr{\Delta_2, B}}
\Infer 2[\rtensor]{\vdash  I_1^\perp,  I_2^\perp,  \spr{\Delta_1,\Delta_2}, \spr{A\tensor B}}
\Infer 1[\rparr]{\vdash  (I_1 \tensor I_2)^\perp,  \spr{\Delta_1,\Delta_2}, \spr{A\tensor B}}
\end{prooftree}
} one gets a solved \CPRIS $\pi'$ such that $\pi \longleftarrow^\star_{\mathsf{cut}} \pi'$ by a commutative reduction of $\rcut$ and a key $\rtensor/\rparr$ case (Indeed, $\mathbb{C}^L=\mathbb{C}^L_1\cup \mathbb{C}^L_2$ and $\mathbb{C}^R = \mathbb{C}^R_1\cup \mathbb{C}^R_2$).

\item[If the last rule is $\rbot$], that is if $\pi = 
\scalebox{0.8}{\begin{prooftree}
\Hypo{\pi^L_1}
\Infer [no rule]1{\vdash \spl{\Gamma}, I_1}
\Hypo{\pi^R_1}
\Infer [no rule]1{\vdash {I_1}^\perp, \spr\Delta}
\Infer 2[\rcut]{\vdash \spl\Gamma, \spr\Delta}
\Infer 1 [\rbot]{\vdash \spl{\bot, \Gamma}, \spr\Delta}\end{prooftree}}
$. 
By setting $I= I_1$, $\pi^L = \scalebox{0.8}{\begin{prooftree}
    \Hypo{\pi^L_1}\Infer [no rule]1{\vdash \spl{\Gamma}, I_1}
    \Infer 1[\rbot]{\vdash \spl{\bot, \Gamma}, I_1}
\end{prooftree}}$ and $\pi^R = \pi^R_1$, one gets 
a solved \CPRIS $\pi' = \scalebox{0.8}{\begin{prooftree}
\Hypo{\pi^L}
\Infer [no rule]1{\vdash \spl{\bot, \Gamma}, I}
\Hypo{\pi^R}
\Infer [no rule]1{\vdash I^\perp, \spr{\Delta}}
\Infer 2[\rcut]{\vdash \spl{\bot, \Gamma}, \spr{\Delta}}
\end{prooftree}}$
such that $\pi'
\cutred 
\pi$
by a cut-commutation case (Indeed, $\mathbb{C}^L,\mathbb{C}^R$ are unchanged).

\item [If the last rule is $\rone$], that is if $\pi = \scalebox{0.8}{\begin{prooftree}\Infer 0 [\rone]{\vdash \spr{\llone}}\end{prooftree}}$ (which is indeed both an initial \CPRIS and an elementary \CPRIS), let $I=\llone$ and 
 $\pi^L_1=\scalebox{0.8}{\begin{prooftree}
        \Infer 0[\rone]{\vdash \llone}
    \end{prooftree}}$ and
       $\pi^R_1 = \scalebox{0.8}{\begin{prooftree}
        \Infer 0 [\rax]{\vdash \bot, \spr{\llone}}
    \end{prooftree}}$. One gets a solved \CPRIS  
    $\pi'$ such that $\pi \longleftarrow^\star_{\mathsf{cut}} \pi'$ by a key $\llone/\bot$ case.

\item [If the last rule is $\roplusi, i\in\{1,2\}$], that is if\\
 $\pi= \scalebox{0.8}{
\begin{prooftree}
\Hypo {\pi^L_1}
\Infer [no rule]1{\vdash \spl{F_i,\Gamma}, I_1}
\Hypo {\pi^R_1}
\Infer [no rule]1 {\vdash {I_1}^\perp, \spr{\Delta}}
\Infer 2[\rcut] {\vdash \spl{F_i, \Gamma}, \spr{\Delta}}
\Infer 1 [\roplusi]{\vdash \spl{F_1\oplus F_2, \Gamma}, \spr{\Delta}}
\end{prooftree}} $, then setting $I=I_1$, 
$\pi^L=\scalebox{0.8}{\begin{prooftree}
\Hypo{\pi^L_1}
\Infer [no rule]1{\vdash \spl{F_i, \Gamma}, I_1}
\Infer 1[\roplusi]{\vdash \spl{F_1\oplus F_2,\Gamma}, I_1}
\end{prooftree}}
$
and
$\pi^R=\pi_1^R$, one gets a solved \CPRIS $\pi'$ such that $\pi \longleftarrow^\star_{\mathsf{cut}} \pi'$ by a commutative reduction of $\rcut$ (Indeed, $\mathbb{C}^L,\mathbb{C}^R$ are unchanged).

\item [If the last rule is $\rwith$], that is if\\ 
$\pi=\scalebox{0.8}{\begin{prooftree}
\Hypo {\pi^L_1}
\Infer [no rule]1{\vdash \spl{F,\Gamma}, I_1}
\Hypo {\pi^R_1}
\Infer [no rule]1 {\vdash {I_1}^\perp, \spr{\Delta}}
\Infer 2[\rcut]{\vdash \spl{F,\Gamma},\spr{\Delta}}
\Hypo {\pi^L_2}
\Infer [no rule]1{\vdash \spl{G,\Gamma}, I_2}
\Hypo {\pi^R_2}
\Infer [no rule]1 {\vdash {I_2}^\perp, \spr{\Delta}}
\Infer 2[\rcut]{\vdash \spl{G,\Gamma},\spr{\Delta}}
\Infer{2}[\rwith]{\vdash \spl{F\with G,\Gamma},\spr{\Delta}}
\end{prooftree}}$, then setting $I = I_1\oplus I_2$ and
$\pi^L = 
\scalebox{0.8}{\begin{prooftree}
\Hypo{\pi^L_1}
\Infer [no rule]1{\vdash \spl{F,\Gamma}, I_1}\Infer 1[\roplusl]{\vdash \spl{F, \Gamma}, I_1\oplus I_2}
\Hypo{\pi^L_2}
\Infer [no rule]1{\vdash \spl{G,\Gamma}, I_2}\Infer 1[\roplusr]{\vdash  \spl{G, \Gamma}, I_1\oplus I_2}
\Infer 2[\rwith]{\vdash \spl{F\with G, \Gamma}, I}
\end{prooftree}}$ 

and $\pi^R = 
\scalebox{0.8}{\begin{prooftree}
\Hypo{\pi^R_1}
\Infer [no rule]1 {\vdash {I_1}^\perp, \spr{\Delta}}
\Hypo{\pi^R_2}
\Infer [no rule]1 {\vdash {I_2}^\perp, \spr{\Delta}}
\Infer 2[\rwith]{\vdash {I_1^\perp}\with{I_2}^\perp, \spr{\Delta}}
\end{prooftree}}$. One observes that: 

$$
\begin{array}{rcl}
&&\scalebox{0.8}{\begin{prooftree}
\Hypo{\pi^L}
\Infer[no rule] 1{\vdash \spl{\Gamma}, I}
\Hypo{\pi^R}
\Infer[no rule] 1{\vdash I^\perp, \spr{\Delta}}
\Infer 2[\rcut]{\vdash \Gamma,\Delta}
\end{prooftree}}\\
&\cutred& 
\scalebox{0.8}{\begin{prooftree}
\Hypo{\pi^L_1}
\Infer [no rule]1{\vdash \spl{F, \Gamma}, I_1}
\Infer 1[\roplusl]{\vdash \spl{F, \Gamma}, I}
\Hypo{\pi^R}
\Infer[no rule] 1{\vdash I^\perp, \spr{\Delta}}\Infer2[\rcut]{\vdash \spl{F, \Gamma}, \spr{\Delta}}
\Hypo{\pi^L_2}
\Infer [no rule]1{\vdash \spl{G, \Gamma}, I_2}
\Infer 1[\roplusr]{\vdash \spl{G, \Gamma}, I}
\Hypo{\pi^R}
\Infer[no rule] 1{\vdash I^\perp, \spr{\Delta}}
\Infer2[\rcut]{\vdash \spl{G, \Gamma}, \spr{\Delta}}
\Infer 2[\rwith]{\vdash \spl{F\with G, \Gamma}, \spr{\Delta}}
\end{prooftree}}\\
&\cutred& 
\scalebox{0.8}{\begin{prooftree}
\Hypo{\pi^L_1}
\Infer [no rule]1{\vdash \spl{F, \Gamma}, I_1}
\Hypo{\pi^R_1}
\Infer [no rule]1 {\vdash {I_1}^\perp, \spr{\Delta}}
\Infer2[\rcut]{\vdash \spl{F, \Gamma}, \spr{\Delta}}
\Hypo{\pi^L_2}
\Infer [no rule]1{\vdash \spl{G, \Gamma}, I_2}\Infer 1[\roplusr]{\vdash \spl{G, \Gamma}, I}
\Hypo{\pi^R}
\Infer[no rule] 1{\vdash I^\perp, \spr{\Delta}}
\Infer2[\rcut]{\vdash \spl{G, \Gamma}, \spr{\Delta}}
\Infer 2[\rwith]{\vdash \spl{F\with G, \Gamma}, \spr{\Delta}}
\end{prooftree}}
\\
&\cutred& 
\scalebox{0.8}{\begin{prooftree}
\Hypo{\pi^L_1}
\Infer [no rule]1{\vdash \spl{F, \Gamma}, I_1}
\Hypo{\pi^R_1}
\Infer [no rule]1 {\vdash {I_1}^\perp, \spr{\Delta}}
\Infer2[\rcut]{\vdash \spl{F, \Gamma}, \spr{\Delta}}
\Hypo{\pi^L_2}
\Infer [no rule]1{\vdash \spl{G, \Gamma}, I_2}
\Hypo{\pi^R_2}
\Infer [no rule]1 {\vdash {I_2}^\perp, \spr{\Delta}}
\Infer2[\rcut]{\vdash \spl{G, \Gamma}, \spr{\Delta}}
\Infer 2[\rwith]{\vdash \spl{F\with G, \Gamma}, \spr{\Delta}}
\end{prooftree}} = \pi\\
\end{array}
$$
so one gets a solved \CPRIS (Indeed, $\mathbb{C}^L=\mathbb{C}^L_1\cup \mathbb{C}^L_2$ and $\mathbb{C}^R = \mathbb{C}^R_1\cup \mathbb{C}^R_2$) such that $\pi$ indeed cut-expands to $\pi'$.

\item [If the last rule is $\rtop$], that is if $\pi = \scalebox{0.8}{\begin{prooftree}
\Infer 0 [\rtop]{\vdash \spl{\top, \Gamma}, \spr{\Delta}}
\end{prooftree}}$, by setting $I = 0$, 
$\pi^L = \scalebox{0.8}{\begin{prooftree}
    \Infer 0 [\rtop]{\vdash \spl{\top, \Gamma}, 0}
\end{prooftree}}$
and 
$\pi^R = \scalebox{0.8}{\begin{prooftree}
    \Infer 0 [\rtop]{\vdash \top, \spr{\Delta}}
\end{prooftree}}
$.
Then $\pi' = 
\scalebox{0.8}{\begin{prooftree}
\Hypo{\pi^L}
\Hypo{\pi^R}
\Infer 2[\rcut]{\vdash \spl{\top, \Gamma}, \spr{\Delta}}
\end{prooftree}}
$ is a solved \CPRIS to which $\pi$ cut-expands by a $\rtop$ commutation case.

\item [If the last rule is $\rder$], that is 
$\pi = \scalebox{0.8}{\begin{prooftree}
\Hypo {\pi^L_1}
\Infer [no rule]1{\vdash \spl{F,\Gamma}, I_1}
\Hypo {\pi^R_1}
\Infer [no rule]1 {\vdash {I_1}^\perp, \spr{\Delta}}
\Infer 2[\rcut]{\vdash \spl{F, \Gamma}, \spr{\Delta}}
\Infer{1}[\rder]{\vdash \spl{\wn F, \Gamma},\spr{\Delta}}
\end{prooftree}}$, then setting $I=I_1$, $\pi^L = \scalebox{0.8}{\begin{prooftree}\Hypo{\pi^L_1}\Infer [no rule]1{\vdash \spl{F,\Gamma}, I_1}\Infer 1[\rder]{\vdash \spl{\wn F, \Gamma}, I_1}\end{prooftree}}$ and $\pi^R=\pi^R_1$ one gets: 
$$
\begin{array}{rcl}
\scalebox{0.8}{\begin{prooftree}
\Hypo {\pi^L}
\Infer [no rule]1{\vdash \spl{\wn F,\Gamma}, I}
\Hypo {\pi^R}
\Infer [no rule]1 {\vdash {I}^\perp, \spr{\Delta}}
\Infer 2[\rcut] {\vdash \wn \spl{F, \Gamma}, \spr{\Delta}}
\end{prooftree}} 
&\cutred&\\
\scalebox{0.8}{\begin{prooftree}
\Hypo {\pi^L_1}
\Infer 1{\vdash \spl{F,\Gamma}, I_1}
\Hypo {\pi^R_1}
\Infer 1 {\vdash {I_1}^\perp, \spr{\Delta}}
\Infer 2[\rcut] {\vdash \spl{F, \Gamma}, \spr{\Delta}}
\Infer 1[\rder] {\vdash \spl{\wn F, \Gamma}, \spr{\Delta}}
\end{prooftree}} &=&\pi
\end{array}
$$
so one gets a solved \CPRIS  (indeed, $\mathbb{C}^L,\mathbb{C}^R$ are unchanged) such that $\pi$ indeed cut-expands to $\pi'$.
\item [If the last rule is $\rprom$], that is 
$\pi = \scalebox{0.8}{\begin{prooftree}
\Hypo {\pi^L_1}
\Infer [no rule]1{\vdash \spl{F,\wn\Gamma}, I_1}
\Hypo {\pi^R_1}
\Infer [no rule]1 {\vdash {I_1}^\perp, \spr{\wn\Delta}}
\Infer 2[\rcut]{\vdash \spl{F, \wn \Gamma}, \spr{\wn \Delta}}
\Infer{1}[\rprom]{\vdash \spl{\oc F, \wn \Gamma},\spr{\wn \Delta}}
\end{prooftree}}$, 
then by setting $I=\wn I_1$, $\pi^L = \scalebox{0.8}{\begin{prooftree}\Hypo {\pi^L_1}
\Infer [no rule]1{\vdash \spl{F,\wn\Gamma}, I_1}\Infer 1[\rder]{\vdash \spl{F, \wn \Gamma}, \wn I_1}
\Infer 1[\rprom]{\vdash \spl{\oc F, \wn \Gamma}, \wn I_1}
\end{prooftree}}$ and 
$\pi^R = 
\scalebox{0.8}{\begin{prooftree}
\Hypo {\pi^R_1}
\Infer [no rule]1 {\vdash {I_1}^\perp, \spr{\wn\Delta}}
\Infer 1[\rprom]{\vdash \oc {I_1}^\perp, \spr{\wn \Delta}}
\end{prooftree}}$, one gets: 
$$
\begin{array}{rcl}
&&\scalebox{0.8}{\begin{prooftree}
\Hypo {\pi^L}
\Infer  [no rule]1{\vdash \spl{\oc F, \wn \Gamma}, \wn I_1}
%\Infer 1{\vdash \oc F,\wn \Gamma', C}
\Hypo {\pi^R}
\Infer  [no rule]1{\vdash \oc {I_1}^\perp, \spr{\wn \Delta}}
%\Infer 1 {\vdash {C}^\perp, \wn \Delta'}
\Infer 2[\rcut]{\vdash \spl{\oc F, \wn \Gamma}, \spr{\wn \Delta}}
\end{prooftree}}\\ 
&\cutred&
\scalebox{0.8}{\begin{prooftree}
\Hypo {\pi^L_1}
\Infer [no rule]1{\vdash \spl{F,\wn \Gamma}, I_1}
\Infer 1[\rder]{\vdash \spl{F,\wn \Gamma}, I}
\Hypo {\pi^R_1}
\Infer [no rule]1 {\vdash {I_1}^\perp, \spr{\wn \Delta}}
\Infer 1 [\rprom]{\vdash {I}^\perp, \spr{\wn \Delta}}
\Infer 2[\rcut]{\vdash \spl{F, \wn \Gamma}, \spr{\wn \Delta}}
\Infer 1 [\rprom]{\vdash \spl{\oc F, \wn \Gamma}, \spr{\wn \Delta}}
\end{prooftree}}
\\
&\cutred&
\scalebox{0.8}{\begin{prooftree}
\Hypo {\pi^L_1}
\Infer [no rule]1{\vdash \spl{F,\wn \Gamma}, I_1}
\Hypo {\pi^R_1}
\Infer [no rule]1 {\vdash {I_1}^\perp, \spr{\wn \Delta}}
\Infer 2[\rcut]{\vdash \spl{F, \wn \Gamma}, \spr{\wn \Delta}}
\Infer 1 [\rprom]{\vdash \spl{\oc F, \wn \Gamma}, \spr{\wn \Delta}}
\end{prooftree}} = \pi\\
\end{array}
$$
so one gets a solved \CPRIS  (indeed, $\mathbb{C}^L,\mathbb{C}^R$ are unchanged) such that $\pi$ indeed cut-expands to $\pi'$.
\item [If the last rule is $\rweak$], that is 
$\pi = \scalebox{0.8}{\begin{prooftree}
\Hypo {\pi^L_1}
\Infer [no rule]1{\vdash \spl{\Gamma}, I_1}
\Hypo {\pi^R_1}
\Infer [no rule]1 {\vdash {I_1}^\perp, \spr{\Delta}}
\Infer 2[\rcut]{\vdash \spl{\Gamma}, \spr{\Delta}}
\Infer{1}[\rweak]{\vdash \spl{\wn F, \Gamma},\spr{\Delta}}
\end{prooftree}}$, then by setting
 $I=I_1$, $\pi^L = \scalebox{0.8}{\begin{prooftree}\Hypo{\pi^L_1}\Infer [no rule]1{\vdash \spl{\Gamma}, I_1}\Infer 1[\rweak]{\vdash \spl{\wn F, \Gamma}, I_1}\end{prooftree}}$ and $\pi^R=\pi^R_1$, 
we ensure that, since $\VOC{I} \subseteq (\VOC{\Gamma}\cup \mathbb{C}^L)\cap (\VOC{\Delta}\cup \mathbb{C}^R)$, 
$\pi'= \scalebox{0.8}{\begin{prooftree}
\Hypo {\pi^L}
\Infer [no rule]1{\vdash \spl{\wn F, \Gamma}, I}
\Hypo {\pi^R}
\Infer [no rule]1 {\vdash {I}^\perp, \spr{\Delta}}
\Infer 2[\rcut] {\vdash \spl{\wn F, \Gamma}, \spr{\Delta}}
\end{prooftree}}
$
is indeed a solved \CPRIS  (indeed, $\mathbb{C}^L,\mathbb{C}^R$ are unchanged) and that
$$
\pi' \quad \cutred\quad 
\scalebox{0.8}{\begin{prooftree}
\Hypo {\pi^L_1}
\Infer [no rule]1{\vdash \spl{\Gamma}, I_1}
\Hypo {\pi^R_1}
\Infer [no rule]1 {\vdash {I_1}^\perp, \spr{\Delta}}
\Infer 2[\rcut] {\vdash \spl{\Gamma}, \spr{\Delta}}
\Infer 1[\rweak] {\vdash \spl{\wn F, \Gamma}, \spr{\Delta}}
\end{prooftree}} \quad =\quad  \pi.
$$
by a cut commutation rule so that 
$\pi$ indeed cut-expands to $\pi'$ as expected.

\item [If the last rule is $\rcontr$], that is 
$\pi = \scalebox{0.8}{\begin{prooftree}
\Hypo {\pi^L_1}
\Infer [no rule]1{\vdash \spl{\wn F, \wn F,\Gamma}, I_1}
\Hypo {\pi^R_1}
\Infer [no rule]1 {\vdash {I_1}^\perp, \spr{\Delta}}
\Infer 2[\rcut]{\vdash \spl{\wn F, \wn F, \Gamma}, \spr{\Delta}}
\Infer{1}[\rcontr]{\vdash \spl{\wn F, \Gamma},\spr{\Delta}}
\end{prooftree}}$
by setting $I=I_1$, $\pi^L = \scalebox{0.8}{\begin{prooftree}\Hypo{\pi^L_1}
\Infer [no rule]1{\vdash \spl{\wn F, \wn F,\Gamma}, I_1}\Infer 1[\rcontr]{\vdash \spl{\wn F, \Gamma}, I}\end{prooftree}}$ and $\pi^R = \pi^R_1$ one gets: 
$$
\begin{array}{rcl}
\pi'=
\scalebox{0.8}{\begin{prooftree}
\Hypo {\pi^L}
\Infer [no rule]1{\vdash \spl{\wn F,\Gamma}, I}
\Hypo {\pi^R}
\Infer [no rule]1 {\vdash {I}^\perp, \spr{\Delta}}
\Infer 2[\rcut] {\vdash \spl{\wn F, \Gamma}, \spr{\Delta}}
\end{prooftree}}
&\cutred&\\
\scalebox{0.8}{\begin{prooftree}
\Hypo {\pi^L_1}
\Infer [no rule]1{\vdash \spl{\wn F, \wn F,\Gamma}, I_1}
\Hypo {\pi^L_1}
\Infer [no rule]1 {\vdash {I_1}^\perp, \spr{\Delta}}
\Infer 2[\rcut] {\vdash \spl{\wn F, \wn F, \Gamma}, \spr{\Delta}}
\Infer 1[\rcontr] {\vdash \spl{\wn F, \Gamma}, \spr{\Delta}}
\end{prooftree}} &=& \pi\\
\end{array}
$$
so that $\pi'$ is indeed a solved \CPRIS  (Indeed, $\mathbb{C}^L,\mathbb{C}^R$ are unchanged) to which $\pi$ cut-expands.

\item [If the last rule is $\rforall$], that is 
$\pi = \scalebox{0.8}{\begin{prooftree}
\Hypo {\pi^L_1}
\Infer [no rule]1{\vdash \spl{F,\Gamma}, I_1}
\Hypo {\pi^R_1}
\Infer [no rule]1 {\vdash {I_1}^\perp, \spr{\Delta}}
\Infer 2[\rcut]{\vdash \spl{F, \Gamma}, \spr{\Delta}}
\Infer{1}[\rforall\quad $x\not\in\mathsf{FV}(\Gamma,\Delta)$]{\vdash \spl{\forall x F, \Gamma},\spr{\Delta}}
\end{prooftree}}$ with  $\VOC {I_1}\subseteq(\VOC {F,\Gamma}\cup \mathbb{C}^L)\cap(\VOC{\Delta}\cup \mathbb{C}^R)$, by setting
 $I=\exists x I_1$, 
 $\pi^L = \scalebox{0.8}{\begin{prooftree}\Hypo{\pi^L_1}
\Infer [no rule]1{\vdash \spl{F,\Gamma}, I_1}\Infer 1[\rexists]{\vdash \spl{F, \Gamma}, \exists x I_1}
\Infer 1[\rforall]{\vdash \spl{\forall x F, \Gamma}, \exists x. I_1}
\end{prooftree}}$ and
$\pi^R = \scalebox{0.8}{\begin{prooftree}\Hypo{\pi^R_1}
\Infer [no rule]1 {\vdash {I_1}^\perp, \spr{\Delta}}
\Infer 1[\rforall]{\vdash \forall x {I_1}^\perp, \spr{\Delta}}
\end{prooftree}}$ 
one gets that: 

$\pi' = 
\scalebox{0.8}{\begin{prooftree}
\Hypo {\pi^L}
\Infer [no rule]1{\vdash \spl{\forall x F, \Gamma}, \exists x. I_1}
%\Infer 1{\vdash \forall x F, \Gamma', \exists x. C'}
\Hypo {\pi^R}
\Infer [no rule]1{\vdash \forall x {I_1}^\perp, \spr{\Delta}}%\Infer 1{\vdash \forall x {C'}^\perp, \Delta}
\Infer 2[\rcut] {\vdash \spl{\forall x F, \Gamma}, \spr{\Delta}}
\end{prooftree}}
$
is a solved \CPRIS  (indeed, $\mathbb{C}^L,\mathbb{C}^R$ are unchanged) to which $\pi$ cut-expands:

$\pi'
\cutred
\scalebox{0.8}{\begin{prooftree}
\Hypo {\pi^L_1}
\Infer [no rule]1{\vdash \spl{F,\Gamma}, I_1}\Infer 1[\rexists]{\vdash \spl{F, \Gamma}, \exists x. I_1}
\Hypo {\pi^R_1}
\Infer [no rule]1 {\vdash {I_1}^\perp, \spr{\Delta}}
\Infer 1[\rforall]{\vdash \forall x {I_1}^\perp, \spr{\Delta}}
\Infer 2[\rcut] {\vdash \spl{F, \Gamma}, \spr{\Delta}}
\Infer 1[\rforall]{\vdash \spl{\forall x F, \Gamma}, \spr{\Delta}}
\end{prooftree}} 
\cutred
\pi.
$

\item [If the last rule is $\rexists$], that is 
$\pi = \scalebox{0.8}{\begin{prooftree}
\Hypo {\pi^L_1}
\Infer [no rule]1{\vdash \spl{F\{y/x\},\Gamma}, I_1}
\Hypo {\pi^R_1}
\Infer [no rule]1 {\vdash {I_1}^\perp, \spr{\Delta}}
\Infer 2[\rcut]{\vdash \spl{F\{y/x\}, \Gamma}, \spr{\Delta}}
\Infer{1}[\rexists]{\vdash \spl{\exists x F, \Gamma},\spr{\Delta}}
\end{prooftree}}$
with $\VOC {I_1}\subseteq(\VOC {F\{y/x\},\Gamma}\cup \mathbb{C}^L)\cap(\VOC{\Delta}\cup \mathbb{C}^R)$.

In this case, note that we treat only the case of a FO language containing no function symbols.

In this case, we reason by case on whether $y$ occurs in $\Gamma,\Delta$:
\begin{itemize}
    \item If $y$ occurs in both, then we simply take $I = I_1$ as interpolant, $\pi^L = \scalebox{0.8}{\begin{prooftree}
        \Hypo{\pi^L_1}\Infer 1[\rexists]{\vdash \exists x F, \Gamma} \end{prooftree}}$ and $\pi^R = \pi^R_1$.
Since $\VOC {I} = \VOC {I_1} \subseteq (\VOC{F, \Gamma}\cup \mathbb{C}^L)\cap(\VOC{\Delta}\cup \mathbb{C}^R) = (\VOC{\exists x F, \Gamma}\cup \mathbb{C}^L)\cap(\VOC{\Delta}\cup \mathbb{C}^R) $, 
we have that 
$\pi' = \scalebox{0.8}{\begin{prooftree}
\Hypo {\pi^L}
\Infer [no rule]1{\vdash \spl{\exists xF,\Gamma}, I}
\Hypo {\pi^R}
\Infer [no rule]1 {\vdash {I}^\perp, \spr{\Delta}}
\Infer 2[\rcut]{\vdash \spl{\exists xF, \Gamma}, \spr{\Delta}}
\end{prooftree}}$
 is a solved \CPRIS to which $\pi$ cut-expands via a cut-commutation rule.
\item If $y$ occurs in $\Gamma$ but not in $\Delta$, then we set $I = \exists y I_1$, 
$\pi^L = \scalebox{0.8}{\begin{prooftree}
\Hypo{\pi^L_1}
\Infer 1[\rexists]{\vdash \spl{\exists x F,\Gamma}, I_1}
\Infer 1[\rexists]{\vdash \spl{\exists x F,\Gamma}, \exists y I_1}
\end{prooftree}}$
and 
$\pi^R = \scalebox{0.8}{\begin{prooftree}
    \Hypo{\pi^R_1}
    \Infer [no rule]1 {\vdash {I_1}^\perp, \spr{\Delta}}
    \Infer 1[\rforall]{\vdash \forall y {I_1}^\perp, \spr{\Delta}}
\end{prooftree}}$
we have that 
$\pi' = \scalebox{0.8}{\begin{prooftree}
\Hypo {\pi^L}
\Infer [no rule]1{\vdash \spl{\exists xF,\Gamma}, I}
\Hypo {\pi^R}
\Infer [no rule]1 {\vdash {I}^\perp, \spr{\Delta}}
\Infer 2[\rcut]{\vdash \spl{\exists xF, \Gamma}, \spr{\Delta}}
\end{prooftree}}$
 is a solved \CPRIS to which $\pi$ cut-expands via a cut-commutation rule and a key $\rexists/\rforall$ rule:
 $$
 \begin{array}{rcl}
\pi' &\cutred&
\scalebox{0.8}{\begin{prooftree}
\Hypo {\pi^L_1}
\Infer [no rule]1{\vdash \spl{F\{y/x\},\Gamma}, I_1}
\Infer 1[\rexists]{\vdash \spl{\exists x F, \Gamma}, I_1}
\Hypo {\pi^R_1}
\Infer 2[\rcut] {\vdash \spl{\exists xF, \Gamma}, \spr{\Delta}}
\end{prooftree}}\\
&\cutred&
\scalebox{0.8}{\begin{prooftree}
\Hypo {\pi^L_1}
\Infer [no rule]1{\vdash \spl{F\{y/x\},\Gamma}, I_1}
\Hypo {\pi^R_1}
\Infer [no rule]1 {\vdash {I_1}^\perp, \spr{\Delta}}
\Infer 2[\rcut] {\vdash \spl{F\{y/x\}, \Gamma}, \spr{\Delta}}
\Infer 1[\rexists]{\vdash \spl{\exists x F, \Gamma}, \spr{\Delta}}
\end{prooftree}} 
= \pi.
\end{array}$$
\item If $y$ occurs in $\Delta$ but not in $\Gamma$, then we set $I = \forall y I_1$, 
$\pi^L = \scalebox{0.8}{\begin{prooftree}
    \Hypo{\pi^L_1}
    \Infer [no rule]1{\vdash \spl{F\{y/x\},\Gamma}, I_1}
    \Infer 1[\rexists]{\vdash \spl{\exists x F, \Gamma}, I_1}
    \Infer 1[\rforall]{\vdash \spl{\exists x F, \Gamma}, \forall yI_1}
\end{prooftree}}$
and $\pi^R = \scalebox{0.8}{\begin{prooftree}
    \Hypo{\pi^R_1}
    \Infer [no rule]1 {\vdash {I_1}^\perp, \spr{\Delta}}
    \Infer 1[\rexists]{\vdash \exists y {I_1}^\perp, \spr{\Delta}}
\end{prooftree}}$.
One gets that $\pi' = \scalebox{0.8}{\begin{prooftree}
\Hypo {\pi^L}
\Infer [no rule]1{\vdash \spl{\exists x F, \Gamma}, \forall yI_1}
%\Infer 1{\vdash \forall x. F, \Gamma', \exists x. C'}
\Hypo {\pi^R}
\Infer [no rule]1{\vdash \exists y {I_1}^\perp, \spr{\Delta}}
%\Infer 1{\vdash \forall x. {C'}^\perp, \Delta}
\Infer 2[\rcut] {\vdash \spl{\exists x F, \Gamma}, \spr{\Delta}}
\end{prooftree}}$ is a solved \CPRIS to which $\pi$ cut-expands:

$$
\begin{array}{rcl}
\pi' 
&\cutred&
\scalebox{0.8}{\begin{prooftree}
\Hypo {\pi^L_1}
\Infer [no rule]1{\vdash \spl{F\{y/x\},\Gamma}, I_1}
\Infer 1[\rexists]{\vdash \spl{\exists x F, \Gamma}, I_1}
\Hypo {\pi^R_1}
\Infer [no rule]1 {\vdash {I_1}^\perp, \spr{\Delta}}
\Infer 2[\rcut] {\vdash \spl{\exists x F, \Gamma}, \spr{\Delta}}
\end{prooftree}} 
\\
&\cutred &
\scalebox{0.8}{\begin{prooftree}
\Hypo {\pi^L_1}
\Infer [no rule]1{\vdash \spl{F\{y/x\},\Gamma}, I_1}
\Hypo {\pi^R_1}
\Infer [no rule]1 {\vdash {I_1}^\perp, \spr{\Delta}}
\Infer 2[\rcut] {\vdash \spl{F\{y/x\}, \Gamma}, \spr{\Delta}}
\Infer 1[\rexists]{\vdash \spl{\exists  x F, \Gamma}, \spr{\Delta}}
\end{prooftree}} 
= \pi.
\end{array}
$$
\end{itemize}

\end{description}

\subsection{Appendix on Section \ref{sec:computational}}
\label{app:systemL}
As usual now, the proof of \Cref{thm:print-cut-systemL} follows straightforwardly from a weak interpolation theorem and checking that the conditions on the cut formulas satisfy the interpolation constraint on the vocabularies: 

\def\VocGammaDeltaALC{\VOC{\Gamma_1,\Delta_1,A, \calC^L}\cap \VOC{\Gamma_2,\Delta_2, \calC^R}}
\def\VocGammaDeltaARC{\VOC{\Gamma_1,\Delta_1, \calC^L}\cap \VOC{\Gamma_2,\Delta_2,A, \calC^R}}
\def\VocGammaDeltaC{\VOC{\Gamma_1,\Delta_1, \calC^L}\cap \VOC{\Gamma_2,\Delta_2, \calC^R}}

\begin{theorem}
    In what follows, $t$ (resp. $e$, resp. $c$) is an L-term (resp. normal L-context, resp. normal L-command).
    et $\calC$ be the set of types of all the $t',e'$ such that $\langle t' \mid e' \rangle$ is a  subcommand of $t$ (resp. $e$, resp. $c$). 
     Let us assume that $\calC$ is split into $\calC^L,\calC^R$ and $\mathbb C^L = \VOC{\calC^L}$ and  $\mathbb C^R = \VOC{\calC^R}$
    The following weak interpolating results hold:
    \begin{enumerate}
        \item If $c: (\spl {\Gamma_1},\spr{\Gamma_2} \vdash  \spl {\Delta_1},\spr{\Delta_2})$, there exist a type $I\in \VocGammaDeltaC$ and $t,e$ such that
        $\spl {\Gamma_1} \vdash t: \spl I \mid \spl {\Delta_1}$ and 
        $\spr{\Gamma_2} \mid e: \spr I \vdash  \spr{\Delta_2}$, and $\langle t \mid e \rangle \longrightarrow^\star c$.        
        \item If $\spl {\Gamma_1},\spr{\Gamma_2} \vdash t: \spl A \mid  \spl {\Delta_1},\spr{\Delta_2}$, there exist a type $I\in \VocGammaDeltaALC$ and $\alpha,t',e'$ such that
        $\spl {\Gamma_1} \vdash t': \spl A \mid \alpha: \spl I, \spl {\Delta_1}$ and 
        $\spr{\Gamma_2} \mid e':\spr I \vdash  \spr{\Delta_2}$, and $t'\{e'/\alpha\} \longrightarrow^\star t$.                        \item If $\spl {\Gamma_1},\spr{\Gamma_2} \mid e: \spl A \vdash   \spl {\Delta_1},\spr{\Delta_2}$, there exist a type $I\in \VocGammaDeltaALC$ and $\alpha,e',e''$ such that
        $\spl {\Gamma_1} \mid e': \spl A \mid \alpha: \spl I, \spl {\Delta_1}$ and 
        $\spr{\Gamma_2}\mid e'' :\spr I \vdash  \spr{\Delta_2}$, and $e'\{e''/\alpha\} \longrightarrow^\star e$.        
        \item If $\spl {\Gamma_1},\spr{\Gamma_2} \vdash t: \spr A \mid  \spl {\Delta_1},\spr{\Delta_2}$, there exist a type $I\in \VocGammaDeltaARC$ and $\alpha,t',t''$ such that
        $\spl {\Gamma_1} \vdash t'': \spl I \mid \spl {\Delta_1}$ and 
        $\spr{\Gamma_2}, x:\spr I \vdash t': \spr A \mid  \spr{\Delta_2}$, and $t'\{t''/x\} \longrightarrow^\star t$.        
        \item If $\spl {\Gamma_1},\spr{\Gamma_2} \mid e: \spr A \vdash  \spl {\Delta_1},\spr{\Delta_2}$, there exist a type $I\in \VocGammaDeltaARC$ and $x,t',e'$ such that
        $\spl {\Gamma_1} \vdash t': \spl I \mid \spl {\Delta_1}$ and 
        $\spr{\Gamma_2}, x: \spr I \mid e':\spr A \vdash  \spr{\Delta_2}$, and $e'\{t'/x\} \longrightarrow^\star e$.        
    \end{enumerate}
\end{theorem}

\begin{proof}
While the proof by Saurin \cite[Thm 20]{Saurin} relies on a long case analysis on both terms, evaluation contexts and commands, it is sufficient to adapt the case of commands as the other cases are straightforward by a simple check that the propagation of the cut-vocabulary constraint works. 
Regarding the case of commands, while Saurin considers only cases  $\langle x \mid e \rangle $  and $\langle t \mid \alpha \rangle $ (due to the assumption on normal terms), we now have full fledged commands which are interpolated in a way close to the tensor case in \cite[Thm 20]{Saurin}. 

Let $c: (\spl {\Gamma_1},\spr{\Gamma_2} \vdash  \spl {\Delta_1},\spr{\Delta_2})$, 
be as in the theorem statement, 
with $c = \langle t \mid e \rangle $.
We want to prove that
there exist a type $K$ such that $\VOC K \in \VocGammaDeltaC$ and $u,f$ such that
        $\spl {\Gamma_1} \vdash u: \spl K \mid \spl {\Delta_1}$ and 
        $\spr{\Gamma_2} \mid f: \spr K \vdash  \spr{\Delta_2}$, and $\langle u \mid f \rangle \longrightarrow^\star c$.

$t$ and $e$ must respectively be a term and a context of type $C$ (that we assume is split to the left in the theorem assumption: $C\in\calC^L$; the other case is symmetrical).
%Let us assume that  type $C$ is split to the left in the hypothesis of the theorem.
We then apply the induction hypothesis to 
$\spl {\Gamma'_1},\spr{\Gamma'_2} \vdash t: \spl{C}  \mid  \spl {\Delta'_1},\spr{\Delta'_2}$
and 
$\spl {\Gamma''_1},\spr{\Gamma''_2}  \mid e: \spl{C}  \vdash  \spl {\Delta''_1},\spr{\Delta''_2}$
and we get (in what follows we can assume that $\alpha \not\in \mathsf{FV}(e',e'')$ and that $\beta \not\in \mathsf{FV}(t',f')$ ): 
\begin{itemize}
\item (i) $\spl {\Gamma'_1} \vdash t': \spl{C}  \mid  \alpha: I, \spl {\Delta'_1}$,
%\item
(ii) $\spr{\Gamma'_2} \mid f': I  \vdash \spr{\Delta'_2}$
and (iii) $t'\{f'/\alpha\} \longrightarrow^\star t$;
\item (i) $\spl {\Gamma''_1} \mid e': \spl{C}  \vdash \beta: J, \spl {\Delta''_1}$,
%\item
(ii)  $\spr{\Gamma''_2}  \mid e'': J  \vdash \spr{\Delta''_2}$
%\item $t'\{f'/\alpha\} \longrightarrow^\star t$
%\item
and (iii)  $e'\{e''/\beta\} \longrightarrow^\star e$.
\end{itemize}
With $\VOC I \subseteq  \VOC{\Gamma'_1,\Delta'_1,C, \calC^L}\cap \VOC{\Gamma'_2,\Delta'_2, \calC^R}$ and
$\VOC J \subseteq  \VOC{\Gamma''_1,\Delta''_1,C, \calC^L}\cap \VOC{\Gamma''_2,\Delta''_2, \calC^R}$

Then, by setting:
%\begin{itemize}
%\item
 $K = I\tensor J$,
%\item
 $u = \mu [\alpha,\beta] \langle t'\mid e'\rangle$
%\item
and  $f = [{f'},{e''}]$
%\end{itemize}
We have  $\VOC K \in \VocGammaDeltaC$ and 
  $\spl {\Gamma_1} \vdash u: \spl K \mid \spl {\Delta_1}$ and 
        $\spr{\Gamma_2} \mid f: \spr K \vdash  \spr{\Delta_2}$ as expected and  
        $\langle u \mid f \rangle \longrightarrow \langle t'\mid e'\rangle\{f'/\alpha, e''/\beta\}
        =  \langle t'\{f'/\alpha\}\mid e'\{e''/\beta\}\rangle  \longrightarrow^\star c$ by induction hypothesis.
\end{proof}

% !TEX root = main.tex
\section{Appendix on Section \ref{sec:proofRelIntCircProof}}

\subsection{Proofs for (ordered) trace manifolds}
\label{appendix:traceManifolds}

Before the proof of Lemmas \ref{lemm:traceManifold} and \ref{lemm:ordTraceManifold}, we introduce the following lemmas and definitons:

\begin{lemma}[Brotherston \cite{brotherston}, Lemma 7.1.3]
    \label{lem:InfPath1}
    Let $\mathcal{P} = (\mathcal{D},\mathcal{R})$ be a \muLLo pre-proof. Then for every infinite path $\pi$ through $\mathcal{G}_{\mathcal{P}}$, there is a non-trivial strongly connected subgraph $S_{\pi} \subseteq \mathcal{G}_{\mathcal{P}}$ and a tail $\pi'$ of $\pi$ such that $\pi'$ is an infinite path through $S_{\pi}$ on which each vertex of $S_{\pi}$ occurs infinitely often. 
\end{lemma}

\begin{definition}[Path composition]
    Let $G$ be a graph, let $\pi_1 = v_0...v_m$ be a finite path in $G$ and let $\pi_2 = v'_0v'_1v'_2...$ be a (possibly infinite) path in $G$. We say that $\pi_1$ \textit{composes} with $\pi_2$ if $v_m = v'_0$ and define the composed path $\pi_1\pi_2$ to be $v_0...v_{m-1}v'_0v'_1v'_2...$
\end{definition}

\begin{lemma}[Brotherston \cite{brotherston}, Lemma 7.1.8]
    \label{lem:PathComp}
    Let $\mathcal{P} = (\mathcal{D}, \mathcal{R})$ be a \muLLo pre-proof and let $S$ be a non-trivial strongly connected subgraph of $\mathcal{G}_{\mathcal{P}}$. Note that $S = \bigcup_{B \in \mathcal{B}_{S}} \mathcal{C}_{B}$ by the Lemma \ref{lem:SCCAsUnion}. Then, any infinite path in $S$, starting from the companion of a bud in $\mathcal{B}_{S}$, is composed of paths (of length $> 0$) each of the form $\mathcal{R}(B_i)...\mathcal{R}(B_j)$, where $B_i, B_j \in \mathcal{B}_{S}$ and $\mathcal{R}(B_i)...\mathcal{R}(B_j)$ is a subpath of some cycle $\mathcal{C}_{B}$.
\end{lemma}

\begin{proof}[Proof of Lemma \ref{lemm:traceManifold}]
        We use Brotherston's proof (\cite{brotherston}, Proposition 7.2.3) with slight modifications to accommodate for the modified properties of the validity condition. Let $\mathcal{P} = (\mathcal{D}, \mathcal{R})$ be a \muLLo pre-proof and consider any infinite path $\pi$ through $\mathcal{G}_{\mathcal{P}}$. By Lemma \ref{lem:InfPath1}, $\pi$ has a tail $\pi'$ that is an infinite path through a non-trivial strongly connected subgraph $S$, where every vertex in $S$ infinitely appears in $\pi'$. Moreover, by lemma \ref{lem:SCCAsUnion}, $S = \bigcup_{B \in \mathcal{B}_{S}} \mathcal{C}_{B}$ where $\mathcal{B}_S \subseteq Bud(\mathcal{D})$ and $\mathcal{B}_S$ is weakly $\leq_{\mathcal{P}}$-connected. We then want to show that we can construct a valid trace on $\pi'$. Without loss of generality, assume that $\mathcal{R}(B_{i_1})$ is the first vertex in $\pi'$, for some $B_{i_1} \in \mathcal{B}_S$. Then by lemma \ref{lem:PathComp}, $\pi'$ is composed of finite paths in $S$ as follows:
        \begin{align*}
            \pi' = \mathcal{R}(B_{i_1})...\mathcal{R}(B_{i_2})...\mathcal{R}(B_{i_3})...
        \end{align*}
        Where for all $n \geq 1$, $B_{i_n} \in \mathcal{B}_S$ and $\mathcal{R}(B_{i_n})...\mathcal{R}(B_{i_{n+1}})$ is a subpath of some basic cycle $\mathcal{C}_{B_{j_n}}$, with $B_{j_n} \in \mathcal{B}_S$. The idea is to form the trace that validates $\pi'$, by associating a segment of a trace $\uptau_{S,B_{j_n}}$, to each subpath $\mathcal{R}(B_{i_n})...\mathcal{R}(B_{i_{n+1}})$. Suppose we have the two composable subpaths $\mathcal{R}(B_{i_n})...\mathcal{R}(B_{i_{n+1}})$ in $\mathcal{C}_{B_{j_n}}$ and $\mathcal{R}(B_{i_{n+1}})...\mathcal{R}(B_{i_{n+2}})$ in $\mathcal{C}_{B_{j_{n+1}}}$. We then associate a segment $\tau_{j_n}$ of the trace $\uptau_{S,B_{j_n}}$, that covers the subpath $\mathcal{R}(B_{i_n})...\mathcal{R}(B_{i_{n+1}})$ and a segment $\tau_{j_{n+1}}$ of the trace $\uptau_{S,B_{j_{n+1}}}$, that covers the subpath $\mathcal{R}(B_{i_{n+1}})...\mathcal{R}(B_{i_{n+2}})$. Notice that $\mathcal{R}(B_{i_{n+1}})$ occurs both in $\mathcal{C}_{B_{j_n}}$ and $\mathcal{C}_{B_{j_{n+1}}}$, hence $B_{i_{n+1}} \leq_{\mathcal{P}} B_{j_n}$ and $B_{i_{n+1}} \leq_{\mathcal{P}} B_{j_{n+1}}$. Then by the property 2, we get that $\uptau_{S,B_{j_n}}(\mathcal{R}(B_{i_{n+1}})) = \uptau_{S,B_{i_{n+1}}}(\mathcal{R}(B_{i_{n+1}})) = \uptau_{S,B_{j_{n+1}}}(\mathcal{R}(B_{i_{n+1}}))$, which then implies that $\uptau_{j_n}$ is composable with $\uptau_{j_{n+1}}$. We then say that the trace $\uptau$ that covers $\pi'$ is the trace formed by using for each subpath $\mathcal{R}(B_{i_n})...\mathcal{R}(B_{i_{n+1}})$ the trace segment $\uptau_{j_n}$ of the trace $\uptau_{S,B_{j_n}}$. We would like to prove that the constructed trace $\uptau$ is a valid trace in the path $\pi'$. To do it, we prove 1) $\uptau$ infinitely visits a $F$-progress point, 2) $\textsf{min}(\textsf{Inf}(\uptau))= F$:
        \begin{enumerate}
            \item By the property 3, there is a $B_{i_k}$ such that $\uptau_{S,B_{i_k}}$ has an $F$-progress point on the basic cycle $\mathcal{C}_{B_{i_k}}$. Moreover, by Lemma \ref{lem:InfPath1} every vertex and edge in $S$ infinitely appears in $\pi'$, which implies that we can find infinite instances of the subpath $\mathcal{R}(B_{i_n})...\mathcal{R}(B_{i_{n+1}})$ in $\pi'$, that includes the vertex with the $F$-progress point. Assume this subpath is contained in the basic cycles $\mathcal{C}_{B_{i_k}}$ and $\mathcal{C}_{B_{i_h}}$. For this reason, we can then assign a portion of the trace $\uptau_{S,B_{i_k}}$ or $\uptau_{S,B_{i_h}}$. We show that we can always assign the progressing trace $\uptau_{S,B_{i_k}}$. Note that $B_{i_n} \leq_{\mathcal{P}} B_{i_k}$ and $B_{i_n} \leq_{\mathcal{P}} B_{i_h}$, therefore $\uptau_{S,B_{i_k}}(\mathcal{R}(B_{i_n})) = \uptau_{S,B_{i_n}}(\mathcal{R}(B_{i_n})) = \uptau_{S,B_{i_h}}(\mathcal{R}(B_{i_n}))$. The same applies for $B_{i_{n+1}}$, so $\uptau_{S,B_{i_k}}(\mathcal{R}(B_{i_{n+1}})) = \uptau_{S,B_{i_{n+1}}}(\mathcal{R}(B_{i_{n+1}})) = \uptau_{S,B_{i_h}}(\mathcal{R}(B_{i_{n+1}}))$. Thus, both $\uptau_{S,B_{i_k}}$ or $\uptau_{S,B_{i_h}}$ follow the subpath and agree at $\mathcal{R}(B_{i_{n}})$ and $\mathcal{R}(B_{i_{n+1}})$. We then always choose the trace $\uptau_{S,B_{i_k}}$ with the progress point, ensuring it infinitely appears in $\pi'$.
            
            \item Let $\uptau_{S,B_{i_1}}, \uptau_{S,B_{i_2}}, ... \uptau_{S,B_{i_n}}$ be the traces  associated to $S$ in the manifold. It is easy to see that since $\tau$ is constructed using the previously mentioned traces, then it must be the case that $ \textsf{Inf}(\uptau) \subseteq \bigcup^{n}_{j=1} \textsf{Form}(\uptau_{S,B_{i_j}})$. Hence, by property 3, $F$ is minimal in $\bigcup^{n}_{j=1} \textsf{Form}(\uptau_{S,B_{i_j}})$, which would also imply that it is minimal in $\textsf{Inf}(\uptau)$. Finally, we now that $F \in \textsf{Inf}(\uptau)$, since $\tau$ infinitely visits the $F$-progress point, thus $\textsf{min}(\textsf{Inf}(\uptau))= F$. 
        \end{enumerate}
\end{proof}

\begin{proof}[Proof of Lemma \ref{lemm:ordTraceManifold}]
We use the same proof as Brotherston's (\cite{brotherston}, Lemma 7.2.8) to prove both directions, with slight modifications to accommodate for the modified properties of the validity condition:
        
    $\Rightarrow)$ Suppose that $\mathcal{P} = (\mathcal{D}, \mathcal{R})$ has a trace manifold. We will show how to construct an induction order $\vartriangleleft$ for $\mathcal{P}$ and ordered trace manifold with respect to $\vartriangleleft$. Take any maximal set $\mathcal{B} \subseteq Bud(\mathcal{D})$ of weakly $\leq_{\mathcal{P}}$-connected buds, i.e. there is no other weakly $\leq_{\mathcal{P}}$-connected set of buds $\mathcal{B}'\subseteq Bud(\mathcal{D})$ such that $\mathcal{B} \subset \mathcal{B'}$, and by Lemma \ref{lem:SCCAsUnion} let $S = \cup_{B\in \mathcal{B}}\mathcal{C}_{B}$. By definition of a trace manifold, there is some $B_k \in \mathcal{B}$ such that the trace $\uptau_{S,B_k}$ progresses. It could be the case that that there are more than one bud that satisfy this property, in which case we pick any of the possible candidates. Then, we choose $B_k$ to be the $\vartriangleleft$-greatest element in $\mathcal{B}$, i.e. for all $B_i \in \mathcal{B}$ then $B_i \vartriangleleft B_k$. Moreover, we define $\uptau_{B_k,B_i} = \uptau_{S,B_i}$ for each $B_i \in \mathcal{B}$. Note that $B_k$ also becomes a $\vartriangleleft$-greatest element in every subset of $\mathcal{B}$ that contains $B_k$. We then consider all the maximal weakly $\leq_\mathcal{P}$-connected subsets of $Bud(\mathcal{D}) \setminus\{B_k\}$ and repeat this process until every weakly $\leq_\mathcal{P}$-connected subset of $Bud(\mathcal{D})$ has a $\vartriangleleft$-greatest element. We check then that the constructed $\vartriangleleft$ order indeed is a partial order and satisfies the two properties to be an induction order:
    \begin{itemize}
        \item $\vartriangleleft$ is reflexive by construction. To prove transitivity, take some $B_i \vartriangleleft B_j$ and $B_j \vartriangleleft B_k$. This implies that there are weakly $\leq_{\mathcal{P}}$-connected subsets $\mathcal{B}_j, \mathcal{B}_k \subseteq Bud(\mathcal{D})$ such that $B_j$ and $B_k$ are $\vartriangleleft$-maximal elements in $\mathcal{B}_j$ and $\mathcal{B}_k$ respectively. Moreover, since $B_j \vartriangleleft B_k$, this implies that $\mathcal{B}_j \subseteq \mathcal{B}_k$, and since $B_i \vartriangleleft B_j$, it is the case that $B_i \in \mathcal{B}_j$. Hence, $B_i \in \mathcal{B}_k$ and $B_i \vartriangleleft B_k$. For anti-symmetry, suppose $B_j \vartriangleleft B_k$ and $B_k \vartriangleleft B_j$. This implies that there are weakly $\leq_{\mathcal{P}}$-connected subsets $\mathcal{B}_j, \mathcal{B}_k \subseteq Bud(\mathcal{D})$ such that $B_j$ and $B_k$ are $\vartriangleleft$-maximal elements in $\mathcal{B}_j$ and $\mathcal{B}_k$ respectively, and  $\mathcal{B}_k = \mathcal{B}_j$ since $\mathcal{B}_j \subseteq \mathcal{B}_k$ and $\mathcal{B}_k \subseteq \mathcal{B}_j$. Therefore $B_j = B_k$, since they are the unique maximum of both subsets.
        
        \item To check that it is forest-like, let $B_i \vartriangleleft B_j$ and $B_i \vartriangleleft B_k$. By construction, there exist weakly $\leq_{\mathcal{P}}$-connected subset $\mathcal{B}_j,\mathcal{B}_k \subseteq Bud(\mathcal{D})$ such that $B_j$ and $B_k$ are $\vartriangleleft$-maximal elements in $\mathcal{B}_j$ and $\mathcal{B}_k$ respectively, where $B_i \in \mathcal{B}_j$ and $B_i \in \mathcal{B}_k$. By construction, it must be the case that either $\mathcal{B}_j \subseteq \mathcal{B}_k$, in which case $B_j \vartriangleleft B_k$, or $\mathcal{B}_k \subseteq \mathcal{B}_j$, in which case $B_k \vartriangleleft B_j$, or $\mathcal{B}_j = \mathcal{B}_k$, in which case $B_j = B_k$. 
        
        \item By construction, it is trivial to see that ever  every weakly $\leq_\mathcal{P}$-connected subset of $Bud(\mathcal{D})$ has a $\vartriangleleft$-greatest element.
    \end{itemize}
    Finally, we check that the constructed traces correspond to a ordered trace manifold, with respect to the induction order $\vartriangleleft$:
    \begin{enumerate}
        \item Let $B_i \vartriangleleft B_j$. By construction, $\uptau_{B_j,B_i} = \uptau_{S,B_i}$ exists and is a trace following the basic cycle $\mathcal{C}_{B_i}$ in $\mathcal{G}_{\mathcal{P}}$ and the same value at both instances of $\mathcal{R}(B_i)$, by the property 1 of the trace manifold.

        \item Let $B_j,B_i \vartriangleleft B_k$ and $B_j \leq_{\mathcal{P}} B_i$. By construction, $\uptau_{B_k,B_j} = \uptau_{S,B_j}$ and $\uptau_{B_k,B_i} = \uptau_{S,B_i}$ for some $S$ such that $S = \cup_{B\in \mathcal{B}}\mathcal{C}_{B}$ and $B_k,B_j,B_i \in \mathcal{B}$ with $B_k$ as $\vartriangleleft$-maximal. By the property 2 of the trace manifold, then $\uptau_{S,B_i}(\mathcal{R}(B_j)) = \uptau_{S,B_j}(\mathcal{R}(B_j))$, hence $\uptau_{B_k,B_i}(\mathcal{R}(B_j)) = \uptau_{B_k,B_j}(\mathcal{R}(B_j))$.

        \item Note that by contruction, For every $k$ and any $j$, $\uptau_{B_k,B_{i_j}} = \uptau_{S,B_{i_j}}$  for some $S$, where $\uptau_{B_k,B_k}$ has an $F$-progress point. Moreover, by property 3 of the trace manifold it is the case that $F$ is minimal in $\bigcup^{n}_{j=1} \textsf{Form}(\uptau_{S,B_{i_j}})$, hence $F$ is minimal in $\bigcup^{n}_{j=1} \textsf{Form}(\uptau_{B_k,B_{i_j}})$.
    \end{enumerate}
        
    $\Leftarrow)$ Let $\vartriangleleft$ be an induction order for $\mathcal{P} = (\mathcal{D}, \mathcal{R})$ and suppose that $\mathcal{P}$ has an ordered trace manifold with respect to $\vartriangleleft$. Now, consider any strongly connected subgraph $S$ such that $S = \cup_{B\in \mathcal{B}_S}\mathcal{C}_{B}$ with $\mathcal{B}_S \subseteq Bud(\mathcal{D})$ and weakly $\leq_{\mathcal{P}}$-connected, by Lemma \ref{lem:SCCAsUnion}. By definition of induction order, $\mathcal{B}_S$ has a $\vartriangleleft$-maximal element $B_k$. We then make $\uptau_{S,B_i} = \uptau_{B_k,B_i}$, for each $i$, and check that this construction satisfies the properties of a trace manifold:
    \begin{enumerate}
        \item By the property 1 of the ordered trace manifold, $\uptau_{S,B_i} = \uptau_{B_k,B_i}$ is a trace following the basic cycle $\mathcal{C}_{B_i}$, and takes the same value at both occurrences of $\mathcal{R}(B_i)$.

        \item Let $B_i, B_j \in \mathcal{B}_S$, where $B_j \leq_{\mathcal{P}} B_i$. Therefore, since $B_k$ is $\vartriangleleft$-maximal, then $B_j, B_i \vartriangleleft B_k$ and by property 2 of the ordered trace manifold, we get that $\uptau_{B_k,B_j}(\mathcal{R}(B_j)) = \uptau_{B_k,B_i}(\mathcal{R}(B_j))$, hence $\uptau_{S,B_j}(\mathcal{R}(B_j)) = \uptau_{S,B_i}(\mathcal{R}(B_j))$.

        \item Note that for any $j$, $\uptau_{S,B_{i_j}} = \uptau_{B_k,B_{i_j}}$ where $\uptau_{S,B_k} = \uptau_{B_k,B_k}$ has an $F$-progress point. Moreover, by property 3 of the ordered trace manifold, it is the case that $F$ is minimal in $\bigcup^{n}_{j=1} \textsf{Form}(\uptau_{B_k,B_{i_j}})$, hence $F$ is minimal in $\bigcup^{n}_{j=1} \textsf{Form}(\uptau_{S,B_{i_j}})$.
    \end{enumerate}
    \end{proof}

%###############################################################
%###############################################################
%###############################################################
%###############################################################

\subsection{Injective bud-companion relation}
\label{appendix:injectiveRel}

Without loss of generality, for all pre-proofs $\mathcal{P} = (\mathcal{D}, \mathcal{R})$ we can consider that $\mathcal{R}$ is an injective function. This can be done by defining the derivable rule $(\textsf{Id})$ which consists of using the mix rule $(\textsf{X})$ without making any change in the order of the formulas in the sequent. Now, let $C$ be any companion node where $C$ is labelled with $\vdash \Gamma$, and $\{B_1, \dots, B_n\}$ is the set of bud nodes such that for all $i$, $\mathcal{R}(B_i) = C$. Then $C$ can be transformed as follows:

\begin{center}
\begin{tabular}{ c c c }

 \begin{prooftree}
    \hypo{\vdash \Gamma'}
    \infer1[(r)]{(\dagger)_{B_1}, \dots (\dagger)_{B_n} \vdash \Gamma}
 \end{prooftree} 
 & $\Rightarrow$ &
 \begin{prooftree}
    \hypo{\vdash \Gamma'}
    \infer1[(r)]{(\dagger)_{B_{p(n)}} \vdash \Gamma}
    \infer1[(\textsf{Id})]{\vdots}
    \infer1[(\textsf{Id})]{(\dagger)_{B_{p(1)}} \vdash \Gamma}
 \end{prooftree}
\end{tabular}
\end{center}

Where $p$ defines a permutation over \{1, \dots, n\}. Note that this change in $\mathcal{P}$ does not alter its validity, in case $\mathcal{P}$ is a valid \muLLo proof. Moreover, in case $\mathcal{P}$ can be coherently decorated, this transformation preserves this property, since the rule \textsf{(Id)} maintains the same split during the ascending phase.

This also implies that for any companion node with multiple bud nodes $B_1, \dots, B_n$, we can choose the order of the back-edges and therefore implicitly choose the relation $\leq_\mathcal{P}$ over $B_1, \dots, B_n$. For this reason, given a tree-compatible pre-proof $\mathcal{P}$, with non-injective $\mathcal{R}$, we can apply this same transformation to obtain a tree-compatible pre-proof $\mathcal{R}$, with injective $\mathcal{R}$. 

The procedure chooses the adequate permutation $p$ as follows: Consider any companion node $C$, with a set of bud nodes $\mathcal{B} = \{B_1, \dots, B_n\}$ such that for all $i$, $\mathcal{R}(B_i) = C$. Since every bud node points to the same companion then the set $\mathcal{B}$ is weekly $\leq_\mathcal{P}$-connected, and by Definition \ref{def:indOrder}, there is a $B_{i} \in \mathcal{B}$ that is a $\vartriangleleft$-maximum for $\mathcal{B}$. Then we assign $p(1) = i$, in order to make $B_{i}$ a $\leq_{\mathcal{P}}$-maximum for $\mathcal{B}$. We then consider the set $\mathcal{B} \setminus \{B_{i}\}$ and repeat the process, until assigning an index in the permutation to every bud node in $\mathcal{B}$. The previously described assignment, clearly preserves tree-compatibility.

%===========================================================
%===========================================================
%===========================================================
%===========================================================

\subsection{Example of non-coherently decorated pre-proof}
\label{appendix:exampleNonCoh}

Let $G = \nu X. \mu Y. X$ and $H = \mu Y. G$. The following pre-proof, cannot be coherently decorated, since the split between its only companion and bud node are different:

\begin{center}
    \begin{prooftree}
    \hypo{(\dagger)}
    \infer1[]{\vdash \spr{G}, \spl{H}}
    \infer1[$(\mu)$]{\vdash \spr{H}, \spl{H}}
    \infer1[(\textsf{X})]{\vdash \spl{H}, \spr{H}}
    \infer1[$(\nu)$]{(\dagger) \vdash \spl{G}, \spr{H}}
\end{prooftree}
\end{center}

However, after the execution of the Algorithm \ref{algo:splitInv}, the resulting pre-proof is:

\begin{center}
    \begin{prooftree}
    \hypo{(\dagger)}
    \infer1[]{\vdash \spl{G}, \spr{H}}
    \infer1[$(\mu)$]{\vdash \spl{H}, \spr{H}}
    \infer1[(\textsf{X})]{\vdash \spr{H}, \spl{H}}
    \infer1[$(\nu)$]{\vdash \spr{G}, \spl{H}}
    \infer1[$(\mu)$]{\vdash \spr{H}, \spl{H}}
    \infer1[(\textsf{X})]{\vdash \spl{H}, \spr{H}}
    \infer1[$(\nu)$]{(\dagger) \vdash \spl{G}, \spr{H}}
\end{prooftree}
\end{center}    

Which can be coherently decorated.

%===========================================================
%===========================================================
%===========================================================
%===========================================================

\subsection{Algorithms}

% !TEX root = main.tex
\begin{figure}[h!]
\centering
\fbox{
\begin{minipage}[t]{0.48\textwidth}
\captionof{algorithm}{Split Invariance}
\label{algo:splitInv}
\scriptsize
\algrenewcommand\alglinenumber[1]{\scriptsize #1:}
\algrenewcommand\algorithmicindent{0.8em}
\input{algorithm1}
\end{minipage}
\hfill
\begin{minipage}[t]{0.48\textwidth}
\captionof{algorithm}{Tree-compatibility \cite{SprengerDam}}
\label{algo:treecomp}
\scriptsize
\algrenewcommand\alglinenumber[1]{\scriptsize #1:}
\algrenewcommand\algorithmicindent{0.8em}
\input{algorithm2}
\end{minipage}
}
\caption{Split Invariance and Tree-compatibility algorithms. Algorithm \ref{algo:treecomp} was originally introduced by Sprenger and Dam in \cite{SprengerDam}. Algorithm \ref{algo:splitInv} can be seen as an extension of Algorithm \ref{algo:treecomp}, that uses the same unfolding procedure.}
\label{fig:algorithms}
\end{figure}

Algorithms \ref{algo:splitInv} and \ref{algo:treecomp}, can be found in the Figure \ref{fig:algorithms}. In the following subsections, we present the proofs for their correctness.

\subsubsection{Correctness of Algorithm \ref{algo:splitInv}}
\label{appendix:correctAlgo1}

\begin{proof}[Proof of Lemma \ref{lem:splitInvariant}]
	To prove the total correctness of the algorithm we show that: 1) the algorithm \ref{algo:splitInv} terminates, 2) $\mathcal{P} = (\mathcal{D}, \mathcal{R})$ is a pre-proof, 3) $\mathcal{P}$ has a trace manifold, and 4) $\mathcal{P}$ can be coherently decorated.
    
    \textbf{(1):} After the initial call (line 12), for all bud node $B \in Bud(\mathcal{D}_0)$, then the algorithm creates the unique path from the root $(v_r, 0)$ in $\mathcal{D}$ to the bud node $(B, 0)$ in $\mathcal{D}$, while also propagating the split of the conclusion (lines 32-39). Moreover, it stores the information of the companion node $\mathcal{R}_0(B)$ and its split $(\Gamma, \Delta)$ in $\mathcal{A}$ (lines 29-30). In the call \texttt{splitInvariance} $(B, 0)$ $(\Gamma', \Delta')$ $\mathcal{A}$, if $(\Gamma', \Delta') = (\Gamma, \Delta)$ then the pair $((B, 0)), (\mathcal{R}_0(B), 0)$ is included in $\mathcal{R}$, and the unfolding is finished for that branch in the derivation tree $\mathcal{D}$ (lines 21-22). On the contrary, if $(\Gamma', \Delta') \neq (\Gamma, \Delta)$ then the node $(B, 0)$ will be replaced by $(\mathcal{R}_0(B), 1)$, and it will serve as the root of the new derivation tree that will be unfolded in the branch (lines 23-26) and the previously explained process is repeated. Therefore, to prove that there is no infinite unfolding of the pre-proof $\mathcal{D}$, we must show that eventually all bud nodes $(B, i)$ in $\mathcal{D}$ are matched with a companion node in $\mathcal{A}$ with the same split. However, note that the number of possible splits of $s_0(B)$ is finite, since $s_0(B)$ is a finite list of formulas. Since the amount of splits is finite and the times we can unfold the proof is infinite, we can ensure that eventually, for some $l$, $(\mathcal{R}_0(B),l)$ will appear in $\mathcal{A}$, with the same split as the bud node $(B,i)$.

    \textbf{(2):} We prove by induction on the recursion tree of the algorithm, that after each call of \texttt{splitInvariance} $(v, k)$ $(\Gamma, \Delta)$, then $\mathcal{P}$ is a pre-proof. In the initial call (line 11), $\mathcal{D}$ contains only the root node $(v_r, 0)$. If $r_0(v_r)$ is a rule with no premises, then no node is added and $\mathcal{P}$ is a pre-proof with a single node. On the contrary, if $r_0(v_r)$ has premises, then each of the nodes corresponding to these premises are added to $\mathcal{D}$ according to the $p_0$ function of $\mathcal{D}_0$ (line 32), which by hypothesis is a well-defined edge and therefore, $\mathcal{P}$ is a pre-proof. The case where $r_0(v_r)$ is undefined, i.e. $v_r$ is a bud node, is impossible since it contradicts the fact that $\mathcal{P}_0$ is pre-proof.

    Now for the inductive step, suppose a call \texttt{splitInvariance} $(v, k)$ $(\Gamma, \Delta)$ $\mathcal{A}$. We know by construction of the algorithm, that $(v,k)$ was already added to $\mathcal{D}$ in the previous step (line 24 or line 32). Moreover, by induction hypothesis we know that before the execution of the recursive call, $\mathcal{P}$ is a pre-proof. It remains to show that after its execution, $\mathcal{P}$ is still a pre-proof. We proceed by cases over $r_0(v)$. If $r_0(v)$ is a rule with or without premises, we use the same arguments as in the base case. If $r_0(v)$ is undefined, there are two cases: if $((\mathcal{R}_0(v), (\Gamma', \Delta')),i)$ is in $\mathcal{A}$ for some $i$, and $(\Gamma, \Delta) = (\Gamma', \Delta')$, then the back edge $((v,k), (\mathcal{R}_0(v), i))$ is added to $\mathcal{R}$ (line 21). Since $s_0(v) = s_0(\mathcal{R}_0(v))$, then $s((v,k)) = s((\mathcal{R}_0(v), i))$, and therefore $\mathcal{P}$ is a pre-proof. Otherwise, if $(\Gamma, \Delta) \neq (\Gamma', \Delta')$ then $(v,k)$ is replaced with the node $(\mathcal{R}_0(v), n)$ with the rule $r_0(\mathcal{R}_0(v))$ and labelling $s_0(\mathcal{R}_0(v))$ (line 24), which constructs a well-defined edge, and therefore $\mathcal{P}$ is a pre-proof.

    \textbf{(3):} Consider any $(B, k) \in Bud(\mathcal{D})$ and $\pi_{B}$ as the unique path from $\mathcal{R}((B, k))$ to $(B, k)$ in $\mathcal{D}$. By inspection on the Algorithm \ref{algo:splitInv}, $\pi_{B}$ is obtained by unfolding a set of bud nodes $U = \{B_1, \dots,B_m\} \subseteq Bud(\mathcal{D}_0)$, represented by the following sequence of recursive calls:
    \begin{align*}
        c = ((\mathcal{R}(B, k)), \mathcal{A}_0) \dots ((B_1, k_1), \mathcal{A}_1) \dots  ((B_m, k_m), \mathcal{A}_m) \dots((B, k), \mathcal{A}_{m+1})   
    \end{align*}
    Where each $B_j \in Bud(\mathcal{D}_0)$, for $1 \leq j \leq m$, is an unpaired bud node that will be replaced by its companion in the line 25. Therefore, the path $\pi_{B}$ will has the form:
    \begin{align*}
            \pi_{B} = \mathcal{R}(B, k) \dots (\mathcal{R}_0(B_1), k'_1) \dots (\mathcal{R}_0(B_m), k'_m) \dots (B, k)
     \end{align*}
     Let $\mathcal{B}_{\pi_{B}} = U \cup \{B\} \subseteq Bud(\mathcal{D}_0)$, and consider the path $\widehat{\pi_{B}} = \mathcal{R}_0(B) \dots \mathcal{R}_0(B_1) \dots \mathcal{R}_0(B_m) \dots B$, obtained after eliminating the numbers of the pairs describing the nodes in the path $\pi_{B}$. From the definition of the Algorithm \ref{algo:splitInv}, one deduces that $\widehat{\pi_{B}}$ describes a cycle contained in the strongly connected subgraph $S_{\pi_B} = \bigcup_{B' \in \mathcal{B}_{\pi_B}} \mathcal{C}_{B'}$, in the graph $\mathcal{G}_{\mathcal{P}_0}$ (here we implicitly use Lemma \ref{lem:SCCAsUnion}). Therefore, by Lemma \ref{lem:PathComp}, we get that:

    \begin{itemize}
         \item The unique path $\pi_{B}$, represented as the basic cycle $\mathcal{C}_{(B,k)}$ in $\mathcal{G}_{\mathcal{P}}$, is formed by the composition of paths (of length $> 0$) each of the form $(\mathcal{R}_0(B_i), k_i) \dots (\mathcal{R}_0(B_j), k_j)$, where $B_i, B_j \in \mathcal{B}_{\pi_B}$ and $\mathcal{R}_0(B_i) \dots \mathcal{R}_0(B_j)$ is a subpath of some cycle $\mathcal{C}_{B''} \subseteq S_{\pi_B} = \bigcup_{B' \in \mathcal{B}_{\pi_B}} \mathcal{C}_{B'}$ in $\mathcal{G}_{\mathcal{P}_0}$. \textbf{(A)}
     \end{itemize}

    With this in mind, we show how to construct the traces for the trace manifold for $\mathcal{P}$. Take any strongly connected subgraph $S$ of $\mathcal{G}_{\mathcal{P}}$. By Lemma \ref{lem:SCCAsUnion} there exists a set $\mathcal{B} \subseteq Bud(\mathcal{D})$ such that $S = \bigcup_{B \in \mathcal{B}} \mathcal{C}_B$. If $B \in \mathcal{B}$, we show how to construct the trace $\uptau_{S,B}$ up the basic cycle obtained from the unique path $\pi_{B}$ from $\mathcal{R}(B)$ to $B$ in $\mathcal{D}$.

    As mentioned before, $\pi_{B}$ is defined by a set of unfolded bud nodes $\mathcal{B}_{\pi_B} \subseteq Bud(\mathcal{D}_0)$ such that $\pi_B$ corresponds to a strongly connected subgraph $S_{\pi_B} = \bigcup_{B' \in \mathcal{B}_{\pi_B}} \mathcal{C}_{B'}$ in $\mathcal{G}_{\mathcal{P}_0}$. Therefore, if we forget of the indexes in the strongly connected subgraph $S$ and we call it $\widehat{S}$, then $\widehat{S} = \bigcup_{B \in \mathcal{B}} S_{\pi_B} = \bigcup_{B \in \mathcal{B}} (\bigcup_{B' \in \mathcal{B}_{\pi_B}} \mathcal{C}_{B'})$. Note that this is the union of intersecting strongly connected subgraphs. Indeed, $\widehat{S}$ describes a cycle in the original graph $\mathcal{G}_{\mathcal{P}_0}$. For this reason:

    \begin{itemize}
         \item For all bud node $B \in \mathcal{B} \subseteq Bud(\mathcal{D})$, and $B' \in \mathcal{B}_{\pi_B} \subseteq Bud(\mathcal{D}_0)$, then $\uptau_{\widehat{S}, B'}$ exists in the trace manifold for $\mathcal{P}_0$, and is a trace following the basic cycle $\mathcal{C}_{B'}$ in $\mathcal{G}_{\mathcal{P}_0}$. \textbf{(B)}
     \end{itemize}

     The trace $\uptau_{S,B}$ for the basic cycle $\mathcal{C}_{B}$ obtained from the path $\pi_B$ is constructed in the following way: take every segment $(\mathcal{R}_0(B_i), k_i) \dots (\mathcal{R}_0(B_j), k_j)$ of $\mathcal{C}_{B}$, as described in (A). The segment $\mathcal{R}_0(B_i) \dots \mathcal{R}_0(B_j)$ is a subpath of some cycle $\mathcal{C}_{B'} \subseteq \bigcup_{B'' \in \mathcal{B}_{\pi_B}} \mathcal{C}_{B''}  \subseteq \widehat{S}$ where $B' \in \mathcal{B}_{\pi_B}$. Therefore, by (B) we can use the trace $\uptau_{\widehat{S}, B'}$ to cover the segment $\mathcal{R}_0(B_i) \dots \mathcal{R}_0(B_j)$, and consequently the segment $(\mathcal{R}_0(B_i), k_i) \dots (\mathcal{R}_0(B_j), k_j)$. In case the segment $\mathcal{R}_0(B_i) \dots \mathcal{R}_0(B_j)$ has a progress point, and it is contained in multiple basic cycles, we always choose the trace containing the formula that progresses, as explained in the proof of Lemma \ref{lemm:traceManifold}.

    Note that these traces can be composed, by the property 2 of the original trace manifold for $\mathcal{P}_0$: consider any segment $(\mathcal{R}_0(B_i), k_i) \dots (\mathcal{R}_0(B_j), k_j) \dots (\mathcal{R}_0(B_h), k_h)$. Such that the segments $\mathcal{R}_0(B_i) \dots \mathcal{R}_0(B_j)$ and $\mathcal{R}_0(B_j) \dots \mathcal{R}_0(B_h)$ are included in the cycles $\mathcal{C}_{B_1}, \mathcal{C}_{B_2} \subseteq \bigcup_{B'' \in \mathcal{B}_{\pi_B}} \mathcal{C}_{B''}$ respectively, in $\mathcal{G}_{\mathcal{P}_0}$. This implies that $B_j \leq_{\mathcal{P}_0} B_1$ and $B_j \leq_{\mathcal{P}_0} B_2$, and therefore $\uptau_{\widehat{S}, B_1}(\mathcal{R}_0(B_j)) = \uptau_{\widehat{S}, B_j}(\mathcal{R}_0(B_j)) = \uptau_{\widehat{S}, B_2}(\mathcal{R}_0(B_j))$.

    We show that this construction satisfies the three properties of trace manifolds. Remember that for all $B = (B_i, k_i) \in Bud(\mathcal{D})$ then $\mathcal{R}(B) = (\mathcal{R}_0(B_i), k_j)$.

     \begin{enumerate}
         \item By construction, the trace $\uptau_{S,B}$ is a trace over the basic cycle $\mathcal{C}_B$ in $\mathcal{G}_\mathcal{P}$. We check that $\uptau_{S,B}$ has a unique value at $\mathcal{R}(B)$. By construction $\uptau_{S,B}(\mathcal{R}(B)) = \uptau_{\widehat{S},B'}(\widehat{\mathcal{R}(B)})$ and $\uptau_{S,B}(\mathcal{R}(B)) = \uptau_{\widehat{S},B''}(\widehat{\mathcal{R}(B)})$ for some $B', B'' \in \mathcal{B}_{\pi_B}$. Note that $\widehat{\mathcal{R}(B)}$ is in $\mathcal{C}_{B'}$ and $\mathcal{C}_{B''}$, implying $\widehat{B} \leq_{\mathcal{P}_0} B'$ and $\widehat{B} \leq_{\mathcal{P}_0} B''$. Then by property 2 of the trace manifold, $\uptau_{\widehat{S},B'}(\widehat{\mathcal{R}(B)}) = \uptau_{\widehat{S},\widehat{B}}(\widehat{\mathcal{R}(B)}) = \uptau_{\widehat{S},B''}(\widehat{\mathcal{R}(B)})$. Thus, $\uptau_{S,B}$ takes the same value at $\mathcal{R}(B)$.

         \item Let $B_1, B_2 \in \mathcal{B}$ such that $B_1 \leq_{\mathcal{P}} B_2$. By construction of $\uptau_{S,B_1}(\mathcal{R}(B_1)) = \uptau_{\widehat{S},B'}(\widehat{\mathcal{R}(B_1)})$ and $\uptau_{S,B_2}(\mathcal{R}(B_1)) = \uptau_{\widehat{S},B''}(\widehat{\mathcal{R}(B_1)})$ for some $B' \in \mathcal{B}_{\pi_{B_1}}$ and $B'' \in \mathcal{B}_{\pi_{B_2}}$. Hence, $\widehat{\mathcal{R}(B_1)}$ is in $\mathcal{C}_{B'}$ and $\mathcal{C}_{B''}$, which implies $\widehat{B_1} \leq_{\mathcal{P}_0} B'$ and $\widehat{B_1} \leq_{\mathcal{P}_0} B''$. Therefore, by property 2 of the trace manifold, $\uptau_{\widehat{S},B'}(\widehat{\mathcal{R}(B_1)}) = \uptau_{\widehat{S},\widehat{B_1}}(\widehat{\mathcal{R}(B_1)}) = \uptau_{\widehat{S},B''}(\widehat{\mathcal{R}(B_1)})$. Thus $\uptau_{S,B_1}(\mathcal{R}(B_1)) = \uptau_{S,B_2}(\mathcal{R}(B_1))$.

         \item Let $\uptau_{S, B_1}, \dots, \uptau_{S, B_n}$ be the traces associated to $S$ in the trace manifold, and consider their respective unique paths $\pi_{B_1}, \dots, \pi_{B_n}$ from companion to bud node. Assume that $\mathcal{B}$ is the set of weakly $\leq_{\mathcal{P}}$-connected bud nodes in $S$. As explained before, we can recover the original strongly connected subgraph in $\mathcal{G}_{\mathcal{P}_0}$ by removing the indexes of the vertices, which gives us $\widehat{S} = \bigcup_{B \in \mathcal{B}} S_{\pi_B} = \bigcup_{B \in \mathcal{B}} (\bigcup_{B' \in \mathcal{B}_{\pi_B}} \mathcal{C}_{B'})$. By property 3 of the original trace manifold, then:

         \begin{itemize}
             \item There exists a bud node $B_i \in \mathcal{B}$ and a bud node $B' \in \mathcal{B}_{\pi_{B_i}}$, such that $\uptau_{\widehat{S}, B'}$ has an $F$-progress point, such that $F$ is minimal in the set $\bigcup_{B'' \in \mathcal{B}_{\pi_{B_1}}} \textsf{Form}(\uptau_{\widehat{S}, B''}) \cup \dots \cup \bigcup_{B'' \in \mathcal{B}_{\pi_{B_n}}} \textsf{Form}(\uptau_{\widehat{S}, B''})$. \textbf{(C)}
         \end{itemize}

         Statement (C) also implies that this $F$-progress point is located in the basic cycle $\mathcal{C}_{B'}$ of $\mathcal{G}_{\mathcal{P}_0}$, say in the vertex $v$. By construction, one can see that this vertex is also included in $S$, with the form $(v,k)$ for some $k$. Since $(v,k)$ is in $S$, it must be the case that there exists some $B_j \in \mathcal{B}$, such that $(v,k)$ is included in $\pi_{B_j}$. By construction of the trace $\uptau_{S,B_j}$, we also now that every time we pass through the segment $(\mathcal{R}_0(B_h), k_h) \dots (\mathcal{R}_0(B_r), k_r)$, containing the vertex $(v,k)$, we used the segment of the trace $\uptau_{\widehat{S}, B'}$ to cover it. For this reason, $\uptau_{S, B_j}$ has an $F$-progress point. We now show that $F$ is minimal in all the traces $\uptau_{S, B_1}, \dots, \uptau_{S, B_n}$.

         Every trace $\uptau_{S, B_h}$  is obtained by the composition of segments of the traces $\uptau_{\widehat{S}, B''}$, such that $B'' \in \mathcal{B}_{\pi_{B_h}}$. Thus, for all $h$ we have that $\textsf{Form}(\uptau_{S, B_h}) \subseteq \bigcup_{B'' \in \mathcal{B}_{\pi_{B_h}}} \textsf{Form}(\uptau_{\widehat{S}, B''})$. Therefore:
         \begin{align*}
             \bigcup_{1 \leq h \leq n} \textsf{Form}(\uptau_{S, B_i}) \subseteq \bigcup_{1 \leq h \leq n}(\bigcup_{B'' \in \mathcal{B}_{\pi_{B_h}}} \textsf{Form}(\uptau_{\widehat{S}, B''}))
         \end{align*}
         Therefore, by (C) and the fact that $\uptau_{S,B_j}$ has the $F$-progress point, one deduces that $F$ is also minimal in $\bigcup_{1 \leq h \leq n} \textsf{Form}(\uptau_{S, B_i})$.
    \end{enumerate}

    \textbf{(4):} After the ascending phase, we obtain a coherent decoration of the derivation tree $\mathcal{D}$. By construction of the algorithm, it is clear that all pairs $(B, \mathcal{R}(B)) \in \mathcal{R}$ added in the line 22, are ensured to have the same split, by the conditional in line 21. Therefore, $\mathcal{P}$ can be coherently decorated.
\end{proof}

\subsubsection{Correctness of Algorithm \ref{algo:treecomp}}
\label{appendix:correctAlgo2}

For the proof of correctness, we briefly introduce the following lemma: 

\begin{definition}[Derivation graph homomorphism]
    Let $G = (V, s, r, p)$ and $G' = (V', s', r', p')$ be two derivation graphs. A derivation graph homomorphism from $G$ to $G'$ is a total function $f: V \rightarrow V'$ satisfying that for any $v \in V$:
    \begin{align*}
        s(v) &= s'(f(v)) & r(v) &\simeq r'(f(v)) & f(p_j(v)) &\simeq p'_j(f(v)) \quad \text{for each } j \in \mathbb{N}
    \end{align*}
    Where $f(x) \simeq g(y)$ means that $f(x)$ is defined iff $g(y)$ is defined and if $f(x)$ is defined then $f(x) = g(y)$. If $G,G'$ are derivations graphs, we write $f:G \rightarrow G'$ to mean that $f$ is a derivation graph homomorphism from $G$ to $G'$.
\end{definition}

\begin{lemma}[Brotherston \cite{brotherston}, Lemma 6.1.3]
    \label{lemm:PathPres}
    Let $G = (V, s, r, p)$ and $G' = (V', s', r', p')$ be derivation graphs such that there exists a derivation graph homomorphism $f: G \rightarrow G'$. Then for every infinite path $v_0v_1v_2...$ in $G$, there exists an infinite path $v'_0v'_1v'_2...$ in $G'$ with the same rule sequent labelling, i.e. $s(v_i) = s'(v'_i)$ and $r(v_i) = r'(v'_i)$ for all $i \geq 0$. Furthermore, if $f$ is surjective the converse holds.
\end{lemma}

The induction order $(Bud(\mathcal{D}),\vartriangleleft)$ for the obtained circular proof $\mathcal{P}$ after the execution of Algorithm \ref{algo:treecomp}, is given by lifting the original induction order $(Bud(\mathcal{D}_0),\vartriangleleft_0)$ of $\mathcal{P}_0$ to $\mathcal{P}$. We define $\widehat{B} = v$ if $B = (v,k)$, where $B \in Bud(\mathcal{D})$ and $\widehat{B} \in Bud(\mathcal{D}_0)$. Then the induction order $(Bud(\mathcal{D}),\vartriangleleft)$ is defined as $B \vartriangleleft B' \ \text{iff} \ \widehat{B} \vartriangleleft_0 \widehat{B'} \ \text{or}  \ (\widehat{B} = \widehat{B'} \ \text{and}  \ k \leq k')$, where $B = (v,k)$ and $B' = (v',k')$.

\begin{proof}[Proof of lemma \ref{lemm:treeComp}]
To prove the total correctness of the algorithm we must show that: 1) $\mathcal{P}$ is a pre-proof, 2) $(Bud(\mathcal{D}),\vartriangleleft)$ is a tree-compatible induction order for $\mathcal{P}$, 3) There exists a ordered trace manifold with respect to $(Bud(\mathcal{D}),\vartriangleleft)$, 4) $\mathcal{P}$ can be coherently decorated, 5) Algorithm \ref{algo:treecomp} terminates. For the item 1), the proof follows the same approach as done for Algorithm \ref{algo:splitInv}, by doing induction on the execution tree for \texttt{treeCompatibility}. For the items 2) and 5) the reader can refer to the proof of Theorem 5 in \cite{SprengerDam}. We prove items 3) and 4):

\textbf{(3):} Consider any bud node $(B,k) \in Bud(\mathcal{D})$ with $\mathcal{R}(B,k) = (\mathcal{R}_0(B), i)$. Moreover, consider the sequence of calls:
    \begin{align*}
        c = ((\mathcal{R}_0(B), i), \mathcal{A}_0) \dots ((\mathcal{R}_0(B_1), k_1), \mathcal{A}_1) \dots ((\mathcal{R}_0(B_m), k_m), \mathcal{A}_m) \dots((B, k), \mathcal{A}_{m+1})   
    \end{align*}
    Where each $B_j \in Bud(\mathcal{D}_0)$, for $1 \leq j \leq m$. This sequence represents the execution calls of the Algorithm \texttt{treeCompatibility}, used to construct the unique path from $(\mathcal{R}_0(B), i)$ to $(B,k)$ in $\mathcal{D}$, where each call $((\mathcal{R}_0(B_j), k_j), \mathcal{A}_j)$ represents a recursive call over a companion node in the original proof $\mathcal{P}_0$, with $1 \leq j \leq m$. Since $\mathcal{R}(B,k) = (\mathcal{R}_0(B), i)$, it must be the case that the pair $(\mathcal{R}_0(B), i)$ is added to $\mathcal{A}_0$ in the first call $c_0$ (line 28) and it appears in all subsequent $\mathcal{A}_j$ with $1 \leq j \leq m+1$. Then, the pair $((B,k), (\mathcal{R}_0(B), i))$ is added to $\mathcal{R}$ in the call $c_{m+1}$ (line 20). Now, consider any bud node $B_j$ appearing in the call $c_j$ with $1 \leq j \leq m$. Since $(\mathcal{R}_0(B), i)$ is in both $\mathcal{A}_j$ and $\mathcal{A}_{j+1}$, and $(\mathcal{R}_0(B_j), k_j)$ is added to $\mathcal{A}_j$ in the call $c_j$ (line 28), it must be the case that:

    \begin{itemize}
        \item If $\mathcal{R}_0(B_j) \notin \operatorname{dom}(\mathcal{A}_j)$, then $B_j \vartriangleleft_0 B$. \textbf{(A)}
    \end{itemize}

    Now consider any $(B, k) \in Bud(\mathcal{D})$ and $\pi_B$ as the unique path from $\mathcal{R}((B, k))$ to $(B, k)$ in $\mathcal{D}$. By inspection on the Algorithm \ref{algo:treecomp}, $\pi_B$ is obtained by unfolding a set of bud nodes $U = \{B_1, \dots,B_m\} \subseteq Bud(\mathcal{D}_0)$, represented by the following sequence of recursive calls:
    \begin{align*}
        c = ((\mathcal{R}(B, k)), \mathcal{A}_0) \dots ((B_1, k_1), \mathcal{A}_1) \dots  ((B_m, k_m), \mathcal{A}_m) \dots((B, k), \mathcal{A}_{m+1})   
    \end{align*}
    Where each $B_j \in Bud(\mathcal{D}_0)$, for $1 \leq j \leq m$, is an unpaired bud node that will be replaced by its companion in the line 23. Therefore, the path $\pi_B$ will have the form:
    \begin{align*}
            \pi_B = \mathcal{R}(B, k) \dots (\mathcal{R}_0(B_1), k'_1) \dots (\mathcal{R}_0(B_m), k'_m) \dots (B, k)
     \end{align*} 
    
     Let $\mathcal{B}_{\pi_{B}} = U \cup \{B\} \subseteq Bud(\mathcal{D}_0)$, and consider the path $\widehat{\pi_{B}} = \mathcal{R}_0(B) \dots \mathcal{R}_0(B_1) \dots \mathcal{R}_0(B_m) \dots B$, obtained after eliminating the numbers of the pairs describing the nodes in the path $\pi_{B}$. From the definition of the Algorithm \ref{algo:treecomp}, one deduces that $\widehat{\pi_{B}}$ describes a cycle contained in the strongly connected subgraph $S_{\pi_B} = \bigcup_{B' \in \mathcal{B}_{\pi_B}} \mathcal{C}_{B'}$, in the graph $\mathcal{G}_{\mathcal{P}_0}$ (here we implicitly use Lemma \ref{lem:SCCAsUnion}). Therefore, by Lemma \ref{lem:PathComp}, we get that:

    \begin{itemize}
         \item The unique path $\pi_{B}$, represented as the basic cycle $\mathcal{C}_{(B,k)}$ in $\mathcal{G}_{\mathcal{P}}$, is formed by the composition of paths (of length $> 0$) each of the form $(\mathcal{R}_0(B_i), k_i) \dots (\mathcal{R}_0(B_j), k_j)$, where $B_i, B_j \in \mathcal{B}_{\pi_B}$ and $\mathcal{R}_0(B_i) \dots \mathcal{R}_0(B_j)$ is a subpath of some cycle $\mathcal{C}_{B''} \subseteq S_{\pi_B} = \bigcup_{B' \in \mathcal{B}_{\pi_B}} \mathcal{C}_{B'}$ in $\mathcal{G}_{\mathcal{P}_0}$. \textbf{(B)}
     \end{itemize}

     With this in mind, we show how to construct the traces for the ordered trace manifold for $\mathcal{P}$. Take any $B, B' \in Bud(\mathcal{D})$ such that $B \vartriangleleft B'$. We will construct the trace $\uptau_{B',B}$ of the ordered trace manifold for $\mathcal{P}$, following the unique path $\pi_B$ from $\mathcal{R}(B)$ to $B$ in $\mathcal{D}$. 

     Take any $B_j$ with $1 \leq j \leq m$. After the call $((B_j, k_j), \mathcal{A}_j)$ in $c$, since $(B_j, k_j)$ is not paired with any companion node, it implies that the conditional in line 19 was not satisfied, and therefore $\mathcal{R}_0(B_j) \notin \operatorname{dom}(\mathcal{A}_j)$. This implies by (A), that $B_j \vartriangleleft_0 \widehat{B}$. Moreover, by definition of $\vartriangleleft$, we have that $\widehat{B} \vartriangleleft_0 \widehat{B'}$ or $\widehat{B} = \widehat{B'}$. Thus, $B_j \vartriangleleft_0 \widehat{B} \vartriangleleft_0 \widehat{B'}$ or $B_j \vartriangleleft_0 \widehat{B} = \widehat{B'}$, in both cases implying $B_j \vartriangleleft_0 \widehat{B'}$. Hence, we have that for all bud node $B'' \in \mathcal{B}_{\pi_B}$, then $B'' \vartriangleleft_0 \widehat{B'}$. Then, by definition of the original ordered trace manifold for $\mathcal{P}_0$: 

     \begin{itemize}
         \item For all bud node $B \vartriangleleft B'$ and $B'' \in \mathcal{B}_{\pi_B}$, then $\uptau_{\widehat{B}', B''}$ exists in the ordered trace manifold for $\mathcal{P}_0$, and is a trace following the basic cycle $\mathcal{C}_{B''}$ in $\mathcal{G}_{\mathcal{P}_0}$. \textbf{(C)}
     \end{itemize}

     The trace $\uptau_{B',B}$ for the path $\pi_B$ is constructed in the following way: take every segment $(\mathcal{R}_0(B_i), k_i) \dots (\mathcal{R}_0(B_j), k_j)$ of $\pi_B$, as described in (B). The segment $\mathcal{R}_0(B_i) \dots \mathcal{R}_0(B_j)$ is a subpath of some cycle $\mathcal{C}_{B''} \subseteq S_{\pi_B}$ where $B'' \in \mathcal{B}_\pi$. Therefore, by (C) we can use the trace $\uptau_{\widehat{B}', B''}$ to cover the segment $\mathcal{R}_0(B_i) \dots \mathcal{R}_0(B_j)$, and consequently the segment $(\mathcal{R}_0(B_i), k_i) \dots (\mathcal{R}_0(B_j), k_j)$. In case the segment $\mathcal{R}_0(B_i) \dots \mathcal{R}_0(B_j)$ has a progress point, and it is contained in multiple basic cycles, we always choose the trace containing the formula that progresses, as explained in the proof of Lemma \ref{lemm:traceManifold}.

     Note that these traces can be composed, by the property 2 of the original ordered trace manifold for $\mathcal{P}_0$: consider any segment $(\mathcal{R}_0(B_i), k_i) \dots (\mathcal{R}_0(B_j), k_j) \dots (\mathcal{R}_0(B_h), k_h)$. Such that the segments $\mathcal{R}_0(B_i) \dots \mathcal{R}_0(B_j)$ and $\mathcal{R}_0(B_j) \dots \mathcal{R}_0(B_h)$ are included in the cycles $\mathcal{C}_{B_1}, \mathcal{C}_{B_2} \subseteq \bigcup_{B'' \in \mathcal{B}_{\pi_B}} \mathcal{C}_{B''}$ respectively, in $\mathcal{G}_{\mathcal{P}_0}$. This implies that $B_j \leq_{\mathcal{P}_0} B_1$ and $B_j \leq_{\mathcal{P}_0} B_2$. Moreover, since $B_1, B_2, B_j \in \mathcal{B}_{\pi_B}$, we know that $B_1, B_2, B_j \vartriangleleft_0 \widehat{B'}$, therefore $\uptau_{\widehat{B'}, B_1}(\mathcal{R}_0(B_j)) = \uptau_{\widehat{B'}, B_j}(\mathcal{R}_0(B_j)) = \uptau_{\widehat{B'}, B_2}(\mathcal{R}_0(B_j))$.

    We show that this construction satisfies the three properties of ordered trace manifolds. Remember that for all $B = (B_i, k_i) \in Bud(\mathcal{D})$ then $\mathcal{R}(B) = (\mathcal{R}_0(B_i), k_j)$.

    \begin{enumerate}
         \item By construction, the trace $\uptau_{B',B}$ is a trace over the basic cycle $\mathcal{C}_B$ in $\mathcal{G}_\mathcal{P}$. We check that $\uptau_{B',B}$ has a unique value at $\mathcal{R}(B)$. By construction $\uptau_{B',B}(\mathcal{R}(B)) = \uptau_{\widehat{B'},B_1}(\widehat{\mathcal{R}(B)})$ and $\uptau_{B',B}(\mathcal{R}(B)) = \uptau_{\widehat{B'},B_2}(\widehat{\mathcal{R}(B)})$ for some $B_1, B_2 \in \mathcal{B}_{\pi_B}$. Note that $\widehat{\mathcal{R}(B)}$ is in $\mathcal{C}_{B_1}$ and $\mathcal{C}_{B_2}$, implying $\widehat{B} \leq_{\mathcal{P}_0} B_1$ and $\widehat{B} \leq_{\mathcal{P}_0} B_2$. Then by property 2 of the ordered trace manifold, $\uptau_{\widehat{B'},B_1}(\widehat{\mathcal{R}(B)}) = \uptau_{\widehat{B'},\widehat{B}}(\widehat{\mathcal{R}(B)}) = \uptau_{\widehat{B'},B_2}(\widehat{\mathcal{R}(B)})$. Thus, $\uptau_{S,B}$ takes the same value at $\mathcal{R}(B)$.

         \item Let $B_1, B_2 \vartriangleleft B'$ such that $B_1 \leq_{\mathcal{P}} B_2$. By construction, $\uptau_{B',B_1}(\mathcal{R}(B_1)) = \uptau_{\widehat{B'},B''}(\widehat{\mathcal{R}(B_1)})$ and $\uptau_{B',B_2}(\mathcal{R}(B_1)) = \uptau_{\widehat{B'},B'''}(\widehat{\mathcal{R}(B_1)})$ for some $B'' \in \mathcal{B}_{\pi_{B_1}}$ and $B''' \in \mathcal{B}_{\pi_{B_2}}$. Hence, $\widehat{\mathcal{R}(B_1)}$ is in $\mathcal{C}_{B''}$ and $\mathcal{C}_{B'''}$, which implies $\widehat{B_1} \leq_{\mathcal{P}_0} B''$ and $\widehat{B_1} \leq_{\mathcal{P}_0} B'''$. Also note that $\widehat{B_1}, B'', B''' \vartriangleleft_0 \widehat{B'}$. Therefore, by property 2 of the ordered trace manifold, $\uptau_{\widehat{B'},B''}(\widehat{\mathcal{R}(B_1)}) = \uptau_{\widehat{B'},\widehat{B_1}}(\widehat{\mathcal{R}(B_1)}) = \uptau_{\widehat{B'},B'''}(\widehat{\mathcal{R}(B_1)})$. Thus, $\uptau_{B',B_1}(\mathcal{R}(B_1)) = \uptau_{B',B_2}(\mathcal{R}(B_1))$.

         \item Let $B_1, \dots, B_n, B' \in Bud(\mathcal{D})$ such that $B_1, \dots, B_n \vartriangleleft B'$, i.e. all bud nodes that are below $B'$ in the $\vartriangleleft$ order (this includes $B'$ itself). Moreover, consider the unique paths $\pi_{B_1}, \dots, \pi_{B_n}$ from companion to bud node for $B_1, \dots, B_2$ respectively. For any $1 \leq i \leq n$, consider the trace $\uptau_{B', B_i}$. By construction, $\uptau_{B', B_i}$  is obtained by the composition of segments of the traces $\uptau_{\widehat{B}', B''}$, such that $B'' \in \mathcal{B}_{\pi_{B_i}}$. Thus, for all $i$, $\textsf{Form}(\uptau_{B', B_i}) \subseteq \bigcup_{B'' \in \mathcal{B}_{\pi_{B_i}}} \textsf{Form}(\uptau_{\widehat{B}', B''})$. Therefore, we have that:
         \begin{align*}
             \bigcup_{1 \leq i \leq n} \textsf{Form}(\uptau_{B', B_i}) \subseteq \bigcup_{1 \leq i \leq n} (\bigcup_{B'' \in \mathcal{B}_{\pi_i}} \textsf{Form}(\uptau_{\widehat{B}', B''}))
         \end{align*}
         Then, the formula $F$ that is minimal in the set $\bigcup_{1 \leq i \leq n} (\bigcup_{B'' \in \mathcal{B}_{\pi_i}} \textsf{Form}(\uptau_{\widehat{B}', B''}))$ by the property 3 in the ordered trace manifold for $\mathcal{P}_0$, is also minimal in the set $\bigcup_{1 \leq i \leq n} \textsf{Form}(\uptau_{B', B_i})$. It remains to be shown that $\uptau_{B',B'}$ contains an $F$-progress point. This can be done by inspection on the path
         $\pi_{B'}$ from $\mathcal{R}(B')$ to $B'$ in $\mathcal{D}$. For every vertex $v$ in the basic cycle $\mathcal{C}_{\widehat{B}'}$ by the definition of the unfolding, one can always find a copy $(v,k)$ of the original vertex in $\mathcal{C}_{B'}$. For this reason, we can use the trace $\uptau_{\widehat{B'},\widehat{B'}}$, to cover the progress point, i.e. $\uptau_{B',B'}(v,k) = \uptau_{\widehat{B'},\widehat{B'}}(v)$, which contains the $F$-progress point.
     \end{enumerate}
     
\textbf{(4):} After the ascending phase, we obtain a coherent decoration of the derivation tree $\mathcal{D}$. It remains to be shown that the bud companion relation $\mathcal{R}$ preserves the coherence of $\mathcal{P}$. Consider then any rooted path $\pi$ in $\mathcal{G}_\mathcal{P}$. As explained before, $\pi$ is obtained by unfolding a set of bud nodes $U = \{B_1, \dots, B_m\} \subseteq Bud(\mathcal{D}_0)$. Therefore, $\pi$ has the form:
\begin{align*}
     \pi = (v_r, 0) \dots (\mathcal{R}_0(B_1), k'_1) \dots (\mathcal{R}_0(B_m), k'_m) \dots (v_f, k)
 \end{align*} 
where $(v_r,0)$ is the root of $\mathcal{D}$ and $(v_f, k)$ is the final vertex of $\pi$. We can define a surjective derivation graph homomorphism from $\mathcal{G}_{\mathcal{P}}$ to $\mathcal{G}_{\mathcal{P}_0}$, by mapping each copy $(v,k)$ in $\mathcal{G}_{\mathcal{P}}$ to its original vertex $v$ in $\mathcal{G}_{\mathcal{P}_0}$. Therefore, by Lemma \ref{lemm:PathPres}, there exists a path $\widehat{\pi}$ with the same rule sequent labelling as $\pi$.

Now consider the function $\rho_0: V_0 \rightarrow \mathcal{S}$, which stores the information of the decoration of the pre-proof $\mathcal{P}_0$, i.e. it maps every node in $\mathcal{D}_0$ with their respective split $(\Gamma, \Delta)$ after the ascending phase over the derivation tree $\mathcal{D}_0$. Similarly, we consider the function $\rho: V \rightarrow \mathcal{S}$, which stores the information of the decoration over the derivation tree $\mathcal{D}$ of the pre-proof $\mathcal{P}$, that is obtained after executing the ascending phase. We would like to prove that:

\begin{itemize}
    \item If $\mathcal{P}_0$ can be coherently decorated, then $\rho_0(v) = \rho(v, i)$ for every vertex $(v,i) \in \pi$ and $v \in \widehat{\pi}$. \textbf{(D)}
\end{itemize}

We proceed by induction on the length $n$ of $\pi$. If $n = 1$, then $\widehat{\pi} = v_r$ and $\pi = (v_r, 0)$. Since both $\mathcal{P}_0$ and $\mathcal{P}$ have the same initial split, then $\rho_0(v_r) = \rho(v_r, 0)$. In the case $n+1$, then $\widehat{\pi} = \widehat{\pi_n}, v_f$ and $\pi = \pi_n, (v_f, k)$. By induction hypothesis, for every $(v',k')$ in $\pi_n$, then $\rho_0(v') = \rho(v,k')$. We will assume that $(v_n, k')$ is the last vertex of the path $\pi_n$. We proceed by cases over the final node $(v_f, k)$. If $(v_f,k)$ is one of the unfolded bud nodes in $\pi$, i.e. $(v_f,k) = (\mathcal{R}_0(B_j), k'_j)$ for $1 \leq j \leq m$, it implies that $\rho(v_f,k) = \rho(\mathcal{R}_0(B_j), k'_j)$ is obtained by propagating the split $\rho(v_n,k')$ using the rule $r(v_n, k')$. Moreover, this implies that in $\mathcal{D}_0$ we have the subpath $\theta = v_n, B_j$. Therefore, $\rho_0(B_j)$ is obtained by propagating the split $\rho_0(v_n)$ using the rule $r_0(v_n)$. Since both $\rho_0(v_n) = \rho(v_n, k')$ by induction hypothesis and $r_0(v_n) = r(v_n, k')$ since $\pi$ and $\widehat{\pi}$ have the same rule sequent labelling, we deduce that $\rho(v_f,k) = \rho(\mathcal{R}_0(B_j), k'_j) = \rho_0(B_j)$. Furthermore, since $\mathcal{P}_0$ can be coherently decorated, it implies that $\rho_0(B_j) = \rho_0(\mathcal{R}_0(B_j))$, hence $\rho(v_f,k) = \rho(\mathcal{R}_0(B_j), k'_j) = \rho_0(\mathcal{R}_0(B_j)) = \rho_0(v_f)$.

Otherwise, if $(v_f,k) \neq (\mathcal{R}_0(B_j), k'_j)$ for $1 \leq j \leq m$ it implies that $\rho(v_f,k)$ is obtained by propagating the split $\rho(v_n,k')$ using the rule $r(v_n, k)$. Moreover, this implies that in $\mathcal{D}_0$ we have the subpath $\theta = v_n, v_f$. Therefore, $\rho_0(v_f)$ is obtained by propagating the split $\rho_0(v_n)$ using the rule $r_0(v_n)$. Since both $\rho_0(v_n) = \rho(v_n, k')$ by induction hypothesis and $r_0(v_n) = r(v_n, k')$ since $\pi$ and $\widehat{\pi}$ have the same rule sequent labelling, we deduce that $\rho(v_f,k) = \rho_0(v_f)$.

Now consider any bud node $(B, k) \in Bud(\mathcal{D})$ and the rooted path $\pi = (v_r,0), \dots, (v_p,k), \mathcal{R}(B,k)$ in $\mathcal{G}_{\mathcal{P}}$, and $\widehat{\pi} = v_r, \dots, v_p, \widehat{\mathcal{R}(B,k)}$ in $\mathcal{G}_{\mathcal{P}_0}$, where $(v_p,k)$ is the predecessor of $(B, k)$ in $\mathcal{D}$, and $\pi$ with $\widehat{\pi}$ have the same rule sequent labelling. By (D), we know that $\rho(v_p, k) = \rho_0(v_p)$. Moreover, there is a subpath $\theta = (v_p,k), (B,k)$ and $\widehat{\theta} = v_p, B$ in $\mathcal{D}$ and $\mathcal{D}_0$ respectively. The split $\rho(B,k)$ is obtained by propagating the split $\rho(v_p,k)$ using the rule $r(v_p, k)$, while the split $\rho_0(B)$ is obtained by propagating the split $\rho_0(v_p)$ using the rule $r_0(v_p)$. Since $\rho(v_p, k) = \rho_0(v_p)$ and $r_0(v_p) = r(v_p, k)$ since $\pi$ and $\widehat{\pi}$ have the same rule sequent labelling, we deduce that $\rho(B,k) = \rho_0(B)$.

Furthermore, by definition of the Algorithm \ref{algo:treecomp} in the line 20, we have that $\mathcal{R}(B,k) = (\mathcal{R}_0(B),t)$ for some $t$. Therefore, by (D) it implies that $\rho(\mathcal{R}(B,k)) = \rho(\mathcal{R}_0(B),t) = \rho_0(\mathcal{R}_0(B))$. Finally, since $\mathcal{P}_0$ can be coherently decorated, then $\rho_0(\mathcal{R}_0(B)) = \rho_0(B)$, which implies $\rho(\mathcal{R}(B,k)) = \rho(\mathcal{R}_0(B),t) = \rho_0(\mathcal{R}_0(B)) = \rho_0(B) = \rho(B,k)$.

\end{proof}

%===========================================================
%===========================================================
%===========================================================
%===========================================================

%===========================================================
%===========================================================
%===========================================================
%===========================================================

\subsection{Proof of Theorem \ref{the:main}}
\label{appendix:proofMainTheorem}

Before the proof of the theorem, we explain the general strategy for proving this theorem. Our main objective is to adapt the proof-relevant interpolation method for circular proofs. However, our goal is not only to produce the interpolated circular pre-proof via cut-introduction, but also to prove that this construction is valid. Our strategy is to certify validity through the construction of ordered trace manifolds for the pre-proofs $\mathcal{P}^l$ and $\mathcal{P}^r$. 

To accomplish this, we follow the transformations induced by the descending phase, using induction on the subtrees of $\mathcal{D}_0$, and maintain as an invariant all the properties needed to construct ordered trace manifolds on both sides $\mathcal{P}^l$ and $\mathcal{P}^r$. Specifically, with these invariants we want to capture the following properties of the interpolated pre-proof $\mathcal{P}$:

\begin{enumerate}
     \item Both $\mathcal{P}_0$ and the right and left pre-proofs $\mathcal{P}^l$ and $\mathcal{P}^r$ have the same ``skeleton'', i.e. their structural connectivity is exactly the same.
     
     \item All traces in the original trace manifold for $\mathcal{P}_0$ live on formulas assigned to one side of the split $\Gamma, \Delta$. Therefore, each trace is transferred to exactly one of the two pre-proofs $\mathcal{P}^l$ and $\mathcal{P}^r$.

     \item The interpolant $I$ and $I^{\perp}$ can define new traces in the right and left pre-proofs $\mathcal{P}^l$ and $\mathcal{P}^r$ respectively, that can be included in the ordered trace manifold for $\mathcal{P}^l$ and $\mathcal{P}^r$. 
 \end{enumerate} 

We will start by formally defining these properties: Consider a pre-proof  $\mathcal{P}_0 = (\mathcal{D}_0 = (V_0, s_0, r_0, p_0),\mathcal{R}_0)$ that can be coherently decorated, with an ordered trace manifold with respect to a tree-compatible induction order $(Bud(\mathcal{D}_0), \vartriangleleft_0)$, where $\mathcal{R}_0$ is injective. Assume $Bud(\mathcal{D}_0) = \{B_1, \dots, B_n\}$. Since $\mathcal{R}_0$ is injective, we define the set of companion nodes of $\mathcal{D}_0$ as $Comp(\mathcal{D}_0) = \{C_{B_1}, \dots, C_{B_n} \}$, where for all $1 \leq i \leq n$, $\mathcal{R}_0(B_i) = C_{B_i}$. Moreover, for any vertex $v \in V_0$, we define $T(v)$ as the derivation subtree in $\mathcal{D}_0$ rooted in $v$. We equally define $Bud(T(v)) \subseteq Bud(\mathcal{D}_0)$ as the set of bud nodes of $\mathcal{D}_0$ appearing in $T(v)$, $Comp(T(v)) \subseteq Comp(\mathcal{D}_0)$ as the set of companion nodes of $\mathcal{D}_0$ appearing in $T(v)$. 

Additionally, for any two vertex $u,v$ in a derivation tree $\mathcal{D}'$, we write $u \preceq_{\mathcal{D}'} v$ to say that there is a unique directed path from $u$ to $v$ of size $m \geq 0$ in $\mathcal{D}'$. Now assume there is a trace $\uptau$ over the path $\alpha$ in a derivation tree $\mathcal{D}'$. If $\alpha'$ is a subpath of $\alpha$, we write $\uptau|_{\alpha'}$, for the trace obtained by restricting $\uptau$ to the subpath $\alpha'$.

The following proposition defines the necessary invariants that are used to describe the structure of the interpolated pre-proof during the execution of the descending phase:

\begin{proposition}
    \label{prop:structInterpol}
    Consider the vertex $v \in V_0$ such that $s(v) = \ \vdash \Gamma, \Delta$. Let $\calC= \{(C_1, C_1^\perp), \dots, (C_n, C_n^\perp)\}$ be the cut formulas of $T(v)$, and let $L,R$ be any partitioning of $\{1,\dots,n\}$ and $\mathbb C^L = \VOC{C_i, i\in L}$ and  $\mathbb C^R = \VOC{C_i, i\in R}$. Then, the ascending and descending phase over $T(v)$ produces two cut-free partial pre-proofs $\mathcal{P}^l = (\mathcal{D}^l, \mathcal{R}^l)$ and $\mathcal{P}^r = (\mathcal{D}^r, \mathcal{R}^r)$ with conclusions $\vdash \Gamma, I$ and $\vdash \Delta, I^{\bot}$, such that $\mathcal{P} = (\mathcal{D}, \mathcal{R} = \mathcal{R}^l \cup \mathcal{R}^r)$, with the form $ \mathcal{P} = \adjustbox{valign=c, scale=0.75}{
            \begin{prooftree}
                    \hypo{\mathcal{P}^l}
                    \infer1[]{\vdash \Gamma, I}
                    \hypo{\mathcal{P}^r}
                    \infer1[]{\vdash \Delta, I^{\bot}}
                    \infer2[\textsf{(Cut)}]{\vdash \Gamma, \Delta}
                \end{prooftree} 
            }$ and satisfying:
    \begin{enumerate}
        \item $Bud(T(v)) = \{B_1, \dots, B_n\}$ if and only if $Bud(\mathcal{D}^l) = \{B^l_1, \dots, B^l_n\}$ and $Bud(\mathcal{D}^r) = \{B^r_1, \dots, B^r_n\}$, i.e. the set of bud nodes in $Bud(\mathcal{D}^l)$ (resp. $Bud(\mathcal{D}^r)$) is an exact copy of the set $Bud(T(v)) = \{B_1, \dots, B_n\}$.

        \item $\VOC{I}\subseteq (\VOC {\Gamma} \cup \mathbb C^L) \cap (\VOC {\Delta}\cup \mathbb C^R)$. Moreover, the set of fixed-point variables of $I$ is given by $Var(I) = \{ X_{B_i} \ | \ B_i \in Bud(T(v)) \}$ while the set of bounded fixed-point variables is $BV(I) = \{ X_{B_i} \ | \ B_i \in Bud(T(v)) \land  C_{B_i} \in Comp(T(v)) \}$.

        \item $\mathcal{R}^l$ and $\mathcal{R}^r$ are injective, and for every $B_i \in Bud(T(v))$ such that $C_{B_i} \in Comp(T(v))$, then $\mathcal{R}^l$ (resp. $\mathcal{R}^r$) is defined over $B^l_i$ (resp. $B^r_i$). We define $C_{B^l_i} = \mathcal{R}^l(B^l_i)$ (resp. $C_{B^r_i} = \mathcal{R}^l(B^r_i)$).

        \item For all $C_{B_i},C_{B_j} \in Comp(T(v))$ and $B_k \in Bud(T(v))$, then:
            \begin{enumerate}
                \item $C_{B_i} \preceq_{T(v)} C_{B_j}$ if and only if $C_{B^l_i} \preceq_{\mathcal{D}^l} C_{B^l_j}$ and $C_{B^r_i} \preceq_{\mathcal{D}^r} C_{B^r_j}$

                \item $C_{B_i} \preceq_{T(v)} B_k$ if and only if $C_{B^l_i} \preceq_{\mathcal{D}^l} B_k^l$ and $C_{B^r_i} \preceq_{\mathcal{D}^r} B_k^r$.
            \end{enumerate}

        \item Assume $v'$ is the node labelled with $\vdash \Gamma, \Delta$ at the root of $\mathcal{P}_0$. Now, consider the original ordered trace manifold for $\mathcal{P}_0$: 
        \begin{align*}
            \{\uptau_{B_j,B_i} \ | \ B_j, B_i \in Bud(\mathcal{D}_0), B_i \vartriangleleft_0 B_j \}
        \end{align*}
        with the induction order $(Bud(\mathcal{D}_0), \vartriangleleft_0)$. Let $\mathcal{B} \subseteq Bud(\mathcal{D}_0)$ be the maximal set of elements such that for all $B' \in \mathcal{B}$, then $B' \vartriangleleft_0 B_k$, for some $B_k \in Bud(\mathcal{D}_0)$, with the associated trace $\uptau_{B_k,B'}$, with $\mathcal{B}$ being weakly $\leq_{\mathcal{P}_0}$-connected. 
        \begin{enumerate}

            \item Assume $B_i \in Bud(T(v))$ and $B_i \in \mathcal{B}$:
            \begin{enumerate}
                \item If $C_{B_i} \notin Comp(T(v))$, then there exists either a trace $\uptau_{B^l_k,B^l_i}$ over the unique path from $v'$ to $B^l_i$ in $\mathcal{D}$, or a trace $\uptau_{B^r_k,B^r_i}$ over the unique path from $v'$ to $B^r_i$ in $\mathcal{D}$, such that $\uptau_{B_k,B_i}(v) = \uptau_{B^l_k,B^l_i}(v')$ and $\uptau_{B_k,B_i}(B_i) = \uptau_{B^l_k,B^l_i}(B^l_i)$ or $\uptau_{B_k,B_i}(v) = \uptau_{B^r_k,B^r_i}(v')$ and $\uptau_{B_k,B_i}(B_i) = \uptau_{B^r_k,B^r_i}(B^r_i)$ depending on the case.

                \item If $C_{B_i} \in Comp(T(v))$, then, there exists either a trace $\uptau_{B^l_j,B^l_i}$ over the unique path from $C_{B^l_i}$ to $B^l_i$ in $\mathcal{D}^l$, or a trace $\uptau_{B^r_j,B^r_i}$ over the unique path from $C_{B^r_i}$ to $B^r_i$ in $\mathcal{D}^r$, such that $\uptau_{B^l_k,B^l_i}(C_{B^l_i}) = \uptau_{B^l_k,B^l_i}(B^l_i)$ or $\uptau_{B^r_k,B^r_i}(C_{B^r_i}) = \uptau_{B^r_k,B^r_i}(B^r_i)$ depending on the case.
            \end{enumerate}

            \item Assume $B_i,B_j \in \mathcal{B}$ (i.e. $B_i,B_j \vartriangleleft_0 B_k$) and $B_i, B_j \in Bud(T(v))$, with $B_j \leq_{\mathcal{P}_0} B_i$ and $C_{B_j} \in Comp(T(v))$. Then:
            \begin{enumerate}
                \item If $C_{B_i} \notin Comp(T(v))$ then the companion node $C_{B^l_j}$ (resp. $C_{B^r_j}$) is in the unique path from $v'$ to $B^l_i$ (resp. $B^r_i$) in $\mathcal{D}$, such that $\uptau_{B^l_k,B^l_i}(C_{B^l_j}) = \uptau_{B^l_k,B^l_j}(C_{B^l_j})$ or $\uptau_{B^r_k,B^r_i}(C_{B^r_j}) = \uptau_{B^r_k,B^r_j}(C_{B^r_j})$.

                \item If $C_{B_i} \in Comp(T(v))$ then companion node $C_{B^l_j}$ (resp. $C_{B^r_j}$) is in the cycle $\mathcal{C}_{B^l_i}$ (resp. $\mathcal{C}_{B^r_i}$) in $\mathcal{G}_{\mathcal{P}^l}$ (resp. $\mathcal{G}_{\mathcal{P}^r}$) such that $\uptau_{B^l_k,B^l_i}(C_{B^l_j}) = \uptau_{B^l_k,B^l_j}(C_{B^l_j})$ or $\uptau_{B^r_k,B^r_i}(C_{B^r_j}) = \uptau_{B^r_k,B^r_j}(C_{B^r_j})$.
            \end{enumerate}

            \item Assume $B_i \in Bud(T(v))$ and $B_i \in \mathcal{B}$. 

            \begin{enumerate}
                \item If $C_{B_i} \notin Comp(T(v))$, consider $\alpha$ as the unique path from $v$ to $B_i$ in $\mathcal{D}_0$. Then $\textsf{Form}(\uptau_{B_k,B_i}|_{\alpha}) = \textsf{Form}(\uptau_{B^l_k,B^l_i})$ (resp. $\textsf{Form}(\uptau_{B_k,B_i}|_{\alpha}) = \textsf{Form}(\uptau_{B^r_k,B^r_i})$). Moreover, if $\uptau_{B_k,B_i}|_{\alpha}$ has an $F$-progress point, then $\uptau_{B^l_k,B^l_i}$ (resp. $\uptau_{B^r_k,B^r_i}$) also has an $F$-progress point.

                \item If $C_{B_i} \in Comp(T(v))$ then $\textsf{Form}(\uptau_{B_k,B_i}) = \textsf{Form}(\uptau_{B^l_k,B^l_i})$ (resp. $\textsf{Form}(\uptau_{B_k,B_i}) = \textsf{Form}(\uptau_{B^r_k,B^r_i})$). Moreover, if $\uptau_{B_k,B_i}$ has an $F$-progress point, then $\uptau_{B^l_k,B^l_i}$ (resp. $\uptau_{B^r_k,B^r_i}$) also has an $F$-progress point.
            \end{enumerate}
        \end{enumerate}

        \item For every $B_i \in Bud(T(v))$, there exist the traces
        $\uprho_{B^l_i}$ and $\uprho_{B^r_i}$, called the interpolant traces of
        $B_i$, satisfying that for any choice of the operators $\sigma_j \in \{\mu, \nu\}$, with the dual operator $\sigma^{\perp}_j$ applied on the right side, performed as the rule of each companion node $C_{B^l_j}$ (resp. $C_{B^r_j}$) in $Comp(T(v))$:
        \begin{enumerate}
            \item We have two cases:
                \begin{enumerate}
                \item If $C_{B_i} \notin Comp(T(v))$, then $\uprho_{B^l_i}$
                (resp. $\uprho_{B^r_i}$) is a trace over the unique path from
                $v'$ to $B^l_i$ (resp. $B^r_i$) in $\mathcal{D}$, such that
                $\uprho_{B^l_i}(v') = I$ (resp.
                $\uprho_{B^r_i}(v') = I^{\perp}$) and
                $\uprho_{B^l_i}(B^l_i) = X_{B_i}$ (resp.
                $\uprho_{B^r_i}(B^r_i) = X_{B_i}$).

                \item If $C_{B_i} \in Comp(T(v))$, then $\uprho_{B^l_i}$ (resp.
                $\uprho_{B^r_i}$) is a trace over the unique path from
                $C_{B^l_i}$ (resp. $C_{B^r_i}$) to $B^l_i$ (resp. $B^r_i$) in
                $\mathcal{D}^l$ (resp. $\mathcal{D}^r$), such that
                $\uprho_{B^l_i}(B^l_i) = \uprho_{B^l_i}(C_{B^l_i})$ (resp.
                $\uprho_{B^r_i}(B^r_i) = \uprho_{B^r_i}(C_{B^r_i})$).
                \end{enumerate}

            \item For every vertex $u$ in the path followed by $\uprho_{B^l_i}$
            (resp. $\uprho_{B^r_i}$), the formula $\uprho_{B^l_i}(u)$ (resp.
            $\uprho_{B^r_i}(u)$) is the unique occurrence of a subformula of $I$
            (resp. $I^{\perp}$) in $s(u)$ containing the occurrence of the
            fixed-point variable $X_{B_i}$ or $\sigma_i X_{B_i}. I_i$ (resp. $\sigma^\perp_i X_{B_i}. I^\perp_i$). In particular, if $B_j \in Bud(T(v))$ with $C_{B_j} \in Comp(T(v))$ and
            $B_j \leq_{\mathcal{P}_0} B_i$, then the companion node $C_{B^l_j}$
            (resp. $C_{B^r_j}$) appears in the path followed by
            $\uprho_{B^l_i}$ (resp. $\uprho_{B^r_i}$), by the property (4), and
            $\uprho_{B^l_i}(C_{B^l_j}) = \sigma_j X_{B_j}. I_j$ (resp.
            $\uprho_{B^r_i}(C_{B^r_j}) = \sigma^{\perp}_j X_{B_j}.
            I^{\perp}_j$), where $\sigma_j X_{B_j}. I_j$ (resp.
            $\sigma^{\perp}_j X_{B_j}. I^{\perp}_j$) is the formula labelling
            the companion node $C_{B^l_j}$ (resp. $C_{B^r_j}$) in the current
            interpolated pre-proof.

            \item If $C_{B_i} \in Comp(T(v))$, then
            $\uprho_{B^l_i}(C_{B^l_i}) = \uprho_{B^l_i}(B^l_i) =
            \sigma_i X_{B_i}. I_i$ (resp. $\uprho_{B^r_i}(C_{B^r_i}) =
            \uprho_{B^r_i}(B^r_i) = \sigma^{\perp}_i X_{B_i}. I^{\perp}_i$),
            and the step of $\uprho_{B^l_i}$ (resp. $\uprho_{B^r_i}$) out of
            $C_{B^l_i}$ (resp. $C_{B^r_i}$) is the Fischer--Ladner unfolding of
            $\sigma_i X_{B_i}. I_i$ (resp.
            $\sigma^{\perp}_i X_{B_i}. I^{\perp}_i$).

            \item For all $C_{B_j}, C_{B_m} \in Comp(T(v))$:
            $C_{B_m} \preceq_{T(v)} C_{B_j}$ if and only if
            $\sigma_m X_{B_m}. I_m$ is a subformula of
            $\sigma_j X_{B_j}. I_j$.
        \end{enumerate}

    \end{enumerate}
\end{proposition}

\begin{proof}
The proof of Proposition \ref{prop:structInterpol}, is done by induction over the tree $T(v)$. The proof of property (6) can be done by reasoning on induction, and making simple observations of how the traces formed by the interpolant change during the descending phase, similar to the proof of the property (5).  

\subsubsection{Case Axiom:}
Assume $T(v)$ has the form $\pi = \adjustbox{valign=c, scale=0.75}{
            \begin{prooftree}
                    \Infer{0}[\rax]{\vdash F, F^\perp}
            \end{prooftree} 
        }$, and assume an arbitrary split $(\Gamma, \Delta)$ of the sequent $\vdash F, F^\perp$. In this case $Bud(T(v)) = Comp(T(v)) = \emptyset$. After the ascending and descending phase we obtain the derivation trees $\mathcal{D}^l = \pi_1^l$ and $\mathcal{D}^r = \pi_1^r$, where $\pi_1^l, \pi_1^r$ are defined in the four transformations for the axiom case in the Appendix \ref{app-main-lemma}, depending on the split $(\Gamma, \Delta)$. 

\begin{enumerate}
    \item We have that $Bud(T(v)) = Bud(\mathcal{D}^l) = Bud(\mathcal{D}^r) = \emptyset$. 
    \item The constraint on the vocabulary of $I$ is satisfied, by observation on all the cases in Appendix \ref{app-main-lemma}. $Var(I) = BV(I) = \emptyset$ since $Bud(T(v)) = \emptyset$.
    \item It is trivially true since $\mathcal{R}^l = \mathcal{R}^r = \emptyset$.
    \item It is trivially true since $ Comp(T(v)) = \emptyset$
    \item It is trivially true since there is no trace passing through the node $v$.
\end{enumerate}

The case for the leafs with the rules $\rtop$ and $\rone$ follow the same reasoning, as for the $\rax$ case.

\subsubsection{Case bud node:}
Assume $T(v)$ has the form $\pi = \ \vdash \Gamma, \Delta$. In this case $v$ is a bud node $B_i \in Bud(\mathcal{D}_0)$, therefore $Bud(T(v)) = \{B_i\}$ and $Comp(T(v)) = \emptyset$. After the ascending and descending phase, we obtain \adjustbox{valign=c, scale=0.75}{\begin{prooftree}
\Hypo{\vdash \Gamma, X_{B_i}}
\Hypo{\vdash \Delta, X_{B_i}}
\Infer2[\textsf{(Cut)}]{\vdash \Gamma,\Delta}
\end{prooftree}}, where we take $\mathcal{D}^l = \ \vdash \Gamma, X_{B_i}$ and $\mathcal{D}^r = \ \vdash \Delta, X_{B_i}$ as the derivation trees containing only one node.
\begin{enumerate}
    \item The bud node $B_i$ labelled with $\vdash \Gamma, \Delta$ is divided into the bud nodes $B^l_i, B^r_i$ labelled as $\vdash \Gamma, X_{B_i}$ and $\vdash \Delta, X_{B_i}$ respectively. Therefore $Bud(\mathcal{D}^l) = \{ B^l_i \}$ and $Bud(\mathcal{D}^r) = \{ B^r_i\}$.
    \item The constraint on the vocabulary of $I$ is true since $\voc(I) = \emptyset$. Moreover, $Var(I) = \{X_{B_i}\}$ and $BV(I) = \emptyset$.
    \item It is trivially true since $\mathcal{R}^l = \mathcal{R}^r = \emptyset$.
    \item It is trivially true since $ Comp(T(v)) = \emptyset$.
    \item In this case $C_{B_i} \notin Comp(T(v))$ and we know that $B^l_i,B^r_i$ exist by (1). Take any trace of the form $\uptau_{B_k,B_i}$ for a $B_k \in Bud(\mathcal{D}_0)$, such that $B_i \vartriangleleft_0 B_k$ in the original ordered trace manifold. Therefore, $\uptau_{B_k,B_i}(v) = G$ with either $G \in \Gamma$ or $G \in \Delta$. Then:
    \begin{enumerate}
         \item If $G \in \Gamma$ take $\uptau_{B_k^l,B^l_i} = (G, G)$ over the path from $v'$ to $B^l_i$, where $\uptau_{B_k,B_i}(v) = \uptau_{B_k^l,B^l_i}(v') = G$ and $\uptau_{B_k,B_i}(B_i) = \uptau_{B_k^l,B^l_i}(B^l_i) = G$. Otherwise, if $G \in \Delta$ then take $\uptau_{B^r_k,B^r_i} = (G, G)$ over the path from $v'$ to $B^r_i$, where $\uptau_{B_k,B_i}(v) = \uptau_{B_k^r,B^r_i}(v') = G$ and $\uptau_{B_k,B_i}(B_i) = \uptau_{B_k^r,B^r_i}(B^r_i) = G$.
         %\item We have no need for induction for proving property (5.b). Refer to the ``Case $n$-ary rule'' for the complete proof.
         \item Trivially true since we have no other bud node.
         \item $\textsf{Form}(\uptau_{B_k,B_i}|_{\alpha_1}) = \textsf{Form}(\uptau_{B^l_k,B^l_i}|_{\alpha_2}) = \{G\}$ (resp. $\textsf{Form}(\uptau_{B_k,B_i}|_{\alpha_1}) = \textsf{Form}(\uptau_{B^r_k,B^r_i}|_{\alpha_2}) = \{G\}$). There are no progress points.
     \end{enumerate} 
\end{enumerate}

\subsubsection{Case n-ary rule:}
Assume the derivation tree $T(v)$ has the form:

\begin{center}
\begin{prooftree}
    \hypo{\pi'_1}
    \hypo{\cdots}
    \hypo{\pi'_n}
    \infer3[(r)]{\vdash \Gamma, \Delta}
\end{prooftree}
\end{center}

After the ascending and descending phase on every subtree $\pi'_i$, we obtain the elementary \CPRIS: 

\begin{center}
\begin{prooftree}
    \hypo{\pi_1^L}
    \infer1{\vdash \Gamma_1, I_1}
    \hypo{\pi_1^R}
    \infer1{\vdash I^{\bot}_1, \Delta_1}
    \infer2[(\textsf{Cut})]{\vdash \Gamma_1, \Delta_1}

    \hypo{\cdots}

    \hypo{\pi_n^L}
    \infer1{\vdash \Gamma_n, I_n}
    \hypo{\pi_n^R}
    \infer1{\vdash I^{\bot}_n, \Delta_n}
    \infer2[(\textsf{Cut})]{\vdash \Gamma_n, \Delta_n}

    \infer3[(r)]{\vdash \Gamma, \Delta}
\end{prooftree}
\end{center}

Where $\pi_1, \dots, \pi_n$ are the subtrees, rooted at the nodes labelled with $(\vdash \Gamma_1, \Delta_1), \dots, (\vdash \Gamma_n, \Delta_n)$ respectively. The cut-introduction step transforms the previous derivation tree into a solved \CPRIS, of the form: 
\begin{center}
\begin{prooftree}
    \hypo{\pi^L}
    \infer1{\vdash \Gamma, I}

    \hypo{\pi^R}
    \infer1{\vdash I^{\bot}, \Delta}

    \infer2[(\textsf{Cut})]{\vdash \Gamma, \Delta}
\end{prooftree}
\end{center}

For each rule $(r)$, the transformation and the derivation trees $\pi^L$ and $\pi^R$ are defined in the Appendix \ref{app-main-lemma} and Figure \ref{fig:newPRIS} (b). Then, we have two cases: either the root $v$ is a companion node or not. If it is not a companion node, then the descending phase concludes with the cut-introduction step specified above. In this case, we define $\mathcal{D}^l = \pi^L$ and $\mathcal{D}^r= \pi^R$.

Otherwise, if the root $v$ is a companion, say $v = C_{B_k}$, then the previous derivation tree is therefore transformed into: 

\begin{center}
    \begin{prooftree}
        \hypo{\pi^L[\sigma X_{B_k}. I/X_{B_k}]}
        \infer1[]{\vdash \Gamma, I[\sigma X_{B_k}. I/X_{B_k}]}
        \infer1[$(\sigma)$]{\vdash \Gamma, \sigma X_{B_k}. I}

        \hypo{\pi^R[\sigma^{\bot} X_{B_k}.I^{\bot}/X_{B_k}]}
        \infer1[]{\vdash \Delta, I^{\bot}[\sigma^{\bot} X_{B_k}.I^{\bot}/X_{B_k}]}
        \infer1[$(\sigma^{\bot})$]{\vdash \Delta, \sigma^{\bot} X_{B_k}.I^{\bot}}
        
        \infer2[\textsf{(Cut)}]{\vdash \Gamma, \Delta}
    \end{prooftree}
\end{center}

In this case, in $\mathcal{R}^l$ (resp. $\mathcal{R}^r$) we pair the bud node $B^l_k$ (resp. $B^r_k$) with the node corresponding to the left (resp. right) premise of the cut. We define $\mathcal{D}^l = \scalebox{.75}{
        \begin{prooftree}
            \hypo{\pi^L[\sigma X_{B_k}. I/X_{B_k}]}
            \infer1[]{\vdash \Gamma, I[\sigma X_{B_k}. I/X_{B_k}]}
            \infer1[$(\sigma)$]{\vdash \Gamma, \sigma X_{B_k}. I}    
        \end{prooftree}
        }$ and $\mathcal{D}^r= \scalebox{.75}{
        \begin{prooftree}
            \hypo{\pi^R[\sigma^{\bot} X_{B_k}.I^{\bot}/X_{B_k}]}
            \infer1[]{\vdash \Delta, I^{\bot}[\sigma^{\bot} X_{B_k}.I^{\bot}/X_{B_k}]}
            \infer1[$(\sigma^{\bot})$]{\vdash \Delta, \sigma^{\bot} X_{B_k}.I^{\bot}}    
        \end{prooftree}
        }$.

\begin{enumerate}
    \item  In both cases, whether $v$ is a companion or not, one observes that: 
    \begin{itemize}
         \item For the unary rules $\{ \parr, \perp, \plus_i, \mathsf{?d}, \mathsf{?w}, \mathsf{?c}, \mathsf{!p}, \nu, \mu \}$, we have that $Bud(T(v)) = Bud(\pi'_1) = \{B_1, \dots, B_m\}$ if and only if $Bud(\pi^L_1) = \{B^l_1, \dots, B^l_m\}$ and $Bud(\pi^R_1) = \{B^r_1, \dots, B^r_m\}$ by the induction hypothesis. Then, by the structure of the derivation trees $\pi^L$ and $\pi^R$, one derives that $Bud(\mathcal{D}^l) = Bud(\pi^L) = Bud(\pi^L_1)$ and $Bud(\mathcal{D}^r) = Bud(\pi^R) = Bud(\pi^R_1)$.

         \item For the binary rules $\{ \tensor, \with, \mathsf{cut} \}$, we have that $Bud(T(v)) = Bud(\pi'_1) \cup Bud(\pi'_2) = \{B_1, \dots, B_m\} \cup \{B'_1, \dots, B'_k\}$ if and only if $Bud(\pi^L_1) = \{B^l_1, \dots, B^l_m\}$, $Bud(\pi^R_1) = \{B^r_1, \dots, B^r_m\}$, $Bud(\pi^L_2) = \{B'^l_1, \dots, B'^l_k\}$, and $Bud(\pi^R_2) = \{B'^r_1, \dots, B'^r_k\}$ by the induction hypothesis. Then, by the structure of the derivation trees $\pi^L$ and $\pi^R$, one derives that $Bud(\mathcal{D}^l) = Bud(\pi^L) = Bud(\pi^L_1) \cup Bud(\pi^L_2)$ and $Bud(\mathcal{D}^r) = Bud(\pi^R) = Bud(\pi^R_1) \cup Bud(\pi^R_2)$.
     \end{itemize} 
    
    \item By the induction hypothesis we have that $\VOC{I_i}\subseteq (\VOC {\Gamma_i} \cup \mathbb C_i^L) \cap (\VOC {\Delta_i}\cup \mathbb C_i^R)$, $Var(I_i) = \{ X_{B_j} \ | \ B_j \in Bud(\pi'_i) \}$, and $BV(I_i) = \{ X_{B_j} \ | \ B_j \in Bud(\pi'_i) \land  C_{B_j} \in Comp(\pi'_i) \}$, for all $i$. We have two cases:
    \begin{itemize}
        \item If $v$ is not a companion node, then by the induction hypothesis on every $I_i$ and observing the transformations for every rule $(r)$, used to construct the interpolant $I$ in the Appendix \ref{app-main-lemma} and Figure \ref{fig:newPRIS} (b), one can observe that $\VOC{I}\subseteq (\VOC {\Gamma} \cup \mathbb C^L) \cap (\VOC {\Delta}\cup \mathbb C^R)$. Moreover, none of these transformations include or bound fixed-point variables in $I$, therefore $Var(I) = \bigcup_{1 \leq i \leq n} Var(I_i) = \bigcup_{1 \leq i \leq n} \{ X_{B_j} \ | \ B_j \in Bud(\pi'_i) \} = \{ X_{B_j} \ | \ B_j \in Bud(T(v))  \}$, and $BV(I) = \bigcup_{1 \leq i \leq n} BV(I_i) = \bigcup_{1 \leq i \leq n} \{ X_{B_j} \ | \ B_j \in Bud(\pi'_i) \land  C_{B_j} \in Comp(\pi'_i) \} = \{ X_{B_j} \ | \ B_j \in Bud(T(v)) \land  C_{B_j} \in Comp(T(v)) \}$.

        \item If $v$ is a companion node, say $C_{B_k}$, then the interpolant is $\sigma X_{B_k}. I$, where $I$ satisfies that $\VOC{I}\subseteq (\VOC {\Gamma} \cup \mathbb C^L) \cap (\VOC {\Delta}\cup \mathbb C^R)$, $Var(I) = \{ X_{B_j} \ | \ B_j \in Bud(T(v)) \}$ and $BV(I) = \{ X_{B_j} \ | \ B_j \in Bud(T(v)) \land  C_{B_j} \in Comp(T(v)) \}$, since we already showed it in the case where $v$ is not a companion. It is then clear that $\voc(\sigma X_{B_k}. I) = \voc(I)$ and $Var(\sigma X_{B_k}. I) = Var(I)$, therefore satisfying both constraints. Finally, since the variable $X_{B_k}$ is bounded in $\sigma X_{B_k}. I$, we get that $BV(\sigma X_{B_k}.I) = BV(I) \cup \{ X_{B_k} \} $, therefore satisfying the property, since $v = C_{B_k} \in Comp(T(v))$.
    \end{itemize}
    
    \item Let $\mathcal{R}_i^l$ and $\mathcal{R}_i^r$ be the partial bud-companion relation for the derivation trees $\pi^L_i$ and $\pi^R_i$ respectively, for all $i$. By the induction hypothesis, we have that $\mathcal{R}_i^l$ and $\mathcal{R}_i^r$ are injective, and for every $B_j \in Bud(\pi_i)$ such that $C_{B_j} \in Comp(\pi_i)$, then $\mathcal{R}_i^l$ (resp. $\mathcal{R}_i^r$) is defined over $B^l_j$ (resp. $B^r_j$). We know $B^l_j$ and $B^r_j$ exist, by induction hypothesis on (1). We have two cases:
    \begin{itemize}
        \item Assume $v$ is not a companion node. As explained for (1), for the unary rules $\{ \parr, \perp, \plus_i, \mathsf{?d}, \mathsf{?w}, \mathsf{?c}, \mathsf{!p}, \nu, \mu \}$, $Bud(\mathcal{D}^l) = Bud(\pi^L_1)$ and $Bud(\mathcal{D}^r) = Bud(\pi^R_1)$. Since $v$ is not a companion, one can also deduce that $Comp(\mathcal{D}^l) = Comp(\pi^L_1)$ and $Comp(\mathcal{D}^r) = Comp(\pi^R_1)$. In this case the relations $\mathcal{R}_1^l$ and $\mathcal{R}_1^r$ are not altered, therefore $\mathcal{R}^l = \mathcal{R}_1^l$ and $\mathcal{R}^r = \mathcal{R}_1^r$, where both relations satisfy (3) by the induction  hypothesis. For the binary rules $\{ \tensor, \with, \mathsf{cut} \}$, the argument is the same: one derives that $Bud(\mathcal{D}^l) = Bud(\pi^L_1) \cup Bud(\pi^L_2)$, $Bud(\mathcal{D}^r) = Bud(\pi^R_1) \cup Bud(\pi^R_2)$, $Comp(\mathcal{D}^l) = Comp(\pi^L_1) \cup Comp(\pi^L_2)$, and $Comp(\mathcal{D}^r) = Comp(\pi^R_1) \cup Comp(\pi^R_2)$. Thus,  both $\mathcal{R}^l = \mathcal{R}_1^l \cup \mathcal{R}_2^l$ and $\mathcal{R}^r = \mathcal{R}_1^r \cup \mathcal{R}_2^r$ satisfy property (3).

        \item In both cases, where $(r)$ is either a unary or binary rule, we have defined $\mathcal{R}^l$ and $\mathcal{R}^r$, both satisfying (3), by the previous proof given where $v$ is not a companion. Now assume $v$ is a companion node, say $C_{B_k}$. We will name $C^l_{B_k}$ (resp. $C^r_{B_k}$) the root of the derivation tree $\mathcal{D}^l$ (resp. $\mathcal{D}^r$), labelled with the sequent $\vdash  \Gamma, \sigma X_{B_k}. I$ (resp. $\vdash \Delta, \sigma^{\bot} X_{B_k}.I^{\bot}$). Then, we update the relations $\mathcal{R}^l$ and $\mathcal{R}^r$ with $\mathcal{R}^l = \mathcal{R}^l \cup \{ (B^l_k, C^l_{B_k}) \}$ and $\mathcal{R}^r = \mathcal{R}^r \cup \{ (B^r_k, C^r_{B_k}) \}$. This update maintains injectivity since $B_k$ has no other companion node assigned, and ensures that $B^l_k \in \operatorname{dom}(\mathcal{R}^l)$ (resp. $B^r_k \in \operatorname{dom}(\mathcal{R}^r)$), therefore satisfying (3).   
    \end{itemize}
    
    \item By the induction hypothesis, we have that for all $C_{B_j}, C_{B_l} \in Comp(\pi'_i)$ and $B_h \in Bud(\pi'_i)$, then:
        \begin{enumerate}
             \item $C_{B_j} \preceq_{\pi'_i} C_{B_l}$ if and only if $C_{B^l_j} \preceq_{\pi_i^L} C_{B^l_l}$ and $C_{B^r_j} \preceq_{\pi_i^R} C_{B^r_l}$

            \item $C_{B_j} \preceq_{\pi'_i} B_h$ if and only if $C_{B^l_j} \preceq_{\pi_i^L} B_h^l$ and $C_{B^r_h} \preceq_{\pi_i^R} B_h^r$.
         \end{enumerate} 
        for every $i$. We can ensure that $B^l_h, C_{B^l_j}, C_{B^l_l}$ (resp. $B^r_h, C_{B^r_j}, C_{B^r_l}$) exist by the induction hypothesis on (1) and (3). Now let $C_{B_j}, C_{B_l} \in Comp(T(v))$ and $B_h \in Bud(T(v))$. We have two cases:
        \begin{itemize}
            \item  Assume $v$ is not a companion node. Therefore, every path starting from a companion in $T(v)$ must be strictly inside some derivation tree $\pi'_i$, for some $i$. In other words, $C_{B_j} \preceq_{T(v)} C_{B_l}$ if and only if $C_{B_j} \preceq_{\pi'_i} C_{B_l}$, and $C_{B_j} \preceq_{T(v)} B_h$ if and only if $C_{B_j} \preceq_{\pi'_i} B_h$, for some $i$. Then, by the induction hypothesis:
        \begin{enumerate}
            \item $C_{B_j} \preceq_{T(v)} C_{B_l}$ if and only if $C_{B^l_j} \preceq_{\pi_i^L} C_{B^l_l}$ and $C_{B^r_j} \preceq_{\pi_i^R} C_{B^r_l}$
            \item $C_{B_j} \preceq_{T(v)} B_h$ if and only if $C_{B^l_j} \preceq_{\pi_i^L} B_h^l$ and $C_{B^r_h} \preceq_{\pi_i^R} B_h^r$.
        \end{enumerate}
            Either if $(r)$ is unary or binary, in both cases we have that the transformations in the Appendix \ref{app-main-lemma} do not introduce new companion nodes, and preserve intact the derivations trees $\pi^L_i$ and $\pi^R_i$. Therefore:
        \begin{enumerate}
            \item $C_{B^l_j} \preceq_{\mathcal{D}^l} C_{B^l_l}$ (resp. $C_{B^r_j} \preceq_{\mathcal{D}^r} C_{B^r_l}$) if and only if $C_{B^l_j} \preceq_{\pi_i^L} C_{B^l_l}$ (resp. $C_{B^r_j} \preceq_{\pi_i^R} C_{B^r_l}$).
            \item $C_{B^l_j} \preceq_{\mathcal{D}^l} B^l_h$ (resp. $C_{B^r_j} \preceq_{\mathcal{D}^r} B^r_h$) if and only if $C_{B^l_j} \preceq_{\pi_i^L} B^l_h$ (resp. $C_{B^r_j} \preceq_{\pi_i^R} B^r_h$).
        \end{enumerate}
        Which implies:
        \begin{enumerate}
            \item $C_{B_j} \preceq_{T(v)} C_{B_l}$ if and only if $C_{B^l_j} \preceq_{\mathcal{D}^l} C_{B^l_l}$ and $C_{B^r_j} \preceq_{\mathcal{D}^r} C_{B^r_l}$

            \item $C_{B_j} \preceq_{T(v)} B_h$ if and only if $C_{B^l_j} \preceq_{\mathcal{D}^l} B_h^l$ and $C_{B^r_h} \preceq_{\mathcal{D}^r} B_h^r$.
        \end{enumerate} 
    \item Assume $v$ is a companion node, say $C_{B_k}$. In the previous case, we have shown that (4) is satisfied by all companion nodes inside some derivation tree $\pi'_i$. It remains to be shown that it is also satisfied by $C_{B_k}$, i.e. 
    \begin{enumerate}
            \item $C_{B_k} \preceq_{T(v)} C_{B_l}$ if and only if $C_{B^l_k} \preceq_{\mathcal{D}^l} C_{B^l_l}$ and $C_{B^r_k} \preceq_{\mathcal{D}^r} C_{B^r_l}$

            \item $C_{B_k} \preceq_{T(v)} B_h$ if and only if $C_{B^l_k} \preceq_{\mathcal{D}^l} B_h^l$ and $C_{B^r_k} \preceq_{\mathcal{D}^r} B_h^r$.
        \end{enumerate} 
    This is true since $C_{B_k}$ is the root of $T(v)$, $C_{B^l_k}$ (resp. $C_{B^r_k}$) is the root of $\mathcal{D}^l$ (resp. $\mathcal{D}^r$), and there is always a path from the root to any vertex in a tree.
    \end{itemize}

    \item Let $\mathcal{B} \subseteq Bud(\mathcal{D}_0)$ be the set of all bud nodes under $B_h \in \mathcal{D}_0$ in the $\vartriangleleft_0$ relation, i.e. for all $B_i \in \mathcal{B}$ then $B_i \vartriangleleft_0 B_h$. We will implicitly use the properties (1) and (3) to refer to the copies of the bud and companion nodes inside every subtree $\pi_i$, with their respective injective relations $\mathcal{R}^l_i$ and $\mathcal{R}^r_i$. Moreover, we name $\alpha$ the unique path from $v$ to $B_i$ in $\mathcal{D}_0$, and if $B_i$ is in $\pi'_k$ with $k \in \{1,2\}$, then we name $\alpha'_k$ the unique path from $v_k$ to $B_i$ in $\pi'_k$, where $v_k$ is the root of $\pi'_k$. We then define $\uptau_{B_h,B_i}|_{\alpha'_k} = \uptau^k_{B_h,B_i}$.
    \begin{enumerate}
         \item Assume $B_i \in Bud(T(v))$ and $B_i \in \mathcal{B}$. We refer to the traces $\uptau^1_{B^l_h, B^l_i}$, $\uptau^1_{B^r_h, B^r_i}$, $\uptau^2_{B^l_h, B^l_i}$, $\uptau^2_{B^r_h, B^r_i}$ as the traces obtained by induction hypothesis over the subtrees $\pi^L_1, \pi^R_1, \pi^L_2, \pi^R_2$ respectively, where $v'_1, v'_2$ are the roots of the subtrees $\pi_1$ and $\pi_2$ respectively. We name $\alpha$ the unique path from $v$ to $B_i$ in $\mathcal{D}_0$.
         \begin{enumerate}
             \item Assume $C_{B_i} \notin Comp(T(v))$. We proceed by cases on the rule $(r)$:
             \begin{itemize}
                 \item \textbf{Case $\rcut$:}
                    Without loss of generality we assume that the cut formulas are in the right split.
                    Note that no trace through $v$ follows the cut formulas $C, C^\perp$: 
                    they do not occur in the conclusion of the rule
                    $\rcut$, and the ancestor relation of the cut relates only the context
                    formulas. Traces following the cut formulas are internal to a premise
                    subtree and are given by the induction hypothesis.
                    \begin{itemize}
                        \item Let $\uptau_{B_h,B_i}|_{\alpha} = (G,G) \circ \uptau^k_{B_h,B_i}$ and
                        $\uptau_{B_h,B_i}(v) = G$ with $G \in \Gamma_k$, for $k \in \{1,2\}$.
                        Then $\uptau^k_{B^l_h,B^l_i}$ exists and
                        $\uptau^k_{B^l_h,B^l_i}(v'_k) = G$ with $G \in \Gamma_k$.
                        We take $\uptau_{B^l_h,B^l_i} = (G, G, G) \circ \uptau^k_{B^l_h, B^l_i}$.

                        \item Let $\uptau_{B_h,B_i}|_{\alpha} = (G,G) \circ \uptau^k_{B_h,B_i}$ and
                        $\uptau_{B_h,B_i}(v) = G$ with $G \in \Delta_k$, for $k \in \{1,2\}$.
                        Then $\uptau^k_{B^r_h,B^r_i}$ exists and
                        $\uptau^k_{B^r_h,B^r_i}(v'_k) = G$ with $G \in \Delta_k$.
                        We take $\uptau_{B^r_h,B^r_i} = (G, G, G, G) \circ \uptau^k_{B^r_h, B^r_i}$.
                    \end{itemize}
            
                  \item \textbf{Case \rparr:} 
                  \begin{itemize}
                      \item Let $\uptau_{B_h,B_i}|_{\alpha} = (G,G) \circ \uptau^1_{B_h,B_i}$ and $\uptau_{B_h,B_i}(v) = G$ with $G \in \Gamma$. Then $\uptau^1_{B^l_h,B^l_i}$ exists and $\uptau^1_{B^l_h,B^l_i}(v'_1) = G$ with $G \in \Gamma$. We take $\uptau_{B^l_h,B^l_i} = (G, G) \circ \uptau^1_{B^l_h, B^l_i}$.
                      \item Let $\uptau_{B_h,B_i}|_{\alpha} = (G,G) \circ \uptau^1_{B_h,B_i}$ and $\uptau_{B_h,B_i}(v) = G$ with $G \in \Delta$. Then $\uptau^1_{B^r_h,B^r_i}$ exists and $\uptau^1_{B^r_h,B^r_i}(v'_1) = G$ with $G \in \Delta$. We take $\uptau_{B^r_h,B^r_i} = (G, G, G) \circ \uptau^1_{B^r_h, B^r_i}$.
                      \item Let $\uptau_{B_h,B_i}|_{\alpha} = (A \parr B, A) \circ \uptau^1_{B_h,B_i}$ and $\uptau_{B_h,B_i}(v) = A \parr B$ with ancestor $A$ after the rule $\rparr$. Then $\uptau^1_{B^r_h,B^r_i}$ exists and $\uptau^1_{B^r_h,B^r_i}(v'_1) = A$. We take $\uptau_{B^r_h,B^r_i} = (A \parr B, A \parr B, A) \circ \uptau^1_{B^r_h, B^r_i}$.
                      \item Let $\uptau_{B_h,B_i}|_{\alpha} = (A \parr B, B) \circ \uptau^1_{B_h,B_i}$ and $\uptau_{B_h,B_i}(v) = A \parr B$ with ancestor $B$ after the rule $\rparr$. Then $\uptau^1_{B^r_h,B^r_i}$ exists and $\uptau^1_{B^r_h,B^r_i}(v'_1) = B$. We take $\uptau_{B^r_h,B^r_i} = (A \parr B, A \parr B, B) \circ \uptau^1_{B^r_h, B^r_i}$.
                  \end{itemize}

                  \item \textbf{Case \rtensor:}
                    \begin{itemize}
                        \item Let $\uptau_{B_h,B_i}|_{\alpha} = (G, G) \circ \uptau^k_{B_h,B_i}$ and
                        $\uptau_{B_h,B_i}(v) = G$ with $G \in \Gamma_k$, for $k \in \{1,2\}$.
                        Then $\uptau^k_{B^l_h,B^l_i}$ exists and
                        $\uptau^k_{B^l_h,B^l_i}(v'_k) = G$ with $G \in \Gamma_k$.
                        We take $\uptau_{B^l_h,B^l_i} = (G, G, G) \circ \uptau^k_{B^l_h, B^l_i}$.

                        \item Let $\uptau_{B_h,B_i}|_{\alpha} = (G, G) \circ \uptau^k_{B_h,B_i}$ and
                        $\uptau_{B_h,B_i}(v) = G$ with $G \in \Delta_k$, for $k \in \{1,2\}$.
                        Then $\uptau^k_{B^r_h,B^r_i}$ exists and
                        $\uptau^k_{B^r_h,B^r_i}(v'_k) = G$ with $G \in \Delta_k$.
                        We take $\uptau_{B^r_h,B^r_i} = (G, G, G, G) \circ \uptau^k_{B^r_h, B^r_i}$.

                        \item Let $\uptau_{B_h,B_i}|_{\alpha} = (A \tensor B, A) \circ \uptau^1_{B_h,B_i}$ and
                        $\uptau_{B_h,B_i}(v) = A \tensor B$ with ancestor $A$ after the rule
                        $\rtensor$. Then $\uptau^1_{B^r_h,B^r_i}$ exists and
                        $\uptau^1_{B^r_h,B^r_i}(v'_1) = A$. We take
                        $\uptau_{B^r_h,B^r_i} = (A \tensor B, A \tensor B, A \tensor B, A) \circ \uptau^1_{B^r_h, B^r_i}$.

                        \item Let $\uptau_{B_h,B_i}|_{\alpha} = (A \tensor B, B) \circ \uptau^2_{B_h,B_i}$ and
                        $\uptau_{B_h,B_i}(v) = A \tensor B$ with ancestor $B$ after the rule
                        $\rtensor$. Then $\uptau^2_{B^r_h,B^r_i}$ exists and
                        $\uptau^2_{B^r_h,B^r_i}(v'_2) = B$. We take
                        $\uptau_{B^r_h,B^r_i} = (A \tensor B, A \tensor B, A \tensor B, B) \circ \uptau^2_{B^r_h, B^r_i}$.
                    \end{itemize}

                \item \textbf{Case \rbot:}
                \begin{itemize}
                    \item Let $\uptau_{B_h,B_i}|_{\alpha} = (G, G) \circ \uptau^1_{B_h,B_i}$ and
                    $\uptau_{B_h,B_i}(v) = G$ with $G \in \Gamma$.
                    Then $\uptau^1_{B^l_h,B^l_i}$ exists and
                    $\uptau^1_{B^l_h,B^l_i}(v'_1) = G$ with $G \in \Gamma$.
                    We take $\uptau_{B^l_h,B^l_i} = (G, G, G) \circ \uptau^1_{B^l_h, B^l_i}$.

                    \item Let $\uptau_{B_h,B_i}|_{\alpha} = (G, G) \circ \uptau^1_{B_h,B_i}$ and
                    $\uptau_{B_h,B_i}(v) = G$ with $G \in \Delta$.
                    Then $\uptau^1_{B^r_h,B^r_i}$ exists and
                    $\uptau^1_{B^r_h,B^r_i}(v'_1) = G$ with $G \in \Delta$.
                    We take $\uptau_{B^r_h,B^r_i} = (G, G) \circ \uptau^1_{B^r_h, B^r_i}$.
                \end{itemize}

                \item \textbf{Case $\roplus_j$:}
                \begin{itemize}
                    \item Let $\uptau_{B_h,B_i}|_{\alpha} = (G, G) \circ \uptau^1_{B_h,B_i}$ and
                    $\uptau_{B_h,B_i}(v) = G$ with $G \in \Gamma$.
                    Then $\uptau^1_{B^l_h,B^l_i}$ exists and
                    $\uptau^1_{B^l_h,B^l_i}(v'_1) = G$ with $G \in \Gamma$.
                    We take $\uptau_{B^l_h,B^l_i} = (G, G, G) \circ \uptau^1_{B^l_h, B^l_i}$.

                    \item Let $\uptau_{B_h,B_i}|_{\alpha} = (G, G) \circ \uptau^1_{B_h,B_i}$ and
                    $\uptau_{B_h,B_i}(v) = G$ with $G \in \Delta$.
                    Then $\uptau^1_{B^r_h,B^r_i}$ exists and
                    $\uptau^1_{B^r_h,B^r_i}(v'_1) = G$ with $G \in \Delta$.
                    We take $\uptau_{B^r_h,B^r_i} = (G, G) \circ \uptau^1_{B^r_h, B^r_i}$.

                    \item Let $\uptau_{B_h,B_i}|_{\alpha} = (F_1 \oplus F_2, F_j) \circ \uptau^1_{B_h,B_i}$ and
                    $\uptau_{B_h,B_i}(v) = F_1 \oplus F_2$ with ancestor $F_j$ after the rule
                    $\roplus_j$. Then $\uptau^1_{B^l_h,B^l_i}$ exists and
                    $\uptau^1_{B^l_h,B^l_i}(v'_1) = F_j$.
                    We take $\uptau_{B^l_h,B^l_i} = (F_1 \oplus F_2, F_1 \oplus F_2, F_j) \circ \uptau^1_{B^l_h, B^l_i}$.
                \end{itemize}

                \item \textbf{Case \rwith:}
                \begin{itemize}
                    \item Let $\uptau_{B_h,B_i}|_{\alpha} = (G, G) \circ \uptau^k_{B_h,B_i}$ and
                    $\uptau_{B_h,B_i}(v) = G$ with $G \in \Gamma$.
                    Then $\uptau^k_{B^l_h,B^l_i}$ exists and
                    $\uptau^k_{B^l_h,B^l_i}(v'_k) = G$ with $G \in \Gamma$.
                    We take $\uptau_{B^l_h,B^l_i} = (G, G, G, G) \circ \uptau^k_{B^l_h, B^l_i}$.

                    \item Let $\uptau_{B_h,B_i}|_{\alpha} = (G, G) \circ \uptau^k_{B_h,B_i}$ and
                    $\uptau_{B_h,B_i}(v) = G$ with $G \in \Delta$.
                    Then $\uptau^k_{B^r_h,B^r_i}$ exists and
                    $\uptau^k_{B^r_h,B^r_i}(v'_k) = G$ with $G \in \Delta$.
                    We take $\uptau_{B^r_h,B^r_i} = (G, G, G) \circ \uptau^k_{B^r_h, B^r_i}$.

                    \item Let $\uptau_{B_h,B_i}|_{\alpha} = (F_1 \with F_2, F_k) \circ \uptau^k_{B_h,B_i}$ and
                    $\uptau_{B_h,B_i}(v) = F_1 \with F_2$ with ancestor $F_k$ after the rule
                    $\rwith$. Then $\uptau^k_{B^l_h,B^l_i}$ exists and
                    $\uptau^k_{B^l_h,B^l_i}(v'_k) = F_k$.
                    We take $\uptau_{B^l_h,B^l_i} = (F_1 \with F_2, F_1 \with F_2, F_k, F_k) \circ \uptau^k_{B^l_h, B^l_i}$.
                \end{itemize}

                \item \textbf{Case $\rder$:}
                \begin{itemize}
                    \item Let $\uptau_{B_h,B_i}|_{\alpha} = (G, G) \circ \uptau^1_{B_h,B_i}$ and
                    $\uptau_{B_h,B_i}(v) = G$ with $G \in \Gamma$.
                    Then $\uptau^1_{B^l_h,B^l_i}$ exists and
                    $\uptau^1_{B^l_h,B^l_i}(v'_1) = G$ with $G \in \Gamma$.
                    We take $\uptau_{B^l_h,B^l_i} = (G, G, G) \circ \uptau^1_{B^l_h, B^l_i}$.

                    \item Let $\uptau_{B_h,B_i}|_{\alpha} = (G, G) \circ \uptau^1_{B_h,B_i}$ and
                    $\uptau_{B_h,B_i}(v) = G$ with $G \in \Delta$.
                    Then $\uptau^1_{B^r_h,B^r_i}$ exists and
                    $\uptau^1_{B^r_h,B^r_i}(v'_1) = G$ with $G \in \Delta$.
                    We take $\uptau_{B^r_h,B^r_i} = (G, G) \circ \uptau^1_{B^r_h, B^r_i}$.

                    \item Let $\uptau_{B_h,B_i}|_{\alpha} = (? F, F) \circ \uptau^1_{B_h,B_i}$ and
                    $\uptau_{B_h,B_i}(v) = ? F$ with ancestor $F$ after the rule
                    $\rder$. Then $\uptau^1_{B^l_h,B^l_i}$ exists and
                    $\uptau^1_{B^l_h,B^l_i}(v'_1) = F$.
                    We take $\uptau_{B^l_h,B^l_i} = (? F, ? F, F) \circ \uptau^1_{B^l_h, B^l_i}$.
                \end{itemize}

                \item \textbf{Case $\rprom$:}
                \begin{itemize}
                    \item Let $\uptau_{B_h,B_i}|_{\alpha} = (G, G) \circ \uptau^1_{B_h,B_i}$ and
                    $\uptau_{B_h,B_i}(v) = G$ with $G \in  ?\Gamma$.
                    Then $\uptau^1_{B^l_h,B^l_i}$ exists and
                    $\uptau^1_{B^l_h,B^l_i}(v'_1) = G$ with $G \in ?\Gamma$.
                    We take $\uptau_{B^l_h,B^l_i} = (G, G, G, G) \circ \uptau^1_{B^l_h, B^l_i}$.

                    \item Let $\uptau_{B_h,B_i}|_{\alpha} = (G, G) \circ \uptau^1_{B_h,B_i}$ and
                    $\uptau_{B_h,B_i}(v) = G$ with $G \in ?\Delta$.
                    Then $\uptau^1_{B^r_h,B^r_i}$ exists and
                    $\uptau^1_{B^r_h,B^r_i}(v'_1) = G$ with $G \in ?\Delta$.
                    We take $\uptau_{B^r_h,B^r_i} = (G, G, G) \circ \uptau^1_{B^r_h, B^r_i}$.

                    \item Let $\uptau_{B_h,B_i}|_{\alpha} = (! F, F) \circ \uptau^1_{B_h,B_i}$ and
                    $\uptau_{B_h,B_i}(v) = ! F$ with ancestor $F$ after the rule
                    $\rprom$. Then $\uptau^1_{B^l_h,B^l_i}$ exists and
                    $\uptau^1_{B^l_h,B^l_i}(v'_1) = F$.
                    We take $\uptau_{B^l_h,B^l_i} = (! F, ! F, F, F) \circ \uptau^1_{B^l_h, B^l_i}$.
                \end{itemize}

                \item \textbf{Case $\rweak$:}
                \begin{itemize}
                    \item Let $\uptau_{B_h,B_i}|_{\alpha} = (G, G) \circ \uptau^1_{B_h,B_i}$ and
                    $\uptau_{B_h,B_i}(v) = G$ with $G \in \Gamma$.
                    Then $\uptau^1_{B^l_h,B^l_i}$ exists and
                    $\uptau^1_{B^l_h,B^l_i}(v'_1) = G$ with $G \in \Gamma$.
                    We take $\uptau_{B^l_h,B^l_i} = (G, G, G) \circ \uptau^1_{B^l_h, B^l_i}$.

                    \item Let $\uptau_{B_h,B_i}|_{\alpha} = (G, G) \circ \uptau^1_{B_h,B_i}$ and
                    $\uptau_{B_h,B_i}(v) = G$ with $G \in \Delta$.
                    Then $\uptau^1_{B^r_h,B^r_i}$ exists and
                    $\uptau^1_{B^r_h,B^r_i}(v'_1) = G$ with $G \in \Delta$.
                    We take $\uptau_{B^r_h,B^r_i} = (G, G) \circ \uptau^1_{B^r_h, B^r_i}$.
                \end{itemize}

                \item \textbf{Case $\rcontr$:} The principal formula $? F$ has two ancestor
                occurrences $? F, ? F$ in the premise of the rule $\rcontr$. Without loss of
                generality, we assume the trace follows the occurrence of $? F$ in the first
                position of the premise sequent, i.e. the leftmost one; the case of the
                rightmost occurrence is identical.
                \begin{itemize}
                    \item Let $\uptau_{B_h,B_i}|_{\alpha} = (G, G) \circ \uptau^1_{B_h,B_i}$ and
                    $\uptau_{B_h,B_i}(v) = G$ with $G \in \Gamma$.
                    Then $\uptau^1_{B^l_h,B^l_i}$ exists and
                    $\uptau^1_{B^l_h,B^l_i}(v'_1) = G$ with $G \in \Gamma$.
                    We take $\uptau_{B^l_h,B^l_i} = (G, G, G) \circ \uptau^1_{B^l_h, B^l_i}$.

                    \item Let $\uptau_{B_h,B_i}|_{\alpha} = (G, G) \circ \uptau^1_{B_h,B_i}$ and
                    $\uptau_{B_h,B_i}(v) = G$ with $G \in \Delta$.
                    Then $\uptau^1_{B^r_h,B^r_i}$ exists and
                    $\uptau^1_{B^r_h,B^r_i}(v'_1) = G$ with $G \in \Delta$.
                    We take $\uptau_{B^r_h,B^r_i} = (G, G) \circ \uptau^1_{B^r_h, B^r_i}$.

                    \item Let $\uptau_{B_h,B_i}|_{\alpha} = (? F, ? F) \circ \uptau^1_{B_h,B_i}$ and
                    $\uptau_{B_h,B_i}(v) = ? F$ with ancestor the leftmost occurrence of
                    $? F$ after the rule $\rcontr$. Then $\uptau^1_{B^l_h,B^l_i}$ exists and
                    $\uptau^1_{B^l_h,B^l_i}(v'_1) = ? F$.
                    We take $\uptau_{B^l_h,B^l_i} = (? F, ? F, ? F) \circ \uptau^1_{B^l_h, B^l_i}$.
                \end{itemize}

                \item \textbf{Case $(\sigma)$, with $\sigma \in \{\rnu, \rmu\}$:}
                \begin{itemize}
                    \item Let $\uptau_{B_h,B_i}|_{\alpha} = (G, G) \circ \uptau^1_{B_h,B_i}$ and
                    $\uptau_{B_h,B_i}(v) = G$ with $G \in \Gamma$.
                    Then $\uptau^1_{B^l_h,B^l_i}$ exists and
                    $\uptau^1_{B^l_h,B^l_i}(v'_1) = G$ with $G \in \Gamma$.
                    We take $\uptau_{B^l_h,B^l_i} = (G, G, G) \circ \uptau^1_{B^l_h, B^l_i}$.

                    \item Let $\uptau_{B_h,B_i}|_{\alpha} = (G, G) \circ \uptau^1_{B_h,B_i}$ and
                    $\uptau_{B_h,B_i}(v) = G$ with $G \in \Delta$.
                    Then $\uptau^1_{B^r_h,B^r_i}$ exists and
                    $\uptau^1_{B^r_h,B^r_i}(v'_1) = G$ with $G \in \Delta$.
                    We take $\uptau_{B^r_h,B^r_i} = (G, G) \circ \uptau^1_{B^r_h, B^r_i}$.

                    \item Let $\uptau_{B_h,B_i}|_{\alpha} = (\sigma X. F, F[\sigma X. F / X]) \circ \uptau^1_{B_h,B_i}$ and
                    $\uptau_{B_h,B_i}(v) = \sigma X. F$ with ancestor $F[\sigma X. F / X]$
                    after the rule $(\sigma)$. Then $\uptau^1_{B^l_h,B^l_i}$ exists and
                    $\uptau^1_{B^l_h,B^l_i}(v'_1) = F[\sigma X. F / X]$.
                    We take $\uptau_{B^l_h,B^l_i} = (\sigma X. F, \sigma X. F, F[\sigma X. F / X]) \circ \uptau^1_{B^l_h, B^l_i}$.
                \end{itemize}
            \end{itemize} 

            In case $v$ is a companion node and $\uptau_{B^l_h,B^l_i}(v') = G$ (resp. $\uptau_{B^r_h,B^r_i}(v') = G$), then we obtain the new trace by concatenating a new instance of $G$ to the trace, due to the newly introduced rule $\sigma$ (resp. $\sigma^{\bot}$), where $\sigma \in \{\nu, \mu\}$. In all cases, $\uptau_{B_h,B_i}(v) = \uptau_{B^l_h,B^l_i}(v')$ (resp. $\uptau_{B_h,B_i}(v) = \uptau_{B^r_h,B^r_i}(v')$) and $\uptau_{B_h,B_i}(B_i) = \uptau_{B^l_h,B^l_i}(B^l_i)$ (resp. $\uptau_{B_h,B_i}(B_i) = \uptau_{B^r_h,B^r_i}(B^r_i)$).

            \item Assume $C_{B_i} \in Comp(T(v))$. There are two cases: If $v \neq C_{B_i}$, it means that $C_{B_i}$ appears in one of the subtrees $\pi'_1$ or $\pi'_2$. If $C_{B_i}$ appears in $\pi'_k$ with $k \in \{1,2\}$, by induction hypothesis there exists either a trace $\uptau_{B^l_h,B^l_i}$ over the unique path from $C_{B^l_i}$ to $B^l_i$ in $\pi_k^L$ or a trace $\uptau_{B^r_h,B^r_i}$ over the unique path from $C_{B^r_i}$ to $B^r_i$ in $\pi_k^R$, satisfying that $\uptau_{B^l_h,B^l_i}(B^l_i) = \uptau_{B^l_h,B^l_i}(C_{B^l_i})$ and $\uptau_{B^r_h,B^r_i}(B^r_i) = \uptau_{B^r_h,B^r_i}(C_{B^r_i})$. In all rules $(r)$ (could be unary or binary), we have that $\uptau_{B^l_h,B^l_i}$ will appear in $\mathcal{D}^l$ and $\uptau_{B^r_h,B^r_i}$ will appear in $\mathcal{D}^r$, since $\pi^L_k$ and $\pi^R_k$ are not modified in the cut-introduction step that defines $\pi^L$ and $\pi^R$.

            We have proven in the previous point (5.a.i), that there exists a trace $\uptau_{B^l_h, B^l_i}$ over the unique path $v'$ to $B^l_i$ in $\mathcal{D}$, or a trace $\uptau_{B^r_h, B^r_i}$ over the unique path $v'$ to $B^r_i$ in $\mathcal{D}$, such that $\uptau_{B_h,B_i}(v) = \uptau_{B^l_h,B^l_i}(v')$ and $\uptau_{B_h,B_i}(B_i) = \uptau_{B^l_h,B^l_i}(B^l_i)$, or $\uptau_{B_h,B_i}(v) = \uptau_{B^r_h,B^r_i}(v')$ and $\uptau_{B_h,B_i}(B_i) = \uptau_{B^r_h,B^r_i}(B^r_i)$. If $v = C_{B_i}$ then $\mathcal{D}$ is obtained by cutting the $\mathcal{D}^l = \scalebox{.75}{
        \begin{prooftree}
            \hypo{\pi^L[\sigma X_{B_k}. I/X_{B_k}]}
            \infer1[]{\vdash \Gamma, I[\sigma X_{B_k}. I/X_{B_k}]}
            \infer1[$(\sigma)$]{\vdash \Gamma, \sigma X_{B_k}. I}    
        \end{prooftree}
        }$ and $\mathcal{D}^r= \scalebox{.75}{
        \begin{prooftree}
            \hypo{\pi^R[\sigma^{\bot} X_{B_k}.I^{\bot}/X_{B_k}]}
            \infer1[]{\vdash \Delta, I^{\bot}[\sigma^{\bot} X_{B_k}.I^{\bot}/X_{B_k}]}
            \infer1[$(\sigma^{\bot})$]{\vdash \Delta, \sigma^{\bot} X_{B_k}.I^{\bot}}    
        \end{prooftree}
        }$. In this case, $s(C_{B^l_i}) = \ (\vdash \Gamma, \sigma X_{B_k}. I)$ and $s(C_{B^r_i}) = \ (\vdash \Delta, \sigma^{\bot} X_{B_k}.I^{\bot})$. It is clear that the rules $\rcut,  (\sigma), (\sigma^{\bot})$ preserve the context, which implies that $\uptau_{B^l_h,B^l_i}(v') = \uptau_{B^l_h,B^l_i}(C_{B^l_i})$ and $\uptau_{B^r_h,B^r_i}(v') = \uptau_{B^l_h,B^l_i}(C_{B^r_i})$. Moreover, by the property 1 of the ordered trace manifold for $\mathcal{P}_0$, we know that $\uptau_{B_h,B_i}(B_i) = \uptau_{B_h,B_i}(C_{B_i})$. Thus:
        \begin{align*}
           \uptau_{B^l_h,B^l_i}(B^l_i) = \uptau_{B_h,B_i}(B_i) = \uptau_{B_h,B_i}(C_{B_i}) = \uptau_{B^l_h,B^l_i}(v') = \uptau_{B^l_h,B^l_i}(C_{B^l_i}) \\
           \uptau_{B^r_h,B^r_i}(B^r_i) = \uptau_{B_h,B_i}(B_i) = \uptau_{B_h,B_i}(C_{B_i}) = \uptau_{B^r_h,B^r_i}(v') = \uptau_{B^r_h,B^r_i}(C_{B^r_i}) 
        \end{align*}
        Therefore $\uptau_{B^l_h,B^l_i}$ or $\uptau_{B^r_h,B^r_i}$ satisfy (5.a.ii). 
         \end{enumerate}

    \item We implicitly use the property (5.a), to refer to the traces $\uptau_{B^l_h,B^l_i}$ or $\uptau_{B^r_h,B^r_i}$, for all $B_i \in \mathcal{B}$. First we proof the following auxiliary propositions:

    \begin{proposition}
        \label{prop:auxPropCaseb}
         Let $\uptau$ be any trace in the ordered trace manifold for $\mathcal{P}_0$. Then either for every vertex $u \in V_0$ with a split $(\Gamma, \Delta)$, then $\uptau(u) \in \Gamma$ or for every vertex $u \in V_0$ with a split $(\Gamma, \Delta)$, then $\uptau(u) \in \Delta$.
    \begin{proof}
        Done by induction on the derivation tree $\mathcal{D}_0$. It holds since every split and trace are propagated using the ancestor relation.
    \end{proof}
    \end{proposition}
    \begin{proposition}
    \label{prop:auxPropCaseb-2}
        For any set $\{B_{i_1}, ..., B_{i_n}\} \subseteq Bud(T(v))$ such that $\{B_{i_1}, ..., B_{i_m}\} \subseteq \mathcal{B}$. Then either $\uptau_{B^l_h,B^l_{i_1}}, \dots \uptau_{B^l_h,B^l_{i_m}}$ exist, or $\uptau_{B^r_h,B^r_{i_1}}, \dots \uptau_{B^r_h,B^r_{i_m}}$ exist (all traces generated from $\mathcal{B}$ are on the same side of the interpolated pre-proof).
    \begin{proof}
        Let $B_i,B_j \in \mathcal{B}$ (i.e. $B_i,B_j \vartriangleleft_0 B_k$). Note that $\mathcal{B}$ is a set of weakly $\leq_{\mathcal{P}_0}$-connected, so either $B_i \leq_{\mathcal{P}_0} B_j$, $B_j \leq_{\mathcal{P}_0} B_i$ or $B_i, B_j$ are not directly related:

        \begin{itemize}
            \item If $B_i \leq_{\mathcal{P}_0} B_j$ then by property 2 of the ordered trace manifold $\uptau_{B_h, B_i}(C_{B_i}) = \uptau_{B_h, B_j}(C_{B_i})$. Therefore, if $s(C_{B_i})$ is split into $(\Gamma', \Delta')$, then either $\uptau_{B_h, B_i}(C_{B_i}), \uptau_{B_h, B_j}(C_{B_i}) \in \Gamma'$ or $\uptau_{B_h, B_i}(C_{B_i}), \uptau_{B_h, B_j}(C_{B_i}) \in \Delta'$. Therefore, by Proposition \ref{prop:auxPropCaseb} for all nodes $u \in V_0$, $\uptau_{B_h, B_i}(u)$ and $\uptau_{B_h, B_j}(u)$ are on the same side of the split of $u$. This implies that for the root vertex $v$, then either $\uptau_{B_h, B_i}(v), \uptau_{B_h, B_j}(v) \in \Gamma$ or $\uptau_{B_h, B_i}(v), \uptau_{B_h, B_j}(v) \in \Delta$. Finally, based on every case of the construction made in (5.a.i), one can derive that $\uptau_{B_h, B_i}(v), \uptau_{B_h, B_j}(v) \in \Gamma$ if and only if $\uptau_{B^l_h, B^l_i}, \uptau_{B^l_h, B^l_j}$ exist, or respectively $\uptau_{B_h, B_i}(v), \uptau_{B_h, B_j}(v) \in \Delta$ if and only if $\uptau_{B^r_h, B^r_i}, \uptau_{B^r_h, B^r_j}$ exist.
            \item The case $B_j \leq_{\mathcal{P}_0} B_i$ is symmetrical to the previous one.
            \item Assume $B_i$ and $B_j$ are not related. Since $\mathcal{B}$ is weakly $\leq_{\mathcal{P}_0}$-connected, then there is a chain of bud nodes $B'_{l_1}, \dots , B'_{l_{m}}$ in $\mathcal{B}$, where $B'_{l_1} = B_i$ and $B'_{l_m} = B_j$ satisfying that for all $B'_{l_{i}}$ and $B'_{l_{i+1}}$ then $B'_{l_{i}} \leq_{\mathcal{P}_0} B'_{l_{i+1}}$ or $B'_{l_{i+1}} \leq_{\mathcal{P}_0} B'_{l_{i}}$, such that $B'_{l_{i}}, B'_{l_{i+1}} \vartriangleleft_0 B_h$. Therefore, by property 2 of the ordered trace manifold for $\mathcal{P}_0$, either $\uptau_{B_h, B'_{l_{i}}}(C_{B'_{l_{i}}}) = \uptau_{B_h, B'_{l_{i+1}}}(C_{B'_{l_{i}}})$ or $\uptau_{B_h, B'_{l_{i}}}(C_{B'_{l_{i+1}}}) = \uptau_{B_h, B'_{l_{i+1}}}(C_{B'_{l_{i+1}}})$. In both cases, by Proposition \ref{prop:auxPropCaseb} one can deduce that for all nodes $u \in V_0$, $\uptau_{B_h, B'_{l_{i}}}(u)$ and $\uptau_{B_h, B'_{l_{i+1}}}(u)$ are on the same side of the split of $u$. Therefore, by transitivity on the chain $B'_{l_1}, \dots , B'_{l_{m}} \in \mathcal{B}$ we derive that for all nodes $u \in V_0$, $\uptau_{B_h, B_i}(u)$ and $\uptau_{B_h, B_j}(u)$ are on the same side of the split of $u$. We conclude by using the same argument used for the case $B_i \leq_{\mathcal{P}_0} B_j$.
        \end{itemize}
    \end{proof}
    \end{proposition} 

    Now assume $B_i, B_j \in \mathcal{B}$ and $B_i, B_j Bud(T(v))$, where $B_j \leq_{\mathcal{P}_0} B_i$ and $C_{B_j} \in Comp(T(v))$. Note that $B_j \leq_{\mathcal{P}_0} B_i$ implies by Proposition \ref{prop:auxPropCaseb-2} that the traces generated by the clause (5.a) are both on the same side, i.e. either $\uptau_{B^l_k, B^l_j}, \uptau_{B^l_k, B^l_i}$ exist, or $\uptau_{B^r_k, B^r_j}, \uptau_{B^r_k, B^r_i}$ exist. We will assume that $\uptau_{B^l_k, B^l_j}, \uptau_{B^l_k, B^l_i}$ exist. The case for the right side is symmetrical. For $k \in \{1,2\}$, we will refer to the root of $\pi_k$ as $v'_k$, and $v^L_k$ (resp. $v^R_k$) to the root of $\pi^L_k$ (resp. $\pi^R_k$).

    \begin{enumerate}
        \item Assume $C_{B_i} \notin Comp(T(v))$. We have two cases: if $v \neq C_{B_j}$ then $C_{B_j}$ is in one of the subtrees $\pi'_k$. By the induction hypothesis, $C_{B^l_j}$ (resp. $C_{B^r_j}$) is in the unique path from $v'_k$ to $B^l_i$ (resp. $B^r_i$) in $\pi_{k}$, such that $\uptau_{B^l_k, B^l_j}(C_{B^l_j}) = \uptau_{B^l_k, B^l_i}(C_{B^l_j})$. This implies that the unique path $\alpha$ from the root $v^L_k$ (resp. $v^R_k$) of $\pi^L_k$ (resp. $\pi^R_k$) to $B^l_i$ (resp. $B^r_i$) in $\pi^L_k$ (resp. $\pi^R_k$) includes the companion $C_{B^l_j}$ (resp. $C_{B^r_j}$). Furthermore, as shown before for all rule $(r)$, the cut-introduction preserve $\pi^L_k$ (resp. $\pi^R_k$). Therefore, the path from $v'$ to $v^L_k$ (resp. $v^R_k$) composed with $\alpha$ which ends in $B^l_i$ (resp. $B^r_i$), satisfies the property.

        If $v = C_{B_j}$, then we have a path from $v'$ to $C_{B^l_j}$ (resp. $C_{B^r_j}$), and since $C_{B^l_j}$ (resp. $C_{B^r_j}$) is the root of $\mathcal{D}^l$ (resp. $\mathcal{D}^r$), it must be the case that there is a path from $C_{B^l_j}$ (resp. $C_{B^r_j}$) to $B^l_j$ (resp. $B^r_j$). Moreover, by property (5.a) we know that $\uptau_{B_h,B_j}(C_{B_j}) = \uptau_{B^l_h,B^l_j}(v')$ and $\uptau_{B_h,B_i}(C_{B_j}) = \uptau_{B^l_h,B^l_i}(v')$. Since the $\rcut$ rule preserve the contexts, then $\uptau_{B_h,B_j}(C_{B_j}) = \uptau_{B^l_h,B^l_j}(C_{B^l_j})$ and $\uptau_{B_h,B_i}(C_{B_j}) = \uptau_{B^l_h,B^l_i}(C_{B^l_j})$. Finally, by property (2) of the ordered trace manifold for $\mathcal{P}_0$, then $\uptau_{B_h,B_j}(C_{B_j}) =\uptau_{B_h,B_i}(C_{B_j})$, which implies $\uptau_{B^l_h,B^l_j}(C_{B^l_j}) = \uptau_{B^l_h,B^l_i}(C_{B^l_j})$.

        \item Assume $C_{B_i} \in Comp(T(v))$. We have two cases: if $v \neq C_{B_i}$ then $C_{B_i}$ and $C_{B_j}$ are in one of the subtrees $\pi'_k$. By the induction hypothesis, $C_{B^l_j}$ (resp. $C_{B^r_j}$) is in the cycle $\mathcal{C}_{C_{B^l_j}}$ (resp. $\mathcal{C}_{C_{B^r_j}}$) in $\mathcal{P}^l_{\pi_k}$ (resp. $\mathcal{P}^r_{\pi_k}$), which is the left (resp. right) partial pre-proof obtained after the descending phase over $\pi'_k$, such that $\uptau_{B^l_k, B^l_j}(C_{B^l_j}) = \uptau_{B^l_k, B^l_i}(C_{B^l_j})$. Note that the derivation tree for $\mathcal{P}^l_{\pi_k}$ (resp. $\mathcal{P}^r_{\pi_k}$) is $\pi^L_k$ (resp. $\pi^R_k$). As shown before, all the cut-introductions preserve $\pi^L_k$ (resp. $\pi^R_k$), and includes it in $\mathcal{D}^l$ (resp. $\mathcal{D}^r$). Thus, the property also holds for $\mathcal{P}^l$ (resp. $\mathcal{P}^r$).

        If $v = C_{B_i}$ then $C_{B_j}$ is in one of the subtrees $\pi'_k$, since $C_{B_i} \neq C_{B_j}$ because $\mathcal{R}_0$ is injective. By the induction hypothesis, $C_{B^l_j}$ (resp. $C_{B^r_j}$) is in the unique path from $v'_k$ to $B^l_i$ (resp. $B^r_i$) in $\pi_{k}$, such that $\uptau_{B^l_k, B^l_j}(C_{B^l_j}) = \uptau_{B^l_k, B^l_i}(C_{B^l_j})$. This implies that the unique path $\alpha$ from the root $v^L_k$ (resp. $v^R_k$) of $\pi^L_k$ (resp. $\pi^R_k$) to $B^l_i$ (resp. $B^r_i$) in $\pi^L_k$ (resp. $\pi^R_k$) includes the companion $C_{B^l_j}$ (resp. $C_{B^r_j}$). Furthermore, as shown before for all rule $(r)$, the cut-introduction preserve $\pi^L_k$ (resp. $\pi^R_k$). Note that in this case $C_{B^l_i}$ (resp. $C_{B^r_i}$) is the root of $\mathcal{D}^l$ (resp. $\mathcal{D}^r$). Therefore, the path from $C_{B^l_i}$ to $v^L_k$ (resp. $v^R_k$) composed with $\alpha$ which ends in $B^l_i$ (resp. $B^r_i$), which represent the cycle $\mathcal{C}_{B^l_i}$ (resp. $\mathcal{C}_{B^r_i}$), and satisfies the property.
    \end{enumerate}

    \item Let $B_i \in Bud(T(v))$ and $B_i \in \mathcal{B}$. We directly refer to the traces $\uptau_{B^l_h,B^l_i}$ or $\uptau_{B^r_h,B^r_i}$, constructed in the proof of the property (5.a). As before, we refer to the traces $\uptau^1_{B^l_h, B^l_i}$, $\uptau^1_{B^r_h, B^r_i}$, $\uptau^2_{B^l_h, B^l_i}$, $\uptau^2_{B^r_h, B^r_i}$ as the traces obtained by induction hypothesis on (5.a) over the subtrees $\pi^L_1, \pi^R_1, \pi^L_2, \pi^R_2$ respectively. Moreover, we name $\alpha$ the unique path from $v$ to $B_i$ in $\mathcal{D}_0$, and if $B_i$ is in $\pi'_k$ with $k \in \{1,2\}$, then we name $\alpha'_k$ the unique path from $v_k$ to $B_i$ in $\pi'_k$, where $v_k$ is the root of $\pi'_k$. We then define $\uptau_{B_h,B_i}|_{\alpha'_k} = \uptau^k_{B_h,B_i}$. 

    \begin{enumerate}
        \item Let $C_{B_i} \notin Comp(T(v))$ and assume $v$ is not a companion.  By induction hypothesis, we know that $\textsf{Form}(\uptau^k_{B_h,B_i}) = \textsf{Form}(\uptau^k_{B^l_h, B^l_i})$ (resp. $\textsf{Form}(\uptau^k_{B_h,B_i}) = \textsf{Form}(\uptau^k_{B^r_h,B^r_i})$), where $\uptau^k_{B^l_h, B^l_i}$ (resp. $\uptau^k_{B^r_h,B^r_i}$) has an $F$-progress point if $\uptau^k_{B_h,B_i}$ has an $F$-progress point. For every rule $(r)$, we can use this induction hypothesis in every case of the proof for (5.a.i) to prove that $\textsf{Form}(\uptau_{B_h,B_i}|_{\alpha}) = \textsf{Form}(\uptau_{B^l_h,B^l_i})$ or $\textsf{Form}(\uptau_{B_h,B_i}|_{\alpha}) = \textsf{Form}(\uptau_{B^r_h,B^r_i})$ holds, depending on whether $\uptau_{B^l_h,B^l_i}$ or $\uptau_{B^r_h,B^r_i}$ exists. If $\uptau_{B_h,B_i}|_{\alpha}$ has a progress point, there are two cases:
        \begin{itemize}
            \item If the progress point is inside $\uptau^k_{B_h,B_i}$, then $\uptau_{B^l_h,B^l_i}$ (resp. $\uptau_{B^r_h,B^r_i}$) also has a progress point since it contains the trace $\uptau^k_{B^l_h, B^l_i}$ (resp. $\uptau^k_{B^r_h, B^r_i}$) for all rule $(r)$.

            \item If the progress point is in $v$, i.e. $(r) = \nu$ and $\uptau_{B_h,B_i}(v)$ is the principal formula $F = \nu X. F'$, by observation on the case $\sigma \in \{\nu,\mu\}$ in the proof for (5.a.i), then this progress point is also included in $\uptau_{B^l_h,B^l_i}$ (resp. $\uptau_{B^r_h,B^r_i}$).
        \end{itemize}
        
        In case $v$ is a companion, the property still is preserved since the rules $\rcut, (\nu), (\mu)$ preserve the context, therefore $\textsf{Form}(\uptau_{B^l_h,B^l_i})$ (resp. $\textsf{Form}(\uptau_{B^r_h,B^r_i})$) reamin the same.

        \item Let $C_{B_i} \in Comp(T(v))$ then there are two cases: if $v \neq C_{B_i}$, then $C_{B_i}$ is in one of the subtrees $\pi'_k$. This case follows directly from the induction hypothesis. 

        If $v = C_{B_i}$, we have already proven in the previous item that $\textsf{Form}(\uptau_{B_h,B_i}|_{\alpha}) = \textsf{Form}(\uptau_{B^l_h,B^l_i})$ (resp. $\textsf{Form}(\uptau_{B_h,B_i}|_{\alpha}) = \textsf{Form}(\uptau_{B^r_h,B^r_i})$), and that $F$-progress points are preserved. However, note that since $v = C_{B_i}$, then $\uptau_{B_h,B_i}|_{\alpha} = \uptau_{B_h,B_i}$. Therefore, the property is satisfied.
    \end{enumerate}
    \end{enumerate} 
\end{enumerate}

With this we conclude the proof of Proposition \ref{prop:structInterpol}.
\end{proof}

%=====================================================================
%=====================================================================
%=====================================================================
%=====================================================================

Given these invariants described in Proposition \ref{prop:structInterpol}, it is possible to deduce the final structure of the pre-proofs $\mathcal{P}^l$ and $\mathcal{P}^r$, which is stated in the Lemma \ref{lemm:finInterpol}. Before defining the lemma, we include the following proposition which generates maximal elements of weakly $\leq_{\mathcal{P}}$-connected set of bud nodes, based on tree-compatibility.

\begin{proposition}
    \label{prop:closestCompanion}
    Let $\mathcal{P} = (\mathcal{D}, \mathcal{R})$ be a \muLLo pre-proof with injective $\mathcal{R}$, let $\vartriangleleft$ be a tree-compatible induction order for $\mathcal{P}$, and let $\mathcal{B} \subseteq Bud(\mathcal{D})$ be a weakly $\leq_{\mathcal{P}}$-connected set of bud nodes. Let $B_i \in \mathcal{B}$ be such that the companion $C_{B_i}$ is closest to the root among the companions in $\mathcal{B}$, i.e. for all $B_j \in \mathcal{B}$, then $C_{B_j}$ does not appear in the path from the root of $\mathcal{D}$ to $C_{B_i}$. Then $B_i$ is a $\vartriangleleft$ maximum for $\mathcal{B}$.
\end{proposition}
\begin{proof}
    We define the sets $\mathcal{L}_0 = \{B_i\}$ and $\mathcal{L}_{n+1} = \mathcal{L}_n \cup \{B \in \mathcal{B} \ | \ B \leq_{\mathcal{P}} B' \text{ or } B' \leq_{\mathcal{P}} B, \text{ for some } B' \in \mathcal{L}_n\}$. We prove by induction on $n$ that for all $B \in \mathcal{L}_n$, then $B \vartriangleleft B_i$, and moreover either $C_B$ appears on the basic cycle $\mathcal{C}_{B_i}$, or $B \leq_{\mathcal{P}} B'$ for some $B' \in \mathcal{L}_{n-1}$. If $n = 0$, then $B = B_i \vartriangleleft B_i$ and $C_{B_i}$ appears on $\mathcal{C}_{B_i}$. If $n > 0$, take $B \in \mathcal{L}_{n} \setminus \mathcal{L}_{n-1}$, related to some $B' \in \mathcal{L}_{n-1}$, where $B' \vartriangleleft B_i$ by the induction hypothesis. We have two cases:

    \begin{itemize}
        \item If $B \leq_{\mathcal{P}} B'$, then by tree-compatibility $B \vartriangleleft B'$, and by transitivity $B \vartriangleleft B_i$. In this case, since $B' \in \mathcal{L}_{n-1}$ the second part of the hypothesis is satisfied.

        \item If $B' \leq_{\mathcal{P}} B$, by multiple applications of the second part of the induction hypothesis, we can form a chain $B' \leq_{\mathcal{P}} B_1 \leq_{\mathcal{P}} B_2 \dots \leq_{\mathcal{P}} B_m$ such that $B_m \leq_{\mathcal{P}} B_i$, and each element in the chain is from a deeper level. Then, $B_{m-1} \leq_{\mathcal{P}} B_m$ imply that $C_{B_{m-1}}$ appears in the path from $C_{B_m}$ to $B_m$, and $B_m \leq_{\mathcal{P}} B_i$ imply that $C_{B_m}$ appears in the path from $C_{B_i}$ to $B_i$. Therefore $B_{m-1} \leq_{\mathcal{P}} B_i$. We can repeat this argument, until reaching $B' \leq_{\mathcal{P}} B_i$. Again, this implies that $C_B'$ is in the path from $C_{B_i}$ to $B_i$. Moreover, since $B' \leq_{\mathcal{P}} B$ then $C_{B'}$ appears in the path from $C_B$ to $B$. However, note that $C_{B_i}$ is the closest companion to the root, for this reason, it must be the case that $B \leq_{\mathcal{P}} B_i$. Then, by tree-compatibility $B \vartriangleleft B_i$.
    \end{itemize}

    Finally note that for some $n$, $\mathcal{B} = \mathcal{L}_n$, hence $B_i$ is a $\vartriangleleft$ maximum for $\mathcal{B}$
\end{proof}

\begin{lemma}
    \label{lemm:finInterpol}
    The ascending and descending phase over $\mathcal{P}_0$ produces two pre-proofs $\mathcal{P}^l = (\mathcal{D}^l, \mathcal{R}^l)$ and $\mathcal{P}^r = (\mathcal{D}^r, \mathcal{R}^r)$ with conclusions $\vdash \Gamma, I$ and $\vdash \Delta, I^{\bot}$, such that $\mathcal{P} = \adjustbox{valign=c, scale=0.75}{
            \begin{prooftree}
                    \hypo{\mathcal{P}^l}
                    \infer1[]{\vdash \Gamma, I}
                    \hypo{\mathcal{P}^r}
                    \infer1[]{\vdash \Delta, I^{\bot}}
                    \infer2[\textsf{(Cut)}]{\vdash \Gamma, \Delta}
                \end{prooftree} 
            }$ and satisfying:
    \begin{enumerate}
        \item $Bud(\mathcal{D}_0) = \{B_1, \dots, B_n\}$ if and only if $Bud(\mathcal{D}^l) = \{B^l_1, \dots, B^l_n\}$ and $Bud(\mathcal{D}^r) = \{B^r_1, \dots, B^r_n\}$.

        \item $I$ is a $\muLLo$ formula such that $\VOC{I}\subseteq (\VOC {\Gamma} \cup \mathbb C^L) \cap (\VOC {\Delta}\cup \mathbb C^R)$, and every fixed-point variable in $I$ is bounded.

        \item $\mathcal{R}^l$ and $\mathcal{R}^r$ are total and injective bud-companion relations. We define $C_{B^l_i} = \mathcal{R}^l(B^l_i)$ (resp. $C_{B^r_i} = \mathcal{R}^l(B^r_i)$).

        \item For all $B_1, B_2 \in Bud(\mathcal{D}_0)$, then $B_1 \leq_{\mathcal{P}_0} B_2$ if and only if $B^l_1 \leq_{\mathcal{P}^l} B^l_2$ and $B^r_1 \leq_{\mathcal{P}^r} B^r_2$. 

        \item Let $(Bud(\mathcal{D}^l), \vartriangleleft_l)$ and $(Bud(\mathcal{D}^r), \vartriangleleft_r)$ be defined as:
        \begin{align*}
            B^l_i \vartriangleleft_l B^l_j \iff B_i \vartriangleleft_0 B_j \ \text{and} \ B^r_i \vartriangleleft_r B^r_j \iff B_i \vartriangleleft_0 B_j
        \end{align*}
        If $(Bud(\mathcal{D}_0), \vartriangleleft_0)$ is a tree-compatible induction order for $\mathcal{P}_0$, then $(Bud(\mathcal{D}^l), \vartriangleleft_l)$ and $(Bud(\mathcal{D}^r), \vartriangleleft_r)$ are tree-compatible induction orders for $\mathcal{P}_l$ and $\mathcal{P}_r$ respectively.

        \item $\mathcal{B} = \{B_1, \dots B_n\} \subseteq Bud(\mathcal{D}_0)$ is the weakly $\leq_{\mathcal{P}_0}$-connected set of all bud nodes such that $B_k \in Bud(\mathcal{D}_0)$ is a $\vartriangleleft_0$ maximum for $\mathcal{B}$ if and only if $\mathcal{B}^l = \{B^l_1, \dots B^l_n\} \subseteq Bud(\mathcal{D}^l)$ (resp. $\mathcal{B}^r = \{B^r_1, \dots B^r_n\} \subseteq Bud(\mathcal{D}^r)$) is the weakly $\leq_{\mathcal{P}^l}$-connected (resp. $\leq_{\mathcal{P}^r}$-connected) set of all bud nodes such that $B^l_k \in Bud(\mathcal{D}^l)$ (resp. $B^r_k \in Bud(\mathcal{D}^r)$) is a $\vartriangleleft_l$ (resp. $\vartriangleleft_r$) maximum for $\mathcal{B}^l$ (resp. $\mathcal{B}^r$). 

        \item Let $\mathcal{B}^l = \{B^l_1, \dots B^l_n\}$ and $\mathcal{B}^r = \{B^r_1, \dots B^r_n\}$ be the sets described in the item (6). Then either the traces $\uptau_{B^l_k, B^l_1} \dots \uptau_{B^l_k, B^l_n}$ or $\uptau_{B^r_k, B^r_1} \dots \uptau_{B^r_k, B^r_n}$ exist, satisfying:
        \begin{enumerate}
            \item For all $B^l_i \in \mathcal{B}^l$ (resp. $B^r_i \in \mathcal{B}^r$), then $\uptau_{B^l_j,B^l_i}$ (resp. $\uptau_{B^r_j,B^r_i}$) is a trace over the unique path from $C_{B^l_i}$ (resp. $C_{B^r_i}$) to $B^l_i$ (resp. $B^r_i$) in $\mathcal{D}^l$ (resp. $\mathcal{D}^r$), such that $\uptau_{B^l_k,B^l_i}(C_{B^l_i}) = \uptau_{B^l_k,B^l_i}(B^l_i)$ (resp. $\uptau_{B^r_k,B^r_i}(C_{B^r_i}) = \uptau_{B^r_k,B^r_i}(B^r_i)$).

            \item Assume $B^l_i,B^l_j \vartriangleleft_l B^l_k$ (resp. $B^r_i,B^r_j \vartriangleleft_r B^r_k$) with $B^l_j \leq_{\mathcal{P}^l} B^l_i$ (resp. $B^r_j \leq_{\mathcal{P}^r} B^r_i$). Then $\uptau_{B^l_k,B^l_i}(C_{B^l_j}) = \uptau_{B^l_k,B^l_j}(C_{B^l_j})$ (resp. $\uptau_{B^r_k,B^r_i}(C_{B^r_j}) = \uptau_{B^r_k,B^r_j}(C_{B^r_j})$).

            \item $\uptau_{B^l_k,B^l_k}$ (resp. $\uptau_{B^r_k,B^r_k}$) has at least one $F$-progress point and $F$ is minimal in $\bigcup^{n}_{i=1} \textsf{Form}(\uptau_{B^l_k,B^l_i})$ (resp. $\bigcup^{n}_{i=1} \textsf{Form}(\uptau_{B^r_k,B^r_i})$) with respect to the subformula relation.
        \end{enumerate}

        \item Let $\mathcal{B}^l = \{B^l_1, \dots B^l_n\}$ and
        $\mathcal{B}^r = \{B^r_1, \dots B^r_n\}$ be the sets described in the
        item (6). Then the interpolant traces
        $\uprho_{B^l_1}, \dots, \uprho_{B^l_n}$ and
        $\uprho_{B^r_1}, \dots, \uprho_{B^r_n}$ exist, satisfying, for any choice
        of the binders $\sigma_i \in \{\mu, \nu\}$ performed at the closure steps
        of the descending phase:
        \begin{enumerate}
            \item If $B^l_k \in \mathcal{B}^l$ (resp. $B^r_k \in \mathcal{B}^r$) is closest to the root among the companions in $\mathcal{B}^l$ (resp. $\mathcal{B}^r$), then $B^l_i$ (resp. $B^r_i$) is a $\vartriangleleft_l$ (resp. $\vartriangleleft_r$) maximum for $\mathcal{B}^l$ (resp. $\mathcal{B}^r$).

            \item For all $B^l_i \in \mathcal{B}^l$ (resp.
            $B^r_i \in \mathcal{B}^r$), then $\uprho_{B^l_i}$ (resp.
            $\uprho_{B^r_i}$) is a trace over the unique path from $C_{B^l_i}$
            (resp. $C_{B^r_i}$) to $B^l_i$ (resp. $B^r_i$) in $\mathcal{D}^l$
            (resp. $\mathcal{D}^r$), such that
            $\uprho_{B^l_i}(C_{B^l_i}) = \uprho_{B^l_i}(B^l_i) =
            \sigma_i X_{B_i}. I_i$ (resp.
            $\uprho_{B^r_i}(C_{B^r_i}) = \uprho_{B^r_i}(B^r_i) =
            \sigma_i^{\perp} X_{B_i}. I_i^{\perp}$), and the step of
            $\uprho_{B^l_i}$ (resp. $\uprho_{B^r_i}$) out of $C_{B^l_i}$ (resp.
            $C_{B^r_i}$) is the Fischer-Ladner unfolding of the $\sigma_i$ (resp. $\sigma^\perp_i$) formula. 
            This step is an $F$-progress point with $F = \sigma_i X_{B_i}. I_i$
            (resp. $F = \sigma_i^{\perp} X_{B_i}. I_i^{\perp}$) if and only if
            $\sigma_i = \nu$ (resp. $\sigma_i = \mu$).

            \item Assume $B^l_i,B^l_j \vartriangleleft_l B^l_k$ (resp. $B^r_i,B^r_j \vartriangleleft_r B^r_k$) with $B^l_j \leq_{\mathcal{P}^l} B^l_i$ (resp. $B^r_j \leq_{\mathcal{P}^r} B^r_i$). Then $\uprho_{B^l_i}(C_{B^l_j}) = \uprho_{B^l_j}(C_{B^l_j})$ (resp.
            $\uprho_{B^r_i}(C_{B^r_j}) = \uprho_{B^r_j}(C_{B^r_j})$).

            \item Consider $B^l_k \in \mathcal{B}^l$ (resp. $B^r_k \in \mathcal{B}^r$) as in the item (a). Then $\uprho_{B^l_k}(C_{B^l_k}) =\sigma_k X_{B_k}. I_k$ (resp. $\uprho_{B^r_k}(C_{B^r_k}) = \sigma_k^{\perp} X_{B_k}. I_k^{\perp}$) is minimal in $\bigcup^{n}_{i=1} \textsf{Form}(\uprho_{B^l_i})$ (resp. $\bigcup^{n}_{i=1} \textsf{Form}(\uprho_{B^r_i})$) with respect to the subformula relation.
        \end{enumerate}
    \end{enumerate}
\end{lemma}

\begin{proof}
    \begin{enumerate}
        \item It follows from Proposition \ref{prop:structInterpol} (1), over the derivation tree $\mathcal{D}_0$.
        
        \item It follows from \ref{prop:structInterpol} (2) over the derivation tree $\mathcal{D}_0$.
        
        \item It follows from Proposition \ref{prop:structInterpol} (3) over the derivation tree $\mathcal{D}_0$.
        
        \item Let $B_1, B_2 \in Bud(\mathcal{D}_0)$. Then: 
        \begin{align*}
         B_1 \leq_{\mathcal{P}_0} B_2  \overset{\text{Def \ref{def:structConnect}}}{\iff}  C_{B_2} \preceq_{\mathcal{D}_0} C_{B_1} \preceq_{\mathcal{D}_0} B_2  \overset{\text{Prop \ref{prop:structInterpol}.4}}{\iff} C_{B^l_2} \preceq_{\mathcal{D}^l} C_{B^l_1} \preceq_{\mathcal{D}^l} B^l_2 \overset{\text{Def \ref{def:structConnect}}}{\iff} B^l_1 \leq_{\mathcal{P}^l} B^l_2.
        \end{align*}
        We can use the same argument to derive $B^r_1 \leq_{\mathcal{P}^r} B^r_2$.

        \item Assume $(Bud(\mathcal{D}_0), \vartriangleleft_0)$ is a tree-compatible induction order. By definition, $\vartriangleleft_l$ and $\vartriangleleft_r$ are also induction orders since they are exact copies of $\vartriangleleft_0$, which is an induction order. Moreover, since the relations $\leq_{\mathcal{P}^l}$ and $\leq_{\mathcal{P}^r}$ are exact copies of $\leq_{\mathcal{P}_0}$ by the item (4), where $\mathcal{R}^l$ and $\mathcal{R}^r$ are injective by the item (3), then tree-compatibility is preserved.

        \item Follows directly from the fact that $\leq_{\mathcal{P}^l}$ and $\leq_{\mathcal{P}^r}$ are exact copies of $\leq_{\mathcal{P}_0}$ by the item (4), and that $\vartriangleleft_l$ and $\vartriangleleft_r$ are also copies of $\vartriangleleft_0$.

        \item The fact that either the traces $\uptau_{B^l_k, B^l_1} \dots \uptau_{B^l_k, B^l_n}$ or $\uptau_{B^r_k, B^r_1} \dots \uptau_{B^r_k, B^r_n}$ exist, follows from Proposition \ref{prop:auxPropCaseb-2}. The properties (a) and (b) are direct consequence of Proposition \ref{prop:structInterpol}.5.a.ii and \ref{prop:structInterpol}.5.b.ii respectively. Note that $\uptau_{B_k, B_k}$ has an $F$-progress point such that $F$ is minimal in $\bigcup^{n}_{j=1} \textsf{Form}(\uptau_{B_k,B_{i_j}})$ with respect to the subformula relation in the original ordered trace manifold. Therefore, by Proposition \ref{prop:structInterpol}.5.c.ii we derive property (c).

        \item Property (a) is a direct application of Proposition \ref{prop:closestCompanion}. The property (b) follows directly from Proposition \ref{prop:structInterpol}.6.a.ii and \ref{prop:structInterpol}.6.c. For property (c) we have that $\uptau_{B^l_k,B^l_i}(C_{B^l_j}) = \sigma_j X_{B_j}. I_j = \uptau_{B^l_k,B^l_i}(C_{B^l_i})$ (resp. $\uptau_{B^r_k,B^r_i}(C_{B^r_j}) = \sigma^\perp_j X_{B_j}. I^\perp_j = \uptau_{B^r_k,B^r_i}(C_{B^r_i})$) by Proposition \ref{prop:structInterpol}.6.b and \ref{prop:structInterpol}.6.c. For the proof of property (d) we have:

        \begin{itemize}
            \item We prove the left case since the right case is symmetrical. First,
            note that by Proposition \ref{prop:structInterpol}.6.b, every formula in
            $\bigcup^{n}_{i=1} \textsf{Form}(\uprho_{B^l_i})$ is either a $\sigma$-formula
            $\sigma_j X_{B_j}. I_j$, taken at a companion node $C_{B^l_j}$ on
            the path followed by some $\uprho_{B^l_i}$, or a 
            decomposition of a formula into its subformulas, containing
            $\sigma$-formula $\sigma_i X_{B_i}. I_i$ as a strict subformula. In the first case
            $B_j \in \mathcal{B}$, since if $C_{B^l_j}$ is on the path of
            $\uprho_{B^l_i}$, it implies $B_j \leq_{\mathcal{P}_0} B_i$ by the item
            (4), and since $B_i \vartriangleleft_0 B_k$, by tree-compatibility
            and transitivity then $B_j \vartriangleleft_0 B_k$.

            Therefore, it suffices to show that no $\sigma$ formula
            $\sigma_j X_{B_j}. I_j$, where $B_j \in \mathcal{B}$ is a strict
            subformula of $\sigma_k X_{B_k}. I_k$. Assume then that
            $\sigma_j X_{B_j}. I_j$ is a strict subformula of
            $\sigma_k X_{B_k}. I_k$, for some $B_j \in \mathcal{B}$. By the
            property Proposition \ref{prop:structInterpol}.6.d, this
            implies $C_{B^l_j} \preceq_{\mathcal{D}^l} C_{B^l_k}$ with
            $C_{B^l_j} \neq C_{B^l_k}$, i.e. the companion $C_{B^l_j}$ is
            strictly closer to the root than $C_{B^l_k}$. This contradicts the
            item (a), since $C_{B^l_k}$ is closest to the root among the
            companions in $\mathcal{B}^l$. Therefore,
            $\sigma_k X_{B_k}. I_k$ is minimal.
    \end{itemize}

    \end{enumerate}
\end{proof}

Intuitively, the construction of the ordered trace manifolds for
$\mathcal{P}^l$ and $\mathcal{P}^r$ can be summarized in the following
series of steps:
\begin{enumerate}
    \item After the descending phase, the structure of
    $\mathcal{P}_0$ is duplicated on both sides. By
    Lemma \ref{lemm:finInterpol}.1, \ref{lemm:finInterpol}.3 and \ref{lemm:finInterpol}.5, every bud node,
    companion node and basic cycle of $\mathcal{P}_0$ has a copy in
    $\mathcal{P}^l$ and a copy in $\mathcal{P}^r$, and by
    Lemma \ref{lemm:finInterpol}.4, the structural connectivity
    $\leq_{\mathcal{P}_0}$ and the induction order $\vartriangleleft_0$
    are transported to both sides. In particular, both $\mathcal{P}^l$
    and $\mathcal{P}^r$ must validate a copy of every cycle of
    $\mathcal{P}_0$.

    \item However, the traces of the original ordered trace manifold are not duplicated.
    This happens since each trace lies entirely on one
    side of the split (Proposition~\ref{prop:auxPropCaseb}). More precisely, by
    Lemma~\ref{lemm:finInterpol}.6, for each maximal weakly
    $\leq_{\mathcal{P}_0}$-connected set $\mathcal{B}$ with the
    associated sets of copies $\mathcal{B}^l$ and $\mathcal{B}^r$, then the
    traces related to $\mathcal{B}$ in the original manifold are
    transferred as traces over the cycles of $\mathcal{B}^l$ in
    $\mathcal{P}^l$, or over the cycles of $\mathcal{B}^r$ in
    $\mathcal{P}^r$, but not both. These traces satisfy the conditions
    of ordered trace manifolds on the side that includes them.

    \item The cycles on the side that did not receive the
    transferred traces of $\mathcal{B}$ must therefore be validated by
    other means: the interpolant traces
    $\uprho_{B^l_i}, \uprho_{B^r_i}$ of
    Lemma~\ref{lemm:finInterpol}.8, which exist on both sides. Since the unfolding step
    at a companion node is a progress point exactly when the binder is
    $\nu$ on that side, choosing each $\sigma_i$ opposite to the
    side receiving the transferred traces of the original ordered trace manifold makes the interpolant traces progress precisely on
    the unsupported side. The minimality condition then follows from
    tree-compatibility, through
    Proposition~\ref{prop:closestCompanion} and Lemma \ref{lemm:finInterpol}.8.d
\end{enumerate}
This procedure is formalized in the following theorem:

\begin{theorem}[Weak interpolation for \muLLo]
    \label{the:preFinal0}
    Let $\mathcal{P}_0 = (\mathcal{D}_0, \mathcal{R}_0)$ be a $\mu\textsf{LL}^{\omega}$ proof with a split conclusion $\vdash \Gamma, \Delta$ and a trace manifold. Moreover, let $\calC= \{(C_1, C_1^\perp), \dots, (C_n, C_n^\perp)\}$ be the cut formulas of $\mathcal{D}_0$, let $L,R$ be any partitioning of $\{1,\dots,n\}$ and $\mathbb C^L = \VOC{C_i, i\in L}$ and  $\mathbb C^R = \VOC{C_i, i\in R}$. After the ascending and descending phase over $\mathcal{P}_0$, we obtain an interpolated pre-proof $\mathcal{P} = \adjustbox{valign=c, scale=0.75}{
        \begin{prooftree}
                \hypo{\mathcal{P}^l}
                \infer1[]{\vdash \Gamma, I}
                \hypo{\mathcal{P}^r}
                \infer1[]{\vdash \Delta, I^{\bot}}
                \infer2[\textsf{(Cut)}]{\vdash \Gamma, \Delta}
            \end{prooftree} 
        }$ satisfying:
\begin{enumerate}
    \item $\mathcal{P}^l = (\mathcal{D}^l, \mathcal{R}^l)$ and $\mathcal{P}^r = (\mathcal{D}^r, \mathcal{R}^r)$ are two pre-proofs with conclusions $\vdash \Gamma, I$ and $\vdash \Delta, I^{\bot}$ respectively, where $\mathcal{R}^l$ (resp. $\mathcal{R}^r$) is injective and $\VOC{I}\subseteq (\VOC {\Gamma} \cup \mathbb C^L) \cap (\VOC {\Delta}\cup \mathbb C^R)$.
    \item There exists an induction order, and an ordered trace manifold for $\mathcal{P}^l$ and $\mathcal{P}^r$ respectively, implying that $\mathcal{P}$ is a valid proof by Lemmas \ref{lemm:traceManifold} and \ref{lemm:ordTraceManifold}. 
    \item $\mathcal{P} \rightarrow^{\omega}_{\rcut} \mathcal{P}_0$.
\end{enumerate}
\end{theorem}

\begin{proof}
    Without loss of generality, we may assume that the bud-companion relations are always injective as explained in the Appendix \ref{appendix:injectiveRel}. Firstly, we transform the trace manifold for $\mathcal{P}_0$ into an ordered trace manifold using the Lemma \ref{lemm:ordTraceManifold}. Then using Algorithm \ref{algo:splitInv} and \ref{algo:treecomp}, we obtain a new proof $\mathcal{P}'_0 = (\mathcal{D}'_0, \mathcal{R}'_0)$ that can be coherently decorated, and has an ordered trace manifold with respect to an induction order $(Bud(\mathcal{D}'_0), \vartriangleleft'_0)$ that is tree-compatible. Therefore, after the ascending and descending phase over $\mathcal{P}'_0$ we obtain an interpolated pre-proof $\mathcal{P} = \adjustbox{valign=c, scale=0.75}{
        \begin{prooftree}
                \hypo{\mathcal{P}^l}
                \infer1[]{\vdash \Gamma, I}
                \hypo{\mathcal{P}^r}
                \infer1[]{\vdash \Delta, I^{\bot}}
                \infer2[\textsf{(Cut)}]{\vdash \Gamma, \Delta}
            \end{prooftree} 
        }$ such that: 
    \begin{enumerate}
        \item By Lemma \ref{lemm:finInterpol}.2 and \ref{lemm:finInterpol}.3, the property (1) is satisfied.
        
        \item Consider the tree-compatible induction orders $(Bud(\mathcal{D}^l), \vartriangleleft^l)$ and $(Bud(\mathcal{D}^r), \vartriangleleft^r)$ for the pre-proofs $\mathcal{P}^l$ and $\mathcal{P}^r$ respectively, given by the Lemma \ref{lemm:finInterpol}.5. We will build an ordered trace manifold for $\mathcal{P}^l$ and $\mathcal{P}^r$ as follows: Consider the bud nodes $B^l_k \in Bud(\mathcal{D}^l)$ and $B^r_k \in Bud(\mathcal{D}^r)$, given by the Lemma \ref{lemm:finInterpol}.1. Now, consider the sets of bud nodes $\mathcal{B}^l = \{B^l \ | \ B^l \in Bud(\mathcal{D}^l) \ \text{and} \ B^l \vartriangleleft^l B^l_k \}$ and $\mathcal{B}^r = \{B^r \ | \ B^r \in Bud(\mathcal{D}^r) \ \text{and} \ B^r \vartriangleleft^r B^r_k \}$, given by the Lemma \ref{lemm:finInterpol}.6. By Lemma \ref{lemm:finInterpol}.7, either for all $B^l_i \in \mathcal{B}^l$ then $\uptau_{B^l_k,B^l_i}$ exists, or for all $B^r_i \in \mathcal{B}^r$ then $\uptau_{B^r_k,B^r_i}$ exists. Assume the first case (the other scenario is symmetrical). Therefore, all the traces associated to $B^l_k$ for the trace manifold in $\mathcal{P}^l$ are covered, and satisfy the ordered trace manifold properties by Lemma \ref{lemm:finInterpol}.7. It remains to be shown that we can construct the traces for $\mathcal{P}^r$ using the interpolant.

        We set $\uptau_{B^r_k,B^r_i} = \uprho_{B^r_i}$, for all $B^r_i \in \mathcal{B}^r$ as specified in Lemma \ref{lemm:finInterpol}.8. We check that these traces satisfy the three properties of the ordered trace manifolds. Properties (1) and (2), they are derived directly from Lemma \ref{lemm:finInterpol}.8.b and Lemma \ref{lemm:finInterpol}.8.c. We now check property (3). First, assume $B^r_h \in \mathcal{B}^r$, has the closest companion to the root of $\mathcal{D}^r$ in the set $\mathcal{B}^r$. By Proposition \ref{prop:closestCompanion}, then $B^r_h$ is a $\vartriangleleft_r$ maximum for $\mathcal{B}^r$. However, this implies that $B^r_h \vartriangleleft_r B^r_k$ and $B^r_k \vartriangleleft_r B^r_h$. Thus, by antisymmetry $B^r_h = B^r_k$, which implies that $B^r_k$ has the closest companion to the root of $\mathcal{D}^r$ in the set $\mathcal{B}^r$. Therefore, $\uprho_{B^r_k}(C_{B^r_k}) = \sigma_k^{\perp} X_{B_k}. I_k^{\perp}$ is minimal in $\bigcup_{B^r_i \in \mathcal{B}^r} \textsf{Form}(\uprho_{B^r_i})$ with respect to the subformula relation by Lemma \ref{lemm:finInterpol}.8.d. Finally, by Lemma \ref{lemm:finInterpol}.8.b if we pick $\sigma_k^{\perp} = \nu$, then $\uprho_{B^r_k}$ has a progress point, which completes the proof of property (3).

        \item For the last part of the theorem one can follow the interpolation proof-sequence backward. \Cref{remark:cut-intro-circ}.(i) ensures that it consists in a cut-elimination reduction sequence while \Cref{remark:cut-intro-circ}.(ii) ensures that the cut-reduction can be performed by unfolding the circular proofs and preserving the synchronisation of the two circular proofs so that at the limit, the proof object is regular and is the infinite unfolding of the original circular proof $\mathcal{P}_0$.
    \end{enumerate}
\end{proof}

    As done in the Section \ref{sec:interpolate-cuts}, if we restrict the vocabulary of the cut formulas in the pre-proof $\mathcal{P}_0$, we obtain an interpolation theorem with cuts.

\begin{theorem}[Interpolation theorem with cuts for \muLLo]
    \label{the:preFinal}
    Let $\mathcal{P}_0 = (\mathcal{D}_0, \mathcal{R}_0)$ be a $\mu\textsf{LL}^{\omega}$ proof with a split conclusion $\vdash \Gamma, \Delta$ and a trace manifold. Moreover, let $\calC= \{(C_1, C_1^\perp), \dots, (C_n, C_n^\perp)\}$ be the cut formulas of $\mathcal{D}_0$, let $L,R$ be any partitioning of $\{1,\dots,n\}$ and $\mathbb C^L = \VOC{C_i, i\in L}$ and  $\mathbb C^R = \VOC{C_i, i\in R}$, such that $(\mathbb C^L\setminus \VOC \Gamma) \cap (\VOC \Delta \cup \mathbb C^R)=\emptyset$ and $(\mathbb C^R\setminus \VOC \Delta) \cap (\VOC \Gamma \cup \mathbb C^L)=\emptyset$. After the ascending and descending phase over $\mathcal{P}_0$, we obtain an interpolated pre-proof $\mathcal{P} = \adjustbox{valign=c, scale=0.75}{
        \begin{prooftree}
                \hypo{\mathcal{P}^l}
                \infer1[]{\vdash \Gamma, I}
                \hypo{\mathcal{P}^r}
                \infer1[]{\vdash \Delta, I^{\bot}}
                \infer2[\textsf{(Cut)}]{\vdash \Gamma, \Delta}
            \end{prooftree} 
        }$ satisfying:
\begin{enumerate}
    \item $\mathcal{P}^l = (\mathcal{D}^l, \mathcal{R}^l)$ and $\mathcal{P}^r = (\mathcal{D}^r, \mathcal{R}^r)$ are two pre-proofs with conclusions $\vdash \Gamma, I$ and $\vdash \Delta, I^{\bot}$ respectively, where $\mathcal{R}^l$ (resp. $\mathcal{R}^r$) is injective and $\voc{(I)}\subseteq \voc {(\Gamma)}\cap \voc {(\Delta)}$.
    \item There exists an induction order, and an ordered trace manifold for $\mathcal{P}^l$ and $\mathcal{P}^r$ respectively, implying that $\mathcal{P}$ is a valid proof by Lemmas \ref{lemm:traceManifold} and \ref{lemm:ordTraceManifold}. 
    \item $\mathcal{P} \rightarrow^{\omega}_{\rcut} \mathcal{P}_0$.
\end{enumerate}
\end{theorem}

In the end, Theorem \ref{the:main} is derived as a corollary of Theorem \ref{the:preFinal}. 

%===========================================================
%===========================================================
%===========================================================
%===========================================================

\end{document}